\documentclass[useAMS,usenatbib]{mn2e}
\usepackage{graphicx}
\title[X-ray binaries: optical/X-ray cross correlations]{High time resolution optical/X-ray cross-correlations for X-ray binaries: anti-correlations and rapid variability}
\author[Durant et al.]{
\parbox{7in}{Martin Durant$^{1,2}$\thanks{E-mail:
durant@astro.ufl.edu}, Tariq Shahbaz$^{1,3}$, Poshak Gandhi$^5$, Remon Cornelisse$^1$,\\  Teodoro Mu\~noz-Darias$^{1, 4}$, Jorge Casares$^1$, 
Vik Dhillon$^7$, Tom Marsh$^8$, \\ Hendrik Spruit$^6$,  Kieran O'Brien$^9$, Danny Steeghs$^8$ and Rob Hynes$^{10}$}\\
~ \\
$^{1}$ Instituto de Astrof\'isica de Canarias, Calle V\'ia L\'actea
  s/n, 38205 La Laguna, Tenerife, Spain\\
$^2$ University of Florida, 211 Bryant Space Center, Gainesville, Florida 32611-2055, USA\\
$^3$ Departamento de Astrof\'isica, Universidad de La Laguna (ULL), 38205
La Laguna, Tenerife, Spain\\
$^4$ Osservatorio Astronomico di Brera, via Brera 28, 20121 Milano, Italy\\
$^5$ Institute of Space and Astronautical Science (ISAS), Japan Aerospace Exploration Agency, 3-1-1 Yoshinodai, chuo-ku, Sagamihara, Kanagawa 229-8510, Japan\\
$^6$ Max Planck Institute for Astrophysics, Box 1317, 85741 Garching, Germany\\
$^7$ Dept. of Physics \& Astronomy, Univ. of Sheffield, Sheffield S3 7RH, UK\\
$^8$ Dept. of Physics, Univ. of Warwick, Coventry CV4 7AL, UK\\
$^9$ European Southern Observaotry, Casilla 19001, Santiago 19, Chile \\
$^{10}$ Department of Physics and Astronomy, Louisiana State University, Baton Rouge, Louisiana 70803, USA\\
\\
  }

\begin{document}
\newcommand{\swift}{SWIFT J1753.5$-$0127}
\newcommand{\gx}{GX 339$-$4}
\newcommand{\sco}{Sco X-1}
\newcommand{\cyg}{Cyg X-2}
\newcommand{\xte}{XTE J1118+480}

\date{August 2010}

\pagerange{\pageref{firstpage}--\pageref{lastpage}} \pubyear{2010}

\maketitle

\label{firstpage}
\begin{abstract}
Using simultaneous observations in X-rays and optical, we have performed a homogeneous analysis of the cross-correlation behaviours of four X-ray binaries: \swift, \gx, \sco, and \cyg. With high time-resolution observations using ULTRACAM and RXTE, we concentrate on the {\em short time-scale}, $\delta t<20$\,s, variability in these sources.
 Here we present our database of observations, with three simultaneous energy bands in both the optical and the X-ray, and multiple epochs of observation for each source, all with $\sim$second or better time resolution.  For the first time, we include a dynamical cross-correlation analysis, i.e., an investigation of how the cross-correlation function changes within an observation. We describe a number of trends which emerge. We include the full dataset of results, and pick a few striking relationships from among them for further discussion.

We find, that the surprising form of X-ray/optical cross-correlation functions, a positive correlation signal preceded by an anti-correlation signal, is seen in all the sources at least some of the time. Such behaviour suggests a mechanism other than reprocessing as being the dominant driver of the short-term variability in the optical emission. This behaviour appears more pronounced when the X-ray spectrum is hard. Furthermore, we find that the cross-correlation relationships themselves are not stable in time, but vary significantly in strength and form. This all hints at dynamic interactions between the emitting components which could be modelled through non-linear or differential relationships. 
\end{abstract}


\section{Introduction}
X-ray binaries (XRBs) are systems in which an ordinary star and a compact object are in close orbit. Matter falls onto the compact object and is accreted by it. The in-falling material releases its binding energy, powering energetic radiation and matter out-flows. In the most typical scenario (and the only one considered here), the donor star is an evolved star at the end of its main sequence life, and over-filling its Roche Lobe. The compact object can be a black hole or a neutron star; white dwarf systems are not normally considered XRBs, since the majority of the luminosity is released at somewhat lower energies, corresponding to the larger radius at which in-falling material is halted. See \citet{1997xrb..book.....L} for a description of X-ray binary phenomenology and theory, and possible models for the accretion process.

Since XRBs are most manifest in the X-ray band, much observation of them has focused here. This is particularly true of timing investigation, since X-ray instrumentation has been capable of high temporal resolution since the start. A wide range of time-domain behaviour have been catalogued for the various binary systems, such as broad-band timing noise and a variety of quasi-periodic oscillations (QPOs). See \citet{2004astro.ph.10551V} for a review of X-ray timing observations.

The X-ray emission corresponds to the bulk of the luminosity and should come from the smallest radii (and correspondingly highest particle energies) around the compact object, where the dynamical time-scale is the shortest. This strongly suggests, therefore, that any optical variability on a similar time-scale to the X-rays must occur by {\em reprocessing}, i.e., the absorption and re-emission of X-ray radiation by cooler material at greater radii in the accretion disc or on the surface of the donor star. Although this idea is not new, \citet{1994A&A...290..133V} modelled this in detail, and presented evidence that the optical luminosities were of a similar magnitude to what might be expected, if reprocessing were dominant in the optical.

Optical emission by reprocessing of higher-energy radiation, e.g. X-rays, has undoubtedly been seen in X-ray bursts (see e.g., \citealt{1978Nature...274...567M}; \citealt{1982ApJ...263..325P}). In these cases, it is believed that the burst originates in an explosive event on, or immediately above, the surface of the neutron star. The optical emission is seen to rise sharply very soon ($<$1\,s) after the X-ray burst, and can be well-modelled as absorption and re-emission of the original flare, mainly in the accretion disc at larger radii, and also from the donor star surface (e.g. \cite{2006ApJ...648.1156H}). The cross-correlation function (CCF) for this case is a single positive peak centred at positive lags. The X-ray light-curve in such a scenario reflects primarily the temperature of the neutron star surface and its short-term behaviour is not connected with the accretion disc or jets. 
Nevertheless, X-ray bursts serve to show that the optical emission, and its rapid variability, can be dominated by reprocessing at least for a short time, given certain specific circumstances.

Reprocessing has also been demonstrated through the technique of Doppler Tomography for individual spectral lines. The known period of the binary orbit is used to find the phase and velocity amplitude of emission components within a strong (emission) line. This can be transformed to map the emission locations within the binary system (see \citealt{2003ApJ...590.1041C}, for an example of this applied to a low mass X-ray binary). Similarly, echo tomography has been attempted in a number of cases using the fast variability near the compact object and the echoes this can induce \citep{2005ASPC..330..237H}, especially for larger-scale accreting systems, the active galactic nuclei (AGNs, e.g., \citealt{2008A&A...480..339R}). In contrast to the optical, infra-red emission may well be contaminated or dominated by a {\em jet}: a relativistic, energetic outflow originating near the compact object and travelling along the polar axes. The emission from a jet is therefore causally connected with the central accretion engine, but fluctuations should happen later than, and in response to, the higher-energy emission in the core. See for example \citet{2008ApJ...678..369E} for the case of a galactic micro-quasar, or \citet{2010MNRAS.404L..21C} for the most recent such observations of one of the systems under consideration in this paper.

Optical high time resolution observation is still a relatively young field, since CCDs capable of fast readout with low noise are relatively new, and the previous generation of detectors suffered from low quantum efficiency and/or high dead-time. Due to the difficulty of coordinating the scheduling of multiple observatories, not many simultaneous sub-second time-resolution, multi-band observations exist in the literature.

\citet{2001Natur.414..180K} observed the galactic black hole XRB \xte\ simultaneously with the {\em Rossi} X-ray Timing Explorer (RXTE) and OPTIMA, an optical single-pixel fast photometer, observing white light. They found two surprising results in the correlation analysis: the optical auto-correlation function (ACF) had a smaller width than X-rays, and the CCF had a small component which was negative and with negative lag (i.e., optical arriving before X-rays). Neither of these factors would be expected for the simplistic reprocessing model. In further analysis, \citet{2002A&A...391..225S} commented that the cross-correlation function was not stable through the observation, and contained two statistically separable principal components, both of similar dip-and-peak shape, but differing in time-scale. A simplistic explanation was presented to explain the {\em precognition dip}, of a local density and vertical height enhancement in the disc obscuring the inner X-ray region, and then causing increased X-ray emission some seconds later as the blob is accreted. No numerical or detailed modelling was done, however. Further discussion on the possible physical processes and time-scales in the system was provided by \citet{2003MNRAS.345..292H}.

\citet{2008ApJ...682L..45D} and \citet{2008MNRAS.390L..29G} increased the number of fast ($\delta t<$1\,s) optical/X-ray cross-correlation analyses in the literature to three. They also suggested that the fast functions seen were not compatible with reprocessing as the dominant source of optical variability. That data is included in our analysis below.

In summary, the reprocessing model is able to explain the energetics of the optical emission for X-ray binaries in general, and many of the details of the spectrum and long-term behaviour that would be expected. It has problems, however, with the fastest time-scale variations, and it is these that we wish to concentrate on in this paper. Time-scales $<2$\,s are shorter than the light-crossing time for a typical accreting binary system (depending, of course, on the binary separation).

Here we analyse multiple ULTRACAM/RXTE observations for four XRBs: \swift, \gx, \sco, and \cyg. For each observation, we consider three energy bands in the optical and three in the X-rays, in a homogeneous way across the dataset. Also, we investigate the dynamical behaviour of the CCF, a technique which is not commonly considered. The total number of presentable relationships are large; we include the full set in Appendix A, and we draw out a few examples to illustrate the types of behaviour we find in the data. What we see suggest that a large fraction of the fast variability seen in optical emission is not produced by reprocessing. Therefore,  some explanation is required for why reprocessed variability is not more pronounced. 

\section{Observations}
Table \ref{log} lists the observations considered as part of this simultaneous observation compilation. As shown, some of this data has been presented before in some form. 

\begin{table*}
\begin{center}
 \centering
  \caption{Log of simultaneous ULTRACAM/RXTE observations}\label{log}
  \begin{tabular}{lccccc}
  \hline
Object & Date & ULTRACAM  & Optical time & Simultaneous & Reference$^1$\\
       &      & site      &resolution (s) & coverage (s)\\
\hline\\
\swift & 2007-06-12 & VLT & 0.14 & 2800 & Durant et al. (2008)\\
 & 2008-08-20 & WHT & 0.2 & 2400\\
\gx & 2007-06-14 & VLT & 0.14 & 2800\\
 & 2007-06-16 & VLT & 0.13 & 2700 & \citet{2008MNRAS.390L..29G}\\
 & 2007-06-18 & VLT & 0.05 & 3300 & \citet{2008MNRAS.390L..29G}\\
\sco & 2004-05-17 & WHT & 0.1 & 950,760,730\\
 & 2004-05-18 & WHT & 0.25-0.5 & 260,700,950& \citet{2007MNRAS.379.1637M}\\
 & 2004-05-19 & WHT & 0.3 & 115\\
\cyg & 2007-10-16 & WHT & 2.0 & 880\\
 & 2007-10-17 & WHT & 2.0$^2$ \\
 & 2007-10-18 & WHT & 2.0$^2$ \\
 & 2007-10-19 & WHT & 2.0 & 600,710 \\
 & 2007-10-20 & WHT & 2.0 & 530 \\
 & 2007-10-21 & WHT & 2.0 & 1050\\
\hline\\
\end{tabular}\\
\end{center}
$^1$ where no reference is given, the data are presented here for the first time.\\
$^2$ the data quality in these observations is too poor for consideration.\\
\end{table*}

ULTRACAM is a high-speed optical camera, which incorporates dichroics to split the input beam into three energy bands. The detectors are fast read-out CCDs with readout noise of only 3.5\,$e-$. Masked data buffer areas on each chip enable an image to be very quickly shuffled from the active region and then to be read while the next is exposing, reducing the dead-time to near zero. By reading only small windows around the target and reference star, sampling up to 500\,Hz is possible. A dedicated GPS receiver provides reliable time-stamping to sub ms accuracy. See \citet{2001NewAR..45...91D} and \citet{2007MNRAS.378..825D}. We analysed each dataset using the custom ULTRACAM pipeline\footnote{see {\tt http://deneb.astro.warwick.ac.uk/phsaap/\\software/ultracam/html/index.html}}, using variable-aperture photometry relative to a bright field star. Slow variations due to the difference in colour of the two stars make negligible effect on the time-scales under investigation. For these observations, ULTRACAM was mounted on either the 8.2\,m Very Large Telescope (VLT), Paranal, Chile, or the 4.2\,m William Herschell Telescope (WHT), La Palma, Spain.

The dichroics in ULTRACAM split the light into the blue, middle and red parts of the spectrum, but filters are typically used before each detector. For \swift\ and \gx, broad r'g'u' filters were used, but for \sco\ and \cyg, narrow filters were used in the r-continuum (i.e., a spectral region without lines), and in the vicinity of the Bowen/He-II blend (the broad u' was retained). The purpose was to try to isolate the emission in the narrow emission lines, but the majority of the flux in each filter is still dominated by the continuum (see \citealt{2007MNRAS.379.1637M} ).

RXTE is an X-ray observatory optimised to timing work. The Proportional Counting Array (PCA, \citealt{1996SPIE.2808...59J}) instrument has a large effective area and is sensitive to photons with energies 2--60\,keV. Here we only consider the more sensitive 2--20\,keV range. Although a number of data collecting settings are possible, the intrinsic time resolution is of the order $\mu$s with full energy resolution for all but the brightest sources. The limiting factor is the telemetry bandwidth. The data were processed using the standard FTOOLS packages\footnote{see {\tt http://heasarc.gsfc.nasa.gov/lheasoft/}}, and three background-subtracted light-curves produced for each dataset: 2--4, 4--8, and 8--20\,keV. Note that for a non-imaging instrument such as the PCA, background subtraction is in general a complicated issue; estimates can be based on empirical measurements, and depend on the spacecraft attitude and recent trajectory. We utilised the latest South Atlantic Anomaly passage history and background models, 2009. For such bright sources however, these make little difference, and the noise at time-scales or order 1\,s or less is dominated by source photon statistics.

\section{Analysis}
Our aim in this paper is to compare the cross-correlation functions between the observed X-ray and optical flux for each source. This is a simplistic determination of the relative time-variability of the two wavebands, i.e.,  to what degree do they vary in similar ways, and what is the relative timing of these variations. It provides a clear, uncomplicated comparison of two signals with respect to time, but is by necessity an averaging process. To go from a significant signal in a CCF to a realistic picture of what the two light-curves are doing is not always straightforward. We endeavour to describe the typical behaviour that gives rise to the CCFs seen. We note that alternative methods for time-series comparison exist, such as the coherence and Fourier cross-spectrum, but in our experience these are no more transparent, and indeed are intimately related to the CCF. We only consider CCFs from here on.

We calculated CCFs for each simultaneous observation by first re-binning the light-curves to the best time resolution of the poorer of the pair. CCFs were calculated in windows of size 50\,s and stepping 25\,s, i.e., overlapping sections where every second CCF in independent. A mean CCF is then derived for each time-series pair for each observation. This acts effectively as a high-pass filter, and variations with time-scales longer than $\sim$50\,s will not appear. For the purposes of this paper, we are only interested in the fastest variability, since this must be predominantly due to fluctuations arising close to the compact object, where the dynamical time is the shortest.

All the CCFs were considered, for each object, observation and pairs of energy bands. In  Section \ref{res} we present a selection of average CCFs, pointing out some similarities between the binary systems, and some trends in the data. The full set of CCFs is presented in Appendix A. Note that, although a comparison star in the field is used in each observation to mitigate atmospheric effects, the weather still greatly affects the signal quality achieved in a CCF, see for example \citep{2010MNRAS.tmp.1137G} and the comparison of successive nights, where we find that the CCF features are most pronounced on the night with best weather 2007-06-18.

\subsection{Dynamic analysis}
In the initial analysis above, the CCFs were the average of many 50\,s sections throughout an individual observation. These segments can, however, be interesting independently. A convenient display for such a set of CCFs, is to stack them above each other, with one line representing a single CCF, and colours along that line giving the strength of the correlation for a given lag, at a given point in the observation. Such a representation is similar in style to the more common dynamic Fourier-gram.
Note that because the segments overlap, adjacent CCFs are not independent, but every second one is. This produces slightly smoother-looking output, and is intended to catch strongly correlated events that would otherwise fall on the border between two segments. No additional numerical smoothing has been applied.

\subsection{CCF noise}\label{ccf_noise}
Aside from intrinsic changes in the signal, two types of noise also contribute to the differences between any two CCFs and to the structures within a single CCF. The first is simple: uncorrelated (white) noise inherent in any measurement. This is dominated in our case by the Poissonian statistics of photon counting, but read noise can also contribute. Since the cross-correlation calculation is a simple multiply-and-sum process, these errors are easy to propagate. Longer time-scale noise is also present in the data, however. These are due both to slow systematics (e.g., seeing variations) and red noise processes intrinsic to the emission processes themselves, and act on a range of time-scales. These make the estimation of the noise contribution to CCFs rather more complicated. 

One simplistic method to establish the significance of an average CCF signal might be to measure the standard deviation of all values for a given lag, compared to the average value. This would only be valid in the case that the CCF is assumed to be stationary, but in our dynamic analysis we find this generally not to be the case, so we do not consider this method further.

The first-order approximation we apply, to see whether there is any significant signal in a given CCF, is to empirically measure the noise in the CCF itself, given by the median of the difference between successive points. This value is a combination of various noise sources, and as such, a useful estimator for analysing a given CCF. Signals falling more than five times the value are likely to be significant. In the averaged CCFs shown in Appendix A, we give both a raw CCF scale, and a $\sigma$ scale ($CCF/\sigma$), so that the y-axis becomes a rough estimate of significance. For any significant average signal, one can inspect the corresponding dynamic CCF plot, to see where in the observation the signal mostly originates from; it may be steady throughout the observation, intermittent, or changeable in height and shape.

Having established a candidate CCF signal, one can ask how well the features within it can be localised, This we apply to the dynamic CCFs, since we find the possible variation in peak height and location interesting.

In order to estimate the confidence in establishing the amplitude and position of a particular CCF peak, we follow the empirical estimation of \citet{2007MNRAS.375.1479S} who show a simple implementation of the theoretical derivation by \citet{Bartlettbook}. This method considers both the correlated and un-correlated noise in each light-curve, as empirically sampled by the auto-correlation functions. It provides the uncertainty on each point in a calculated CCF. Since  the method is only applicable to a stable CCF relationship, our approach here is necessarily simplistic. To estimate the confidence interval on an individual peak, one can fit a simple function (e.g., a Gaussian) with a standard chi-squared process incorporating these uncertainties. Such a fit is only a good estimator for the case that one has good reason to believe the Gaussian function fit is a good description of the data. The CCF functions are clearly more complicated than this, but we nevertheless think this approach is a good  approximation for characterising peaks.

\section{Results}\label{res}
We include the complete set of cross correlations, averaged and dynamic, in Figures A1 to A34, in Appendix \ref{data}. They serve to illustrate, that the range of behaviours that we find are not particular to a small subset of the sample, but general features. The figures are shown with consistent  colour mapping. For each averaged CCF, we give both significance levels, based on our empirical estimate (see above), and absolute CCF values. In this Section, we focus on  a few interesting CCFs, in order  to best illustrate some of the trends in the data.

\begin{figure*}
\includegraphics[width=0.7\hsize]{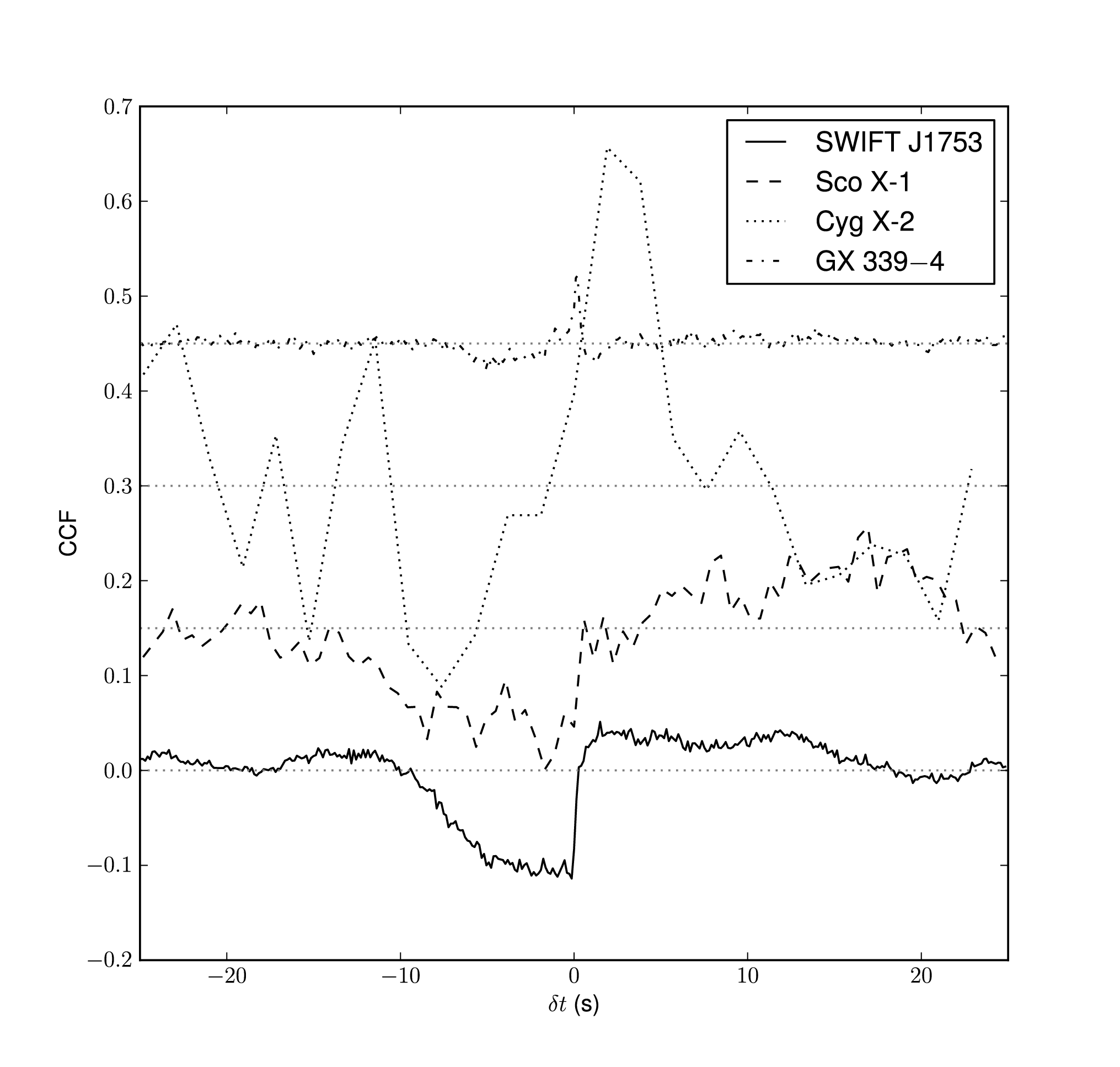}
\caption{Example average CCFs for our sample of XRBs, demonstrating the ubiquity an anti-correlation component at negative lags in each case. These are each in the optical red band and 2-4\,keV for X-rays. There is a vertical offset of 0.15 between successive curves, for clarity.}\label{fourbinaries}  
\end{figure*}

\begin{figure*}
\includegraphics[width=0.49\hsize]{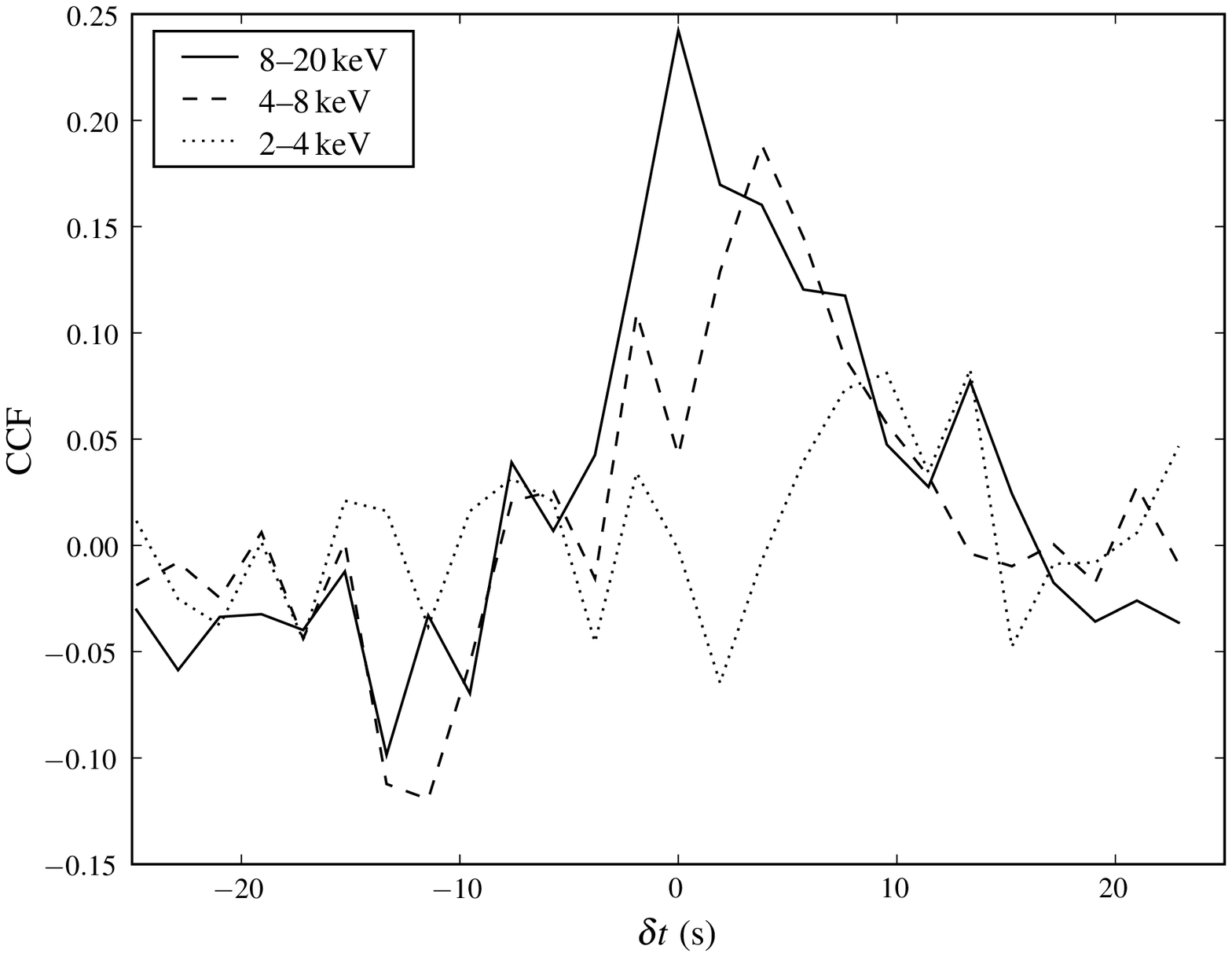}
\includegraphics[width=0.49\hsize]{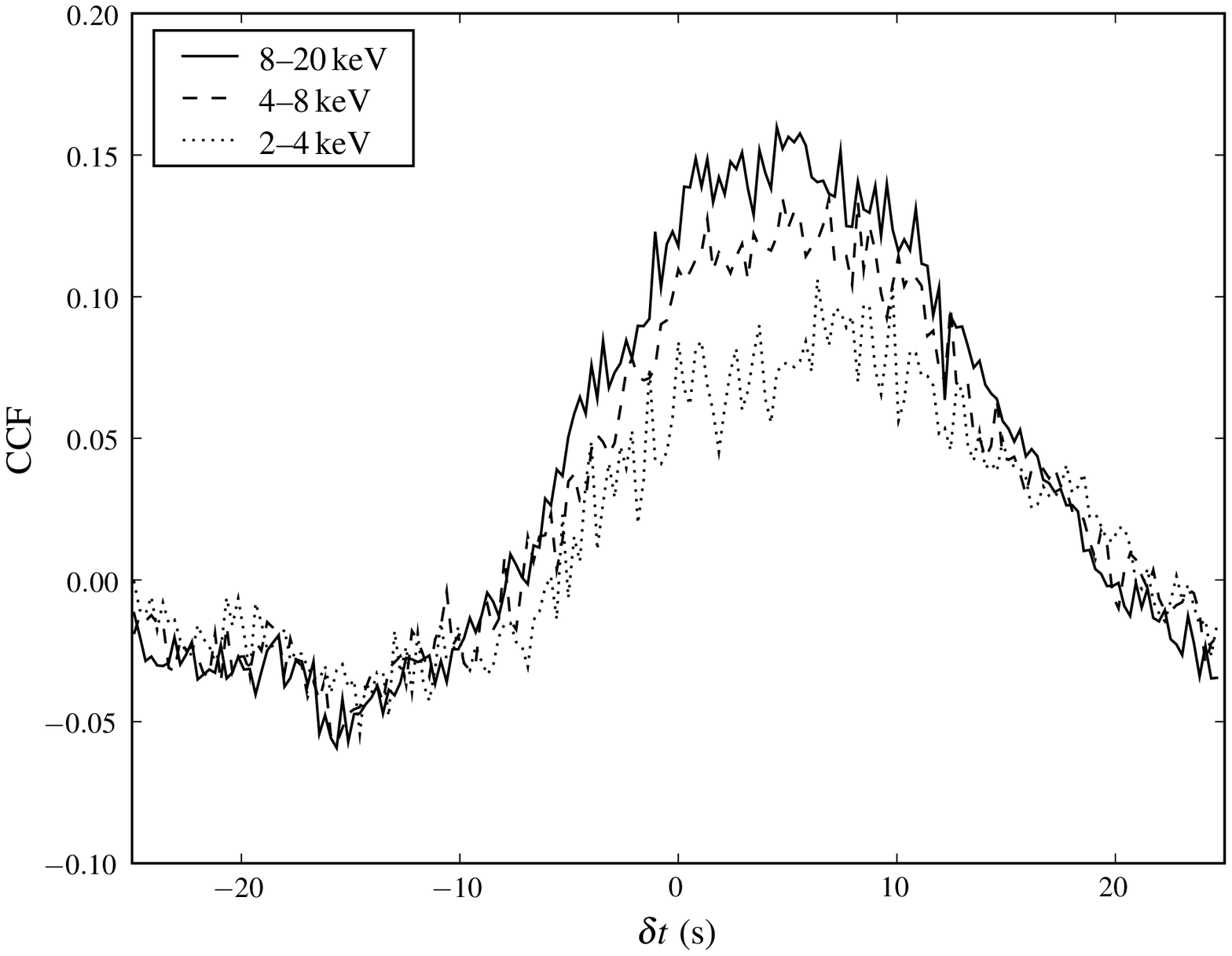}
\caption{Example average CCFs for single observations (left: \cyg, 2007-10-20, u'; right: \sco, 2004-05-28, green) demonstrating trends with X-ray energy band.}\label{Xenergies}  
\end{figure*}

\begin{figure*}
\includegraphics[width=0.49\hsize]{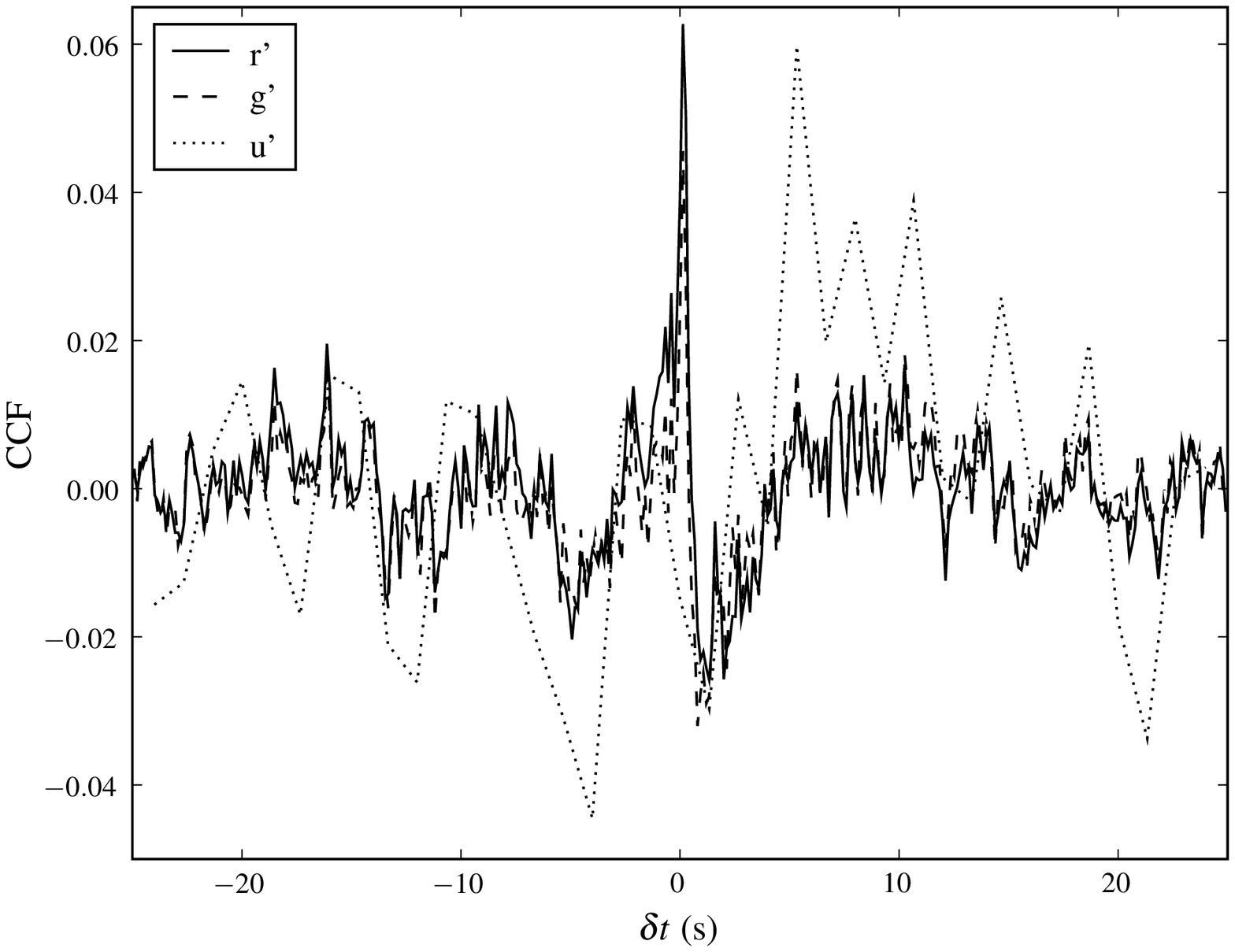}
\includegraphics[width=0.49\hsize]{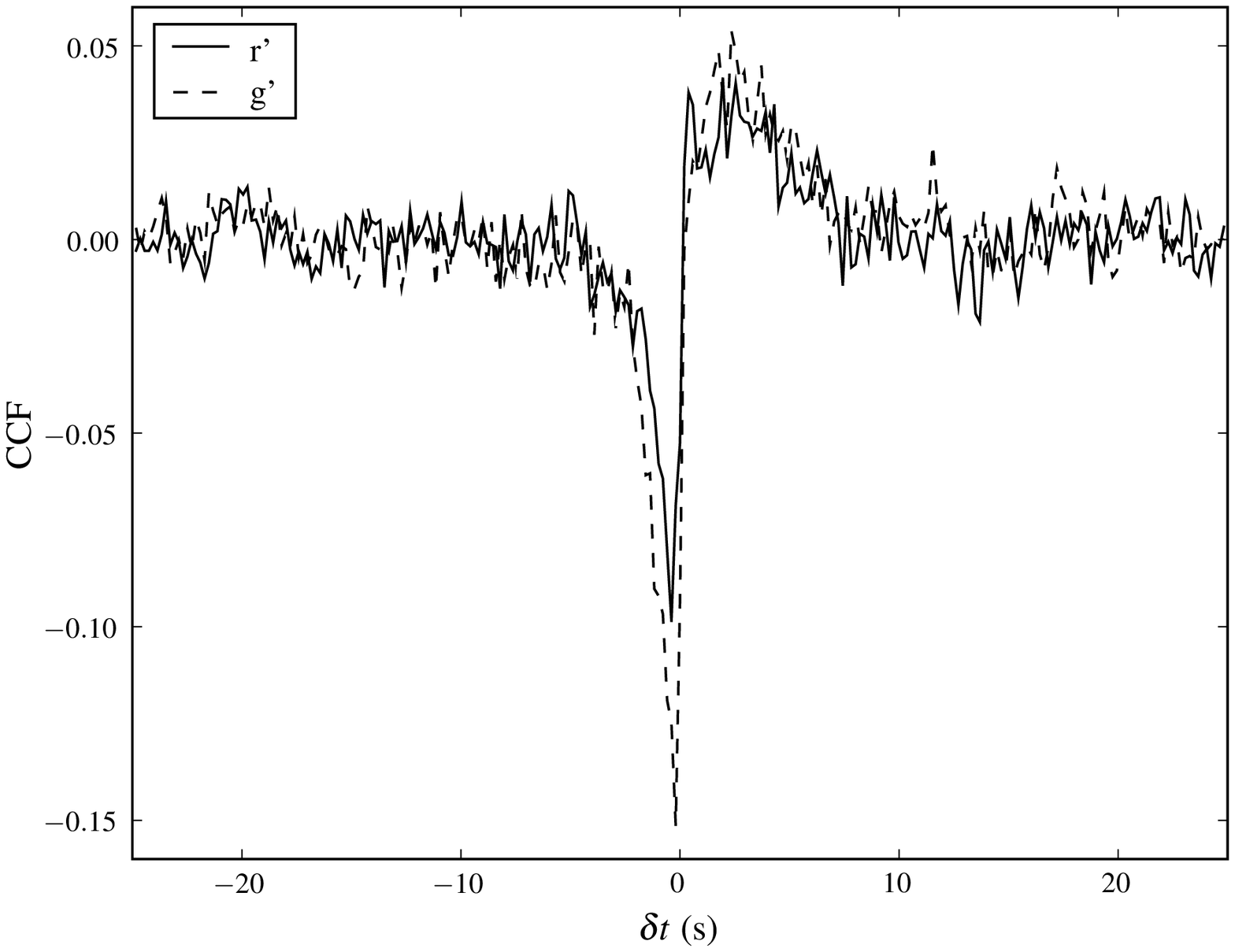}
\caption{Example average CCFs for single observations (left: \gx, 2007-06-16, 2--4\,keV; right: \swift, 2008-08-20, 4--8\,keV) demonstrating trends with optical band. }\label{Oenergies}  
\end{figure*}

\begin{figure*}
\includegraphics[width=0.49\hsize]{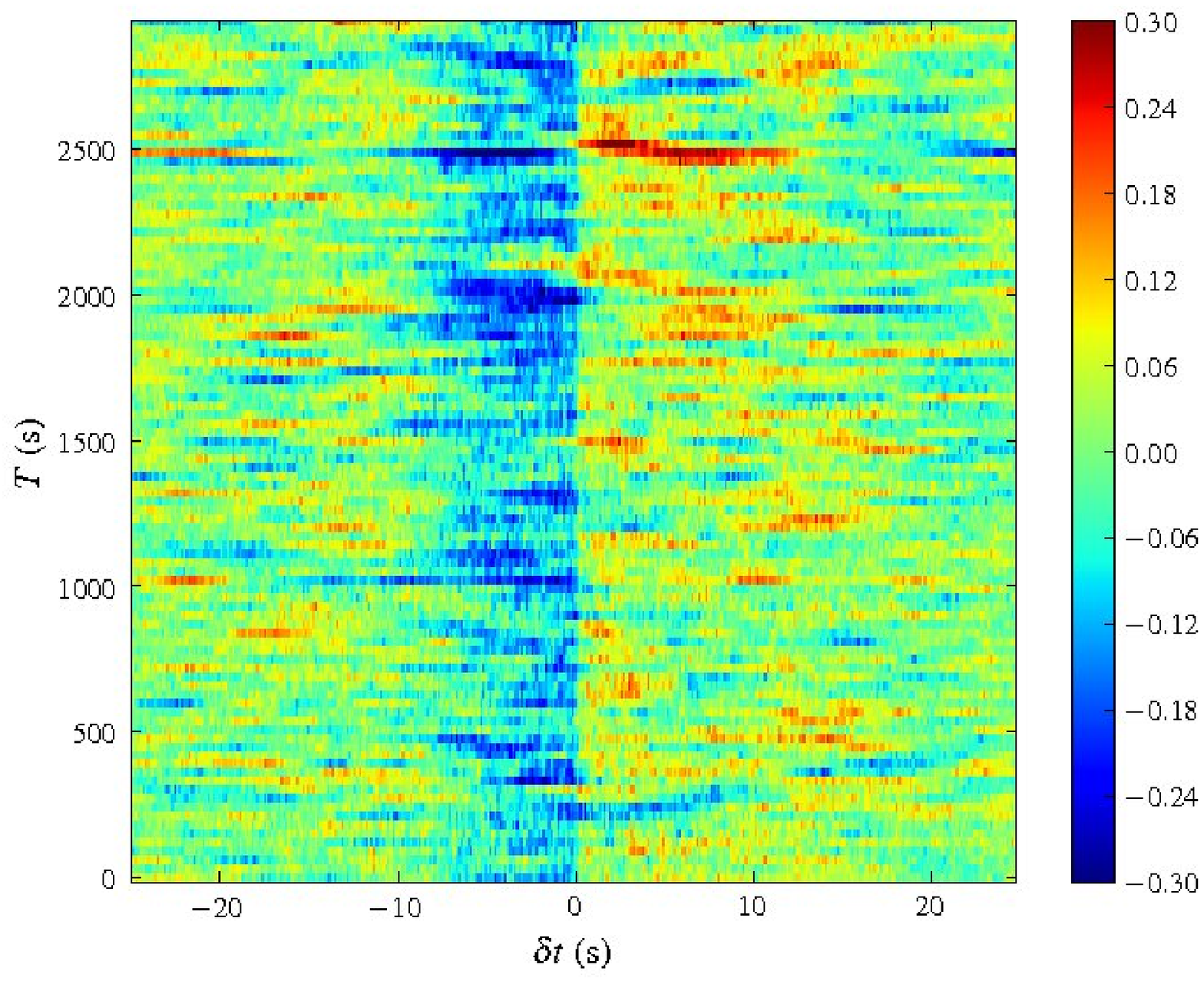}
\includegraphics[width=0.49\hsize]{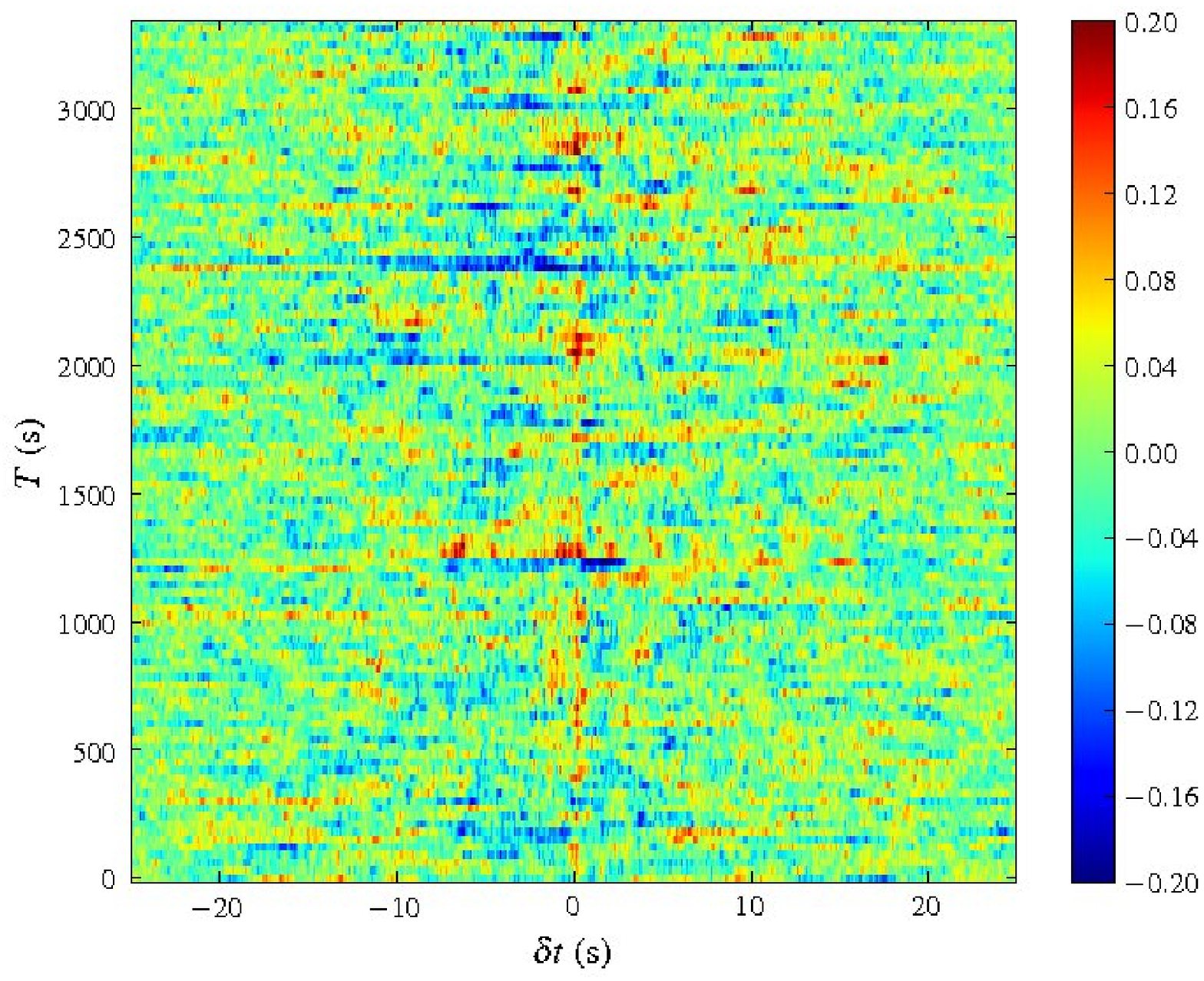}
\includegraphics[width=0.49\hsize]{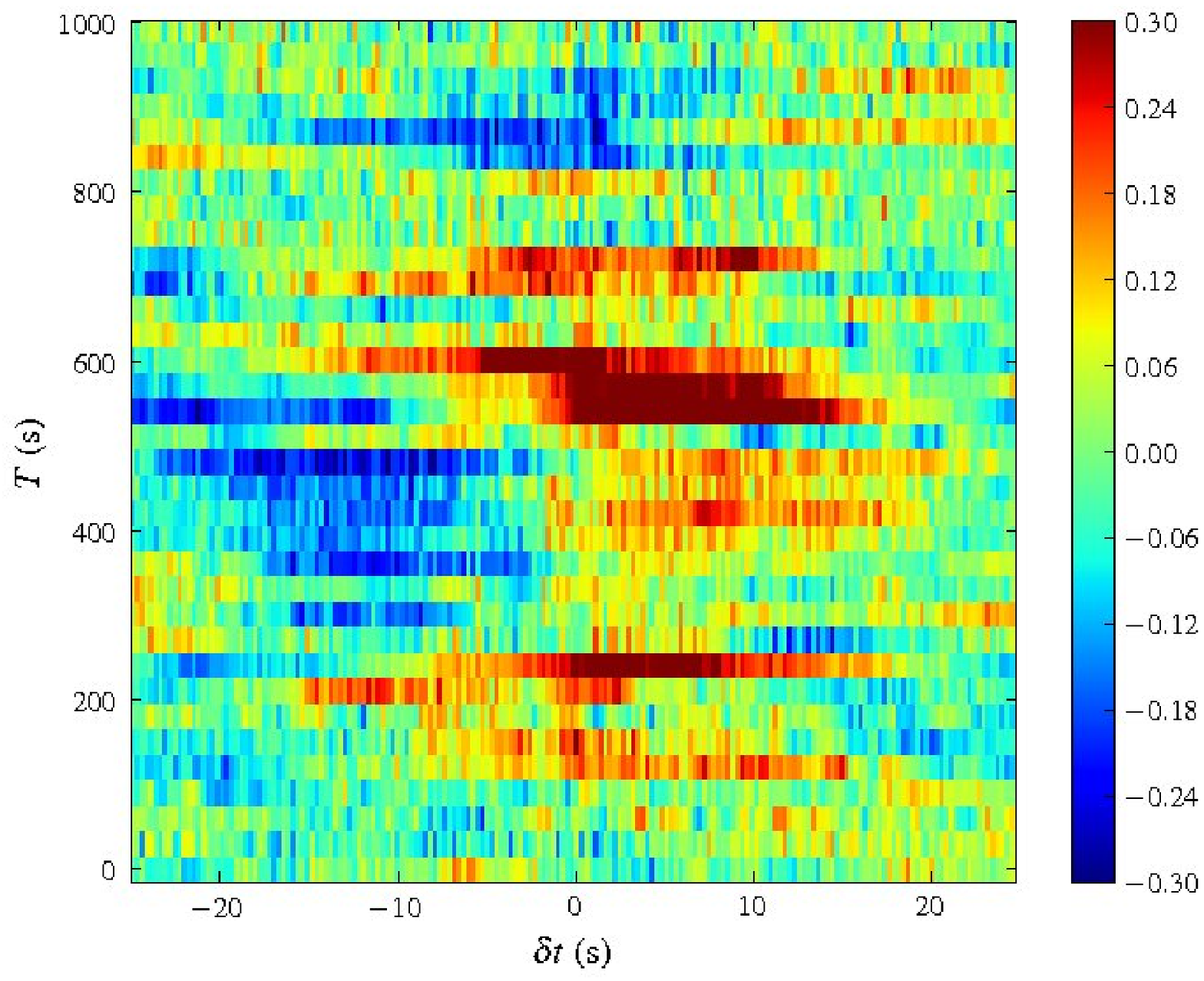}
\caption{Example dynamic CCFs (upper left: \swift, 2007-06-12 (93119-02-02-00); upper right: \gx, 2007-06-18 (93119-01-04); bottom: \sco, 2004-05-18 (90020-01-02-02); all r/2--4\,keV, Obs IDs in parentheses). Each horizontal line is a single CCF from a 50\,s segment, with the colours representing negative (blue) or positive (red) correlation. Successive lines overlap, see text. }\label{dynamic}  
\end{figure*}

In Figures \ref{fourbinaries} to \ref{dynamic} we present a selection of the CCFs calculated for our sample of objects, energies and epochs. The style of observation and the analysis have been conducted in a homogeneous manner. We extract a few trends that are apparent in the data. First we present typical CCFs for each source, then look at the trends with X-ray and optical energies, and finally display some examples of the dynamical analysis. The latter requires a little more careful consideration, since the changes seen in the CCF through a single observation could in principle be caused by noise processes alone - see Section \ref{dyn_beh}. Although in Section \ref{conc} we attempt to draw conclusions from what we find, and speculate on the physical scenario behind them, principally the results below stand as a phenomenological study, for interpretation in future theoretical works.

For the purposes of these example CCFs, we have picked plots that show particular features clearly. The full set are included in Appendix A, and the reader is encouraged to look through these to see the frequency with which the various features and trends outlined below occur.

\subsection{Anti-correlations}
Figure \ref{fourbinaries} shows example CCFs for each of the four binaries of our sample. They show the following typical form: negative correlation at negative lags ($CCF<0$, $\delta t<0$) and somewhat weaker positive correlation at positive lags. The sharp slopes seen around $\delta t=0$ indicate that in each case, much of the signal is due to variations at times-scales comparable to the the shortest measured. Even shorter time-resolution might of course be present, but is unobservable in these data. The most important point, is that of the four systems observed, all showed a similar CCF form at least some of the time. 

Not all of the observations show the typical CCF form described above. In a number of observations, the level of noise, predominantly Poissonian (with some contribution from, for instance, seeing and transparency variation), make any CCF signal relatively weak or unobservable. CCFs showing a single, strong positive peak at positive lags, as would be expected from reprocessing, are notable by their rarity. Furthermore the optical ACFs (not shown) are consistently of similar width or sometimes narrower compared to the X-ray ACFs. It would thus seem that the majority of the fast variability in the optical is not caused by reprocessing. The existence of a CCF signal indicates, however, that the two light-curves are not independent, but connected. As noted by \citet{2008MNRAS.390L..29G}, it may be possible to describe a complicated dip-and-peak CCF as a broader trough and simple peak, both centred near $\delta t=0$. This was motivated for the case of \gx\ by the similarity between the width one would ascribe to the broad trough and the width of the optical auto-correlation function. 

The sources presented here have various emission states in their epoch to epoch evolution, typically classified by their location on an X-ray flux versus {\em hardness} (i.e., flux ratio of a higher and lower energy band) diagram. The observations here all refer to times when the sources were in a low-hard state or flaring states, with a luminosity well below the maximum known, and the spectral slope in the X-rays harder. For example, \swift\ has since its discovery, never been seen to leave the low-hard state \citep{2009MNRAS.392..309D}. For \sco, where the hardness of the source changed markedly during the observation campaign, the strongest CCF signatures correspond to times when the X-ray emission was hardest, although this generally also coincided with times when the overall flux was the highest. Note that none of the systems was seen in the classical High/Soft state during these observations (e.g., \citealt{2007ApJ...667..411D} for a typical evolution between various states for \sco); in the high/soft state, one expects copious thermal emission to dominate the luminosity, and fast variability to be much less apparent (e.g., \citealt{2007ApJ...663.1225B}).

\subsection{Energy dependence}
Figure \ref{Xenergies} shows two examples of a systematic trend in the CCFs with X-ray energy band. For the \cyg\ data (left panel), one sees that the positive component becomes stronger and peaks closer to zero lag for successively higher energies. The change in strength, from $\sim$0 to 0.2 is dramatic. These curves were generated from the same observation. The right-hand panel for \sco\ shows a similar result: the curve becomes stronger and peaks closer to zero lag with higher energy, while the negative component shows little change. The change in correlation strength in this case is more subtle.

Figure \ref{Oenergies} shows two examples of a systematic trend in the CCFs with optical band. For \gx\ (left panel), one sees that both r' and g' show the same structure, anti-correlation near zero lag, except for a sharp peak very close to zero lag, but that the relationship is stronger for r', particularly for the positive part. u' shows something different, broad negative correlation at negative lags and positive correlation at positive lags. Note that, due to the lower time-resolution of the u' data, the sharp peak at $\delta t=0$ for r' and g' is not excluded in the u' CCF. The right-hand panel for \swift\ shows a different result: the g' CCF curve has the same shape, but stronger than the r' one.

\subsection{Dynamic behaviour}\label{dyn_beh}
Figure \ref{dynamic} shows three examples of the results of our analysis. Each figure represents a set of CCFs from two time-series in one particular observation. Each horizontal line is a single CCF as measured at the time $T$ in the observation. This time is given as an offset since the start of simultaneous coverage. The colours along the horizontal line represent the strength of correlation for different values of lag, $\delta t$, along the x-axis. The colour scales are given for each panel individually, and can be compared with the values on the y-axes of Figures \ref{fourbinaries} to \ref{Oenergies}, which are average CCFs, or to the whole set of CCFs in Appendix A.

By eye, a number of features are apparent:
\begin{itemize}
\item{for \gx, the positive component seen in the average CCF appears for very short bursts and relatively strongly, with a low duty cycle through the observation. The negative component appears to be present whenever the positive component is not.}
\item{for \swift, the negative component is ever-present and strong, but the centroid appears to shift, and somewhat cross to $\delta t>0$. Likewise, the weaker positive component is sometimes centred at $\delta t<0$. The strength and width of both components changes markedly, but without any obvious trend through the observation.}
\item{for \sco, the form of the CCF seems to be constant, but the strength and location of the peak changes remarkably throughout the observation.}
\end{itemize}

The dynamic CCF diagrams clearly show a lot of variation with time. One must consider  whether such variations could be caused by the noise processes inherent to the light-curves under analysis. Here we present the application of such an investigation to a single dynamic CCF diagram which shows marked variations, as an illustration.
We wish to test the significance of the peak variations for case of \sco\ (2004-05-18), because of all the observations, this is the one that shows the most spectacular and unexpected CCF variations. Using the Bartlett formalism described in Section \ref{ccf_noise} to assign uncertainties to every CCF bin value, we have performed fitting to find the peaks in each line of the dynamic CCF. We fitted a single Gaussian function to the positive component of each CCF line and minimised the $\chi^2$ statistic. The uncertainty of the location of that peak is then the value at which $\chi^2$ increases by 1. 

\begin{figure*}
\includegraphics[width=0.95\hsize]{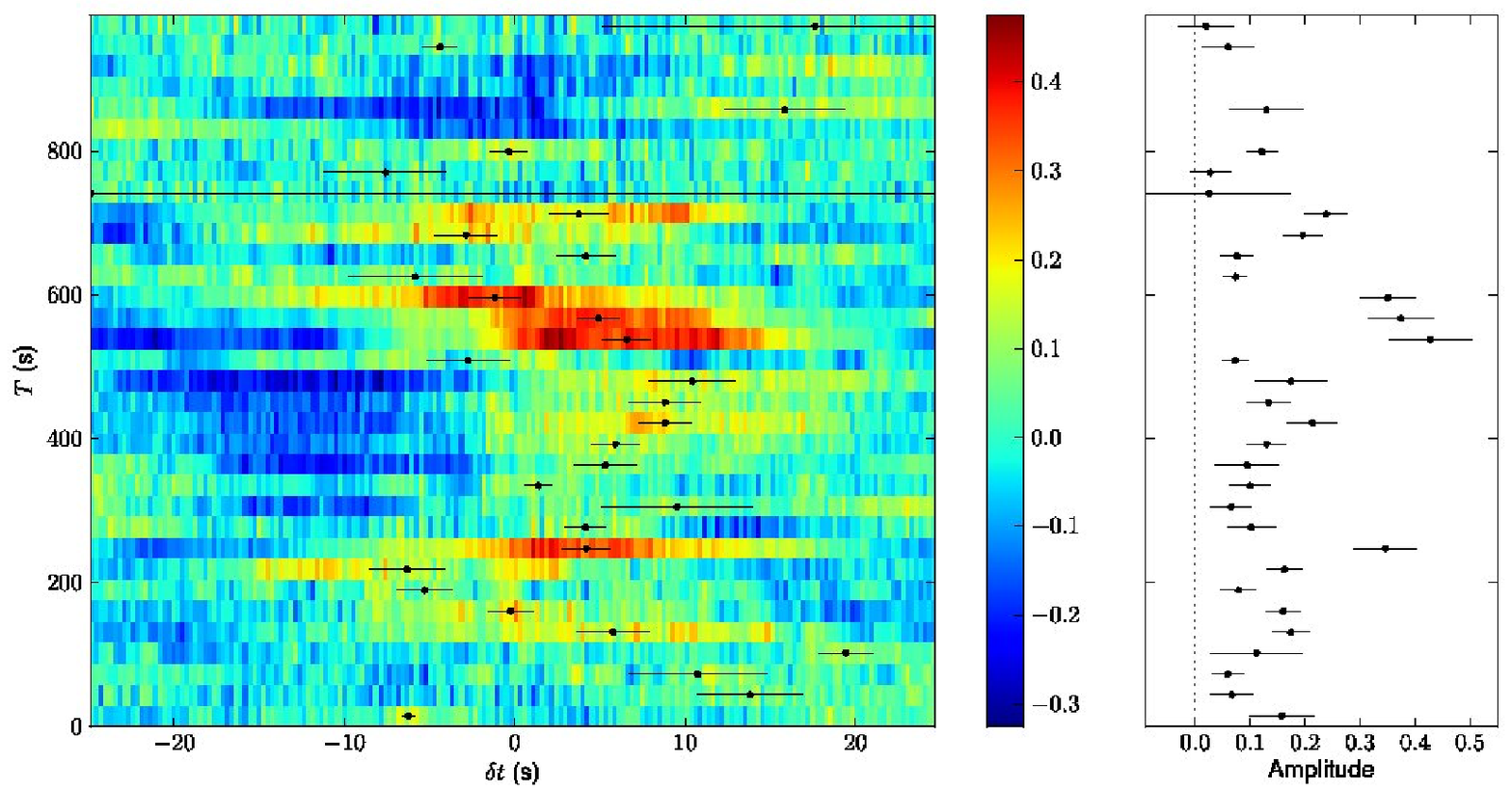}
\caption{Evolution of the positive correlation component of the dynamic CCF for \sco, 2004-05-18 (Figure \ref{dynamic}, lower panel). A single, positive Gaussian has been fitted to each line, as a way to estimate the centre of the positive correlation peak. The uncertainties on this fit have been found using realistic errors (see text) and the error-bars represent 1-$\sigma$ confidence limits. On the left panel, we show the best-fit location in $\delta t$ of the Gaussian, and it's uncertainty, and on the right panel we show the fitted amplitude and its uncertainty for each CCF line. Where the marker and associated error bar are missing, the Gaussian fit failed. }\label{error}  
\end{figure*}

The result of such an analysis are shown in Figure \ref{error}. We find that the shift in the location of the positive peak is highly significant, and a peak lag of $\delta t<0$ at some times is marginally significant. Note that here we are assuming that a single Gaussian function is a reasonable description of the data, which is clearly not the case; it is sufficient, however, for illustrative purposes.

Mu\~noz-Darias et al.'s (2007) analysis was performed on a 180\,s window centred at around $T=450$\,s on this diagram. Had we restricted our analysis to just this section of the data, we would also have concluded that the major feature of the CCF is a positive peak with a centre at around $\delta t= 5..10$\,s. Note however that these authors were attempting to measure the differences between CCFs, as a way to probe the emission lines which are present in the spectral passband of one of the filters but not the other.

Using a similar process for the rest of the dynamical CCFs, we find that the features noted above for Figure \ref{dynamic} are significant.

\section{Conclusions/Discussion}\label{conc}
We have presented a number of optical/X-ray CCFs from simultaneous observations of four XRBs, and deeper analysis of the energy- and temporal dependence of these. The most obvious and important result is, that for such a heterogeneous sample, optical-leading anti-correlations appear to be typical of the emission behaviour. Currently the accepted standard picture is that the optical emission arising from X-ray binaries is dominated by reprocessing of the X-rays in the outer accretion disc, or on the surface of the donor star. One would then expect that the cross-correlation of optical and X-ray flux shows that any variability in the X-rays is mirrored in the optical flux, but arriving to the observer a little later to account for the light-travel time across the binary system and the absorption/emission time-scale. In other words, the CCF should show a single strong, positive peak at positive lags ($\delta t>0$, optical arriving after X-rays.). 

Our observations show that the CCFs one typically sees are more complicated, and in particular, that signals of anti-correlation and optical-leading are ubiquitous, at least in the relatively low states in which our binaries were observed. It is thus difficult to reconcile the origin of the fastest variability in optical being due to X-ray reprocessing. This suggests that there are additional processes involved in accretion, that are not part of the standard model, the accretion process is more complex than previously thought. 

This does not mean, however, that reprocessing is absent altogether: if X-rays are illuminating cooler material, it is natural to ask why the reprocessing CCF signal is {\em not} seen. One simple answer could be that whatever is giving rise to the signals we observe is swamping the reprocessing signal; but it could also be that the range of light-travel times or the reprocessing time-scale is longer than previously thought, and that therefore the variable component of reprocessing is washed out on the time-scales we are probing. Decomposing the CCF into its fourier components and phase lags may be a way to  get around this difficulty (\citealt{2003A&A...407..335M}, \citealt{2010MNRAS.tmp.1137G}).  Certainly, there are strong fluorescence lines in some of the spectra, e.g., known since as early as \citet{1971ApJ...163L..69M}, but  notably absent for \swift\ \citep{2009MNRAS.392..309D}. Using fast spectroscopy, it would be possible to measure the variability in lines compared to X-rays, as opposed to the variability in the continuum (or the relative variability of lines and continuum).

Since the column density required to reach an optical depth of 1 (i.e., attenuation by a factor $e$) for photon energies $>$2\,keV is of the order $>10^{22}$\,cm$^{-1}$, and higher for more energetic photons, it could be simply that the penetration depth is so deep that the optical reprocessed light takes time to emerge (e.g. \cite{2003MNRAS.345.1039M}). With the X-ray spectra moderately hard as here, a far smaller proportion of the luminosity is released in the thermal, 0.5--2\,keV X-ray band than for soft sources. Conversely, irradiation by higher-energy photons may be expected to produce a stronger temperature change in the surface layers of this disc \citep{2001MNRAS.320..177W}. It would be interesting to do a similar study at thermal energies ($<$2\,keV), and also at much higher energies. Furthermore, it is not necessarily justified to assume that the X-ray light-curve reaching us, is the same as that 'seen' by the material which is producing the optical light reaching us. A detailed three-dimensional model would be required to describe this process; however we have found that the light-curves {\em are} related, but not as we might have expected. 

In Section 1 we noted that for the particular circumstances of an X-ray burst, optical emission by reprocessing is clearly observed.
We also mentioned the use of Doppler Tomography to map the sites of emission using the variations of line profiles with orbital phase, for particular, strong emission lines. Such an analysis shows that both some of the short-term and of the persistent optical flux originates in the accretion disc, in-fall stream, surface of the donor star etc. The total amount of flux contained within the emission lines is typically small compared to the continuum, and one would not expect the line flux to dominate even in a narrow filter centred around a stronger line. Given that there are typically not many such fluorescence lines in the spectrum, one would certainly not expect the flux measured in a broad-band filter to be affected by lines. 
Therefore, the fast variability measured here is indicative of the continuum emission and not of the fluorescence lines, and there is no contradiction between the majority of the CCF signal being incompatible with reprocessing, and the lines which are indeed necessarily produced by reprocessing. Furthermore, we have not considered slow variability ($\delta t>20$\,s) here at all, where reprocessing is known to occur.

\begin{figure}
\includegraphics[width=\hsize]{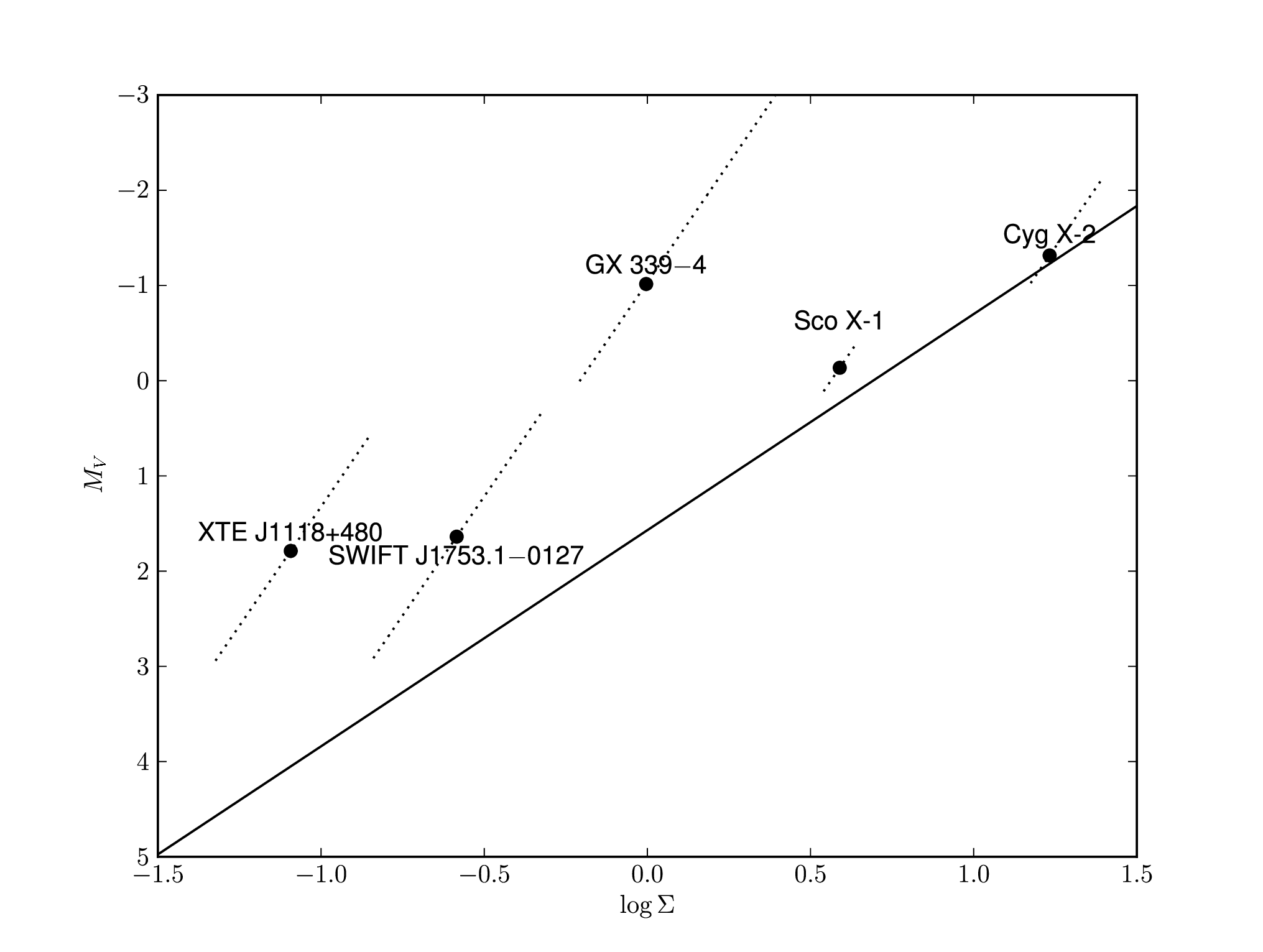}
\caption{Expected absolute magnitudes predicted by \citet{1994A&A...290..133V} (solid line) compared to the measured magnitudes for the X-ray binaries of interest (markers). The dotted lines indicate a range of plausible distances for each object. $\Sigma$ is as defined by these authors, the expression in square brackets in Equation \ref{mags}.}\label{figmags}
\end{figure}

\citet{1994A&A...290..133V} made some theoretical predictions on the amount of luminosity to be expected from reprocessing in an X-ray binary. They propose a linear relationship for the visual luminosity $L_V \propto L_X^{1/2} P^{2/3}$ ($P$ is the binary period and $L_X$ the X-ray luminosity responsible for reprocessing). In a linear fit to several well-studied LMXBs, they find absolute magnitude 
\begin{equation}
M_V = 1.57(24)- 2.27(32) \log\left[ \left( \frac{P}{1\,hr} \right)^{2/3} \left( \frac{L_X}{L_{Edd}} \right)^{1/2} \right]\label{mags}
\end{equation}
where $L_{Edd}$ is the Eddington luminosity.
Applying the same method, and making reasonable assumptions for the unknown $L_X/L_{Edd}$ and using a range of possible distances, we have plotted the four objects in our study, plus \xte\ next to the relationship given in Equation \ref{mags}, see Figure \ref{figmags} (first discussed by \citealt{2010MNRAS.tmp.1137G} for the specific case of \gx). For the purposes of this plot, the luminosity is in the 1--100\,keV range (as extrapolated from spectral fits), and assumed all to contribute to reprocessing, and the magnitudes are corrected for reddening, which is an additional source of uncertainty. Only for \cyg\ would one expect reprocessing to be a significant fraction of the optical emission. Perhaps it should not be a surprise, then, that reprocessing does not appear to make much of a difference to the CCFs; but conversely a new source of the optical variability, and its connection to X-rays is required. Note that the systems showing the most unusual timing properties (and the fastest variability), \xte, \gx\ and \swift\ are also the ones lying furthest from the \citet{1994A&A...290..133V} relation.  It is interesting to note that these are black hole systems, whereas the neutron stars systems lie much closer to the relationship.

\citet{2009ApJ...697L.167G} found that a relationship exists between the RMS and flux for the three binary systems \gx, \swift\ and \xte. The two quantities can be adequately described by a linear trend (with significant scatter), that does not pass through zero, for various time-scales. Such behaviour had been found in X-rays by \citet{2005MNRAS.359..345U}, and implies that the light-curves cannot be described by a simple {\em shot} process of independent flares, but rather that the size and duration of events is related, and probably connected to their location in the accretion disc. The emission thus cannot originate from many independent launching sites. Thus any theory wishing to describe the CCF behaviour above must also consider this important result.

In our observations, variability is seen down to the shortest time-scales sampled for every source, and the resulting CCFs have such sharp gradients that some of the correlation must occur at these time-scales or faster.
One wonders how much more structure would be discovered with higher time resolution, high signal-to-noise observations.
Variations of the CCFs with X-ray or optical band is an indication that the varying component responsible for the CCF signatures has a different spectrum to the average spectrum. Differences between the CCFs of the various bands will be affected by the difference signal-to-noise of the individual light-curve, differences that are hard to model.

A number of qualitative models have appeared since the initial discovery in \citet{2001Natur.414..180K} of the unexpected correlations for \xte:- 
\begin{itemize}
\item{The first attempt was by the same authors in more detailed analysis \citep{2002A&A...391..225S}. Their simplistic description was of blobs of dense material in the accretion disc, having a higher density and therefore larger vertical extent. As they travel towards the central source, they would partly obscure the X-ray light-source, just before being accreted and causing an increase in X-ray luminosity. Unfortunately, the model did not include detailed predictions. We would expect, however, either X-ray energy independent behaviour (for partial covering by a dense, opaque blob) or variability much stronger in softer X-ray bands (for moderate optical depth).}
\item{\citet{2004MNRAS.351..253M} produced an alternative toy model to explain \xte's CCF. This involves a magnetic energy reservoir, where the energy input is a stochastic process, and the output can be channelled either to X-ray luminosity or the jet. Thus the emission at a moment depends not only on the current accretion rate, but on the past history; the emission modes are therefore linked in a differential relationship. Such a model can quantitatively explain the cross- and auto-correlation function, although how to maintain such an energetic magnetised zone is not addressed. Coupling between the different emission zones, at various time-scales is required by the flux-RMS relationship mentioned above. \citep{2005ApJ...620..905Y} developed a quantitative model along the ideas of multiple contributing components to the optical flux.}
\item{\citet{2009ApJ...703L..63C} provide yet another explanation, this time questioning the assumed synchrotron spectrum expected from a magnetised outflow. Typically a broad-band flat $F_\nu$ spectrum is assumed, breaking and falling in the IR-range; but this assumes a special magnetic field/outflow configuration, developed to match flat radio spectra. For suitable magnetic fields, it can be that the synchrotron spectrum had a bump near the optical range, but also provides the bulk of the X-ray flux. Thus the emission in the two waveband is intrinsically connected, and the relative timing depends on the magnetic field strength and geometry in the jet.}
\item{Conversely, part of the X-ray emission could be directly caused by Comptonisation of the synchrotron jet component, e.g. \citet{2005ApJ...635.1203M}. This could explain a number of the longer-term correlation trends in various X-ray states, e.g., for \gx\ \citep{2009MNRAS.400..123C}. What the consequence of this would be on shorter time-scales is less clear, particularly since the modelling mainly concerned the infra-red emission. Fast variability from a jet-like component has been studied in detail by \citet{2008ApJ...678..369E} for the case of the micro-quasar GRS 1915+105. }
\item{In any scenario, and indeed in some of the ones above, one requires a source of instability to drive emission enhancements. Such instabilities may naturally arise from the interactions of magnetic fields and matter, and act as a primary channel to emit energy from the system. The presence of jets and of high viscosity in the accretion discs are often seen as evidence for strong large-scale fields, although more direct evidence (such as Zeeman lines or cyclotron harmonics) is missing \citep{2009ApJ...701..885L}. Note that for the case of \swift, humps were seen in the spectrum that may suggest cyclotron emission \citep{2009MNRAS.392..309D}.}
\end{itemize}
In summary, there are several plausible ideas for the interaction of low-energy and high-energy emission in accretion systems. Typically they have been formulated to explain one particular set of observations, and so lack generality or predictions.

As mentioned in some of the models which provide possible explanations for our CCFs, multiple spectral component may be involved, some of which span large ranges of the available EM spectrum. Both the optical range and 2--20\,keV in X-rays are relatively small windows, and it is not at present possible to disentangle spectral components, at least for the time-varying portion.
In order to separate the spectral components within the emission, much wider spectral coverage is required. Since fast IR devices are now starting to be available, coupled with wide spectral range X-ray satellites such as {\em SWIFT} and {\em Suzaku}, it is possible to extend this work. Naturally, observations of more sources in more states will help answer the ubiquity or otherwise of the behaviour presented here.

It is clear that anti-correlations in CCFs are not simply a curiosity restricted to a couple of sources. Here we have shown their appearance (for at least some of the time) in two black hole binaries and two NS binaries. Given the relatively sparse sampling and short datasets examined so far, their presence here may suggest that this phenomenon is far more common than previously thought. Longer coverage during various X-ray states is required to test this hypothesis.

\section*{Acknowledgments}
MD was supported by the Spanish Ministry of Science under grant
AYA2007-66887 during part of this work, and partly
supported through the US National Science Foundation grant No.\  0908733; RC, TS and JC received
some support from Consolider-Ingeniero 2010 Program grant
CSD2006-00070; RC acknowledges a Ramon y Cajal
fellowship (RYC-2007-01046).
RXTE is operated by NASA, and ULTRACAM was used as a visitor instrument at ESO and La Palma, Spain.

\bibliography{database}

\begin{thebibliography}{}

\bibitem[\protect\citeauthoryear{{Bartlett}}{{Bartlett}}{1955}]{Bartlettbook}
{Bartlett} M.~S.,  1955, {An Introduction to Stochastic Processes}.
CUP

\bibitem[\protect\citeauthoryear{{Bradshaw}, {Titarchuk} \&
  {Kuznetsov}}{{Bradshaw} et~al.}{2007}]{2007ApJ...663.1225B}
{Bradshaw} C.~F.,  {Titarchuk} L.,    {Kuznetsov} S.,  2007, ApJ, 663, 1225

\bibitem[\protect\citeauthoryear{{Casares}, {Steeghs}, {Hynes}, {Charles} \&
  {O'Brien}}{{Casares} et~al.}{2003}]{2003ApJ...590.1041C}
{Casares} J.,  {Steeghs} D.,  {Hynes} R.~I.,  {Charles} P.~A.,    {O'Brien} K.,
   2003, ApJ, 590, 1041

\bibitem[\protect\citeauthoryear{{Casella}, {Maccarone}, {O'Brien}, {Fender},
  {Russell}, {van der Klis}, {Pe'Er}, {Maitra}, {Altamirano}, {Belloni},
  {Kanbach}, {Klein-Wolt}, {Mason}, {Soleri}, {Stefanescu}, {Wiersema} \&
  {Wijnands}}{{Casella} et~al.}{2010}]{2010MNRAS.404L..21C}
{Casella} P.,  {Maccarone} T.~J.,  {O'Brien} K.,  {Fender} R.~P.,  {Russell}
  D.~M.,  {van der Klis} M.,  {Pe'Er} A.,  {Maitra} D.,  {Altamirano} D.,
  {Belloni} T.,  {Kanbach} G.,  {Klein-Wolt} M.,  {Mason} E.,  {Soleri} P.,
  {Stefanescu} A.,  {Wiersema} K.,    {Wijnands} R.,  2010, MNRAS, 404, L21

\bibitem[\protect\citeauthoryear{{Casella} \& {Pe'er}}{{Casella} \&
  {Pe'er}}{2009}]{2009ApJ...703L..63C}
{Casella} P.,  {Pe'er} A.,  2009, ApJ, 703, L63

\bibitem[\protect\citeauthoryear{{Coriat}, {Corbel}, {Buxton}, {Bailyn},
  {Tomsick}, {K{\"o}rding} \& {Kalemci}}{{Coriat}
  et~al.}{2009}]{2009MNRAS.400..123C}
{Coriat} M.,  {Corbel} S.,  {Buxton} M.~M.,  {Bailyn} C.~D.,  {Tomsick} J.~A.,
  {K{\"o}rding} E.,    {Kalemci} E.,  2009, MNRAS, 400, 123

\bibitem[\protect\citeauthoryear{{D'A{\'{\i}}}, {{\.Z}ycki}, {Di Salvo},
  {Iaria}, {Lavagetto} \& {Robba}}{{D'A{\'{\i}}}
  et~al.}{2007}]{2007ApJ...667..411D}
{D'A{\'{\i}}} A.,  {{\.Z}ycki} P.,  {Di Salvo} T.,  {Iaria} R.,  {Lavagetto}
  G.,    {Robba} N.~R.,  2007, ApJ, 667, 411

\bibitem[\protect\citeauthoryear{{Dhillon} \& {Marsh}}{{Dhillon} \&
  {Marsh}}{2001}]{2001NewAR..45...91D}
{Dhillon} V.,  {Marsh} T.,  2001, New Astronomy Review, 45, 91

\bibitem[\protect\citeauthoryear{{Dhillon}, {Marsh}, {Stevenson}, {Atkinson},
  {Kerry}, {Peacocke}, {Vick}, {Beard}, {Ives}, {Lunney}, {McLay}, {Tierney},
  {Kelly}, {Littlefair}, {Nicholson}, {Pashley}, {Harlaftis} \&
  {O'Brien}}{{Dhillon} et~al.}{2007}]{2007MNRAS.378..825D}
{Dhillon} V.~S.,  {Marsh} T.~R.,  {Stevenson} M.~J.,  {Atkinson} D.~C.,
  {Kerry} P.,  {Peacocke} P.~T.,  {Vick} A.~J.~A.,  {Beard} S.~M.,  {Ives}
  D.~J.,  {Lunney} D.~W.,  {McLay} S.~A.,  {Tierney} C.~J.,  {Kelly} J.,
  {Littlefair} S.~P.,  {Nicholson} R.,  {Pashley} R.,  {Harlaftis} E.~T.,
  {O'Brien} K.,  2007, MNRAS, 378, 825

\bibitem[\protect\citeauthoryear{{Durant}, {Gandhi}, {Shahbaz}, {Fabian},
  {Miller}, {Dhillon} \& {Marsh}}{{Durant} et~al.}{2008}]{2008ApJ...682L..45D}
{Durant} M.,  {Gandhi} P.,  {Shahbaz} T.,  {Fabian} A.~P.,  {Miller} J.,
  {Dhillon} V.~S.,    {Marsh} T.~R.,  2008, ApJ, 682, L45

\bibitem[\protect\citeauthoryear{{Durant}, {Gandhi}, {Shahbaz}, {Peralta} \&
  {Dhillon}}{{Durant} et~al.}{2009}]{2009MNRAS.392..309D}
{Durant} M.,  {Gandhi} P.,  {Shahbaz} T.,  {Peralta} H.~H.,    {Dhillon} V.~S.,
   2009, MNRAS, 392, 309

\bibitem[\protect\citeauthoryear{{Eikenberry}, {Patel}, {Rothstein},
  {Remillard}, {Pooley} \& {Morgan}}{{Eikenberry}
  et~al.}{2008}]{2008ApJ...678..369E}
{Eikenberry} S.~S.,  {Patel} S.~G.,  {Rothstein} D.~M.,  {Remillard} R.,
  {Pooley} G.~G.,    {Morgan} E.~H.,  2008, ApJ, 678, 369

\bibitem[\protect\citeauthoryear{{Gandhi}}{{Gandhi}}{2009}]{2009ApJ...697L.167%
G}
{Gandhi} P.,  2009, ApJ, 697, L167

\bibitem[\protect\citeauthoryear{{Gandhi}, {Dhillon}, {Durant}, {Fabian},
  {Kubota}, {Makishima}, {Malzac}, {Marsh}, {Miller}, {Shahbaz}, {Spruit} \&
  {Casella}}{{Gandhi} et~al.}{2010}]{2010MNRAS.tmp.1137G}
{Gandhi} P.,  {Dhillon} V.~S.,  {Durant} M.,  {Fabian} A.~C.,  {Kubota} A.,
  {Makishima} K.,  {Malzac} J.,  {Marsh} T.~R.,  {Miller} J.~M.,  {Shahbaz} T.,
   {Spruit} H.~C.,    {Casella} P.,  2010, MNRAS, pp 1137--+

\bibitem[\protect\citeauthoryear{{Gandhi}, {Makishima}, {Durant}, {Fabian},
  {Dhillon}, {Marsh}, {Miller}, {Shahbaz} \& {Spruit}}{{Gandhi}
  et~al.}{2008}]{2008MNRAS.390L..29G}
{Gandhi} P.,  {Makishima} K.,  {Durant} M.,  {Fabian} A.~C.,  {Dhillon} V.~S.,
  {Marsh} T.~R.,  {Miller} J.~M.,  {Shahbaz} T.,    {Spruit} H.~C.,  2008,
  MNRAS, 390, L29

\bibitem[\protect\citeauthoryear{{Grindley}, {McClintock}, {Canizares}, {van
  Paradijs}, {Cominsky}, {LI} \& {Lewin}}{{Grindley}
  et~al.}{1978}]{1978Nature...274...567M}
{Grindley} J.~E.,  {McClintock} J.~E.,  {Canizares} C.~R.,  {van Paradijs} J.,
  {Cominsky} L.,  {LI} F.~K.,    {Lewin} W.~H.~G.,  1978, Nature, 274, 567

\bibitem[\protect\citeauthoryear{{Hynes}}{{Hynes}}{2005}]{2005ASPC..330..237H}
{Hynes} R.~I.,  2005, in {J.-M.~Hameury \& J.-P.~Lasota} ed., The Astrophysics
  of Cataclysmic Variables and Related Objects Vol.~330 of Astronomical Society
  of the Pacific Conference Series, {Correlated X-ray and Optical Variability
  in X-ray Binaries}.
p.~237

\bibitem[\protect\citeauthoryear{{Hynes}, {Haswell}, {Cui}, {Shrader},
  {O'Brien}, {Chaty}, {Skillman}, {Patterson} \& {Horne}}{{Hynes}
  et~al.}{2003}]{2003MNRAS.345..292H}
{Hynes} R.~I.,  {Haswell} C.~A.,  {Cui} W.,  {Shrader} C.~R.,  {O'Brien} K.,
  {Chaty} S.,  {Skillman} D.~R.,  {Patterson} J.,    {Horne} K.,  2003, MNRAS,
  345, 292

\bibitem[\protect\citeauthoryear{{Hynes}, {Horne}, {O'Brien}, {Haswell},
  {Robinson}, {King}, {Charles} \& {Pearson}}{{Hynes}
  et~al.}{2006}]{2006ApJ...648.1156H}
{Hynes} R.~I.,  {Horne} K.,  {O'Brien} K.,  {Haswell} C.~A.,  {Robinson} E.~L.,
   {King} A.~R.,  {Charles} P.~A.,    {Pearson} K.~J.,  2006, ApJ, 648, 1156

\bibitem[\protect\citeauthoryear{{Jahoda}, {Swank}, {Giles}, {Stark},
  {Strohmayer}, {Zhang} \& {Morgan}}{{Jahoda}
  et~al.}{1996}]{1996SPIE.2808...59J}
{Jahoda} K.,  {Swank} J.~H.,  {Giles} A.~B.,  {Stark} M.~J.,  {Strohmayer} T.,
  {Zhang} W.,    {Morgan} E.~H.,  1996, in {O.~H.~Siegmund \& M.~A.~Gummin}
  ed., Society of Photo-Optical Instrumentation Engineers (SPIE) Conference
  Series Vol.~2808 of Presented at the Society of Photo-Optical Instrumentation
  Engineers (SPIE) Conference, {In-orbit performance and calibration of the
  Rossi X-ray Timing Explorer (RXTE) Proportional Counter Array (PCA)}.
pp 59--70

\bibitem[\protect\citeauthoryear{{Kanbach}, {Straubmeier}, {Spruit} \&
  {Belloni}}{{Kanbach} et~al.}{2001}]{2001Natur.414..180K}
{Kanbach} G.,  {Straubmeier} C.,  {Spruit} H.~C.,    {Belloni} T.,  2001,
  Nature, 414, 180

\bibitem[\protect\citeauthoryear{{Lewin}, {van Paradijs} \& {van den
  Heuvel}}{{Lewin} et~al.}{1997}]{1997xrb..book.....L}
{Lewin} W.~H.~G.,  {van Paradijs} J.,    {van den Heuvel} E.~P.~J.,  1997,
  {X-ray Binaries}.
CUP

\bibitem[\protect\citeauthoryear{{Lovelace}, {Rothstein} \&
  {Bisnovatyi-Kogan}}{{Lovelace} et~al.}{2009}]{2009ApJ...701..885L}
{Lovelace} R.~V.~E.,  {Rothstein} D.~M.,    {Bisnovatyi-Kogan} G.~S.,  2009,
  ApJ, 701, 885

\bibitem[\protect\citeauthoryear{{Malzac}, {Belloni}, {Spruit} \&
  {Kanbach}}{{Malzac} et~al.}{2003}]{2003A&A...407..335M}
{Malzac} J.,  {Belloni} T.,  {Spruit} H.~C.,    {Kanbach} G.,  2003, A\&Ap,
  407, 335

\bibitem[\protect\citeauthoryear{{Malzac}, {Merloni} \& {Fabian}}{{Malzac}
  et~al.}{2004}]{2004MNRAS.351..253M}
{Malzac} J.,  {Merloni} A.,    {Fabian} A.~C.,  2004, MNRAS, 351, 253

\bibitem[\protect\citeauthoryear{{Markoff}, {Nowak} \& {Wilms}}{{Markoff}
  et~al.}{2005}]{2005ApJ...635.1203M}
{Markoff} S.,  {Nowak} M.~A.,    {Wilms} J.,  2005, ApJ, 635, 1203

\bibitem[\protect\citeauthoryear{{McGowan}, {Charles}, {O'Donoghue} \&
  {Smale}}{{McGowan} et~al.}{2003}]{2003MNRAS.345.1039M}
{McGowan} K.~E.,  {Charles} P.~A.,  {O'Donoghue} D.,    {Smale} A.~P.,  2003,
  MNRAS, 345, 1039

\bibitem[\protect\citeauthoryear{{Mook}, {Hiltner} \& {Lynds}}{{Mook}
  et~al.}{1971}]{1971ApJ...163L..69M}
{Mook} D.,  {Hiltner} W.~A.,    {Lynds} R.,  1971, ApJ, 163, L69+

\bibitem[\protect\citeauthoryear{{Mu{\~n}oz-Darias}, {Mart{\'{\i}}nez-Pais},
  {Casares}, {Dhillon}, {Marsh}, {Cornelisse}, {Steeghs} \&
  {Charles}}{{Mu{\~n}oz-Darias} et~al.}{2007}]{2007MNRAS.379.1637M}
{Mu{\~n}oz-Darias} T.,  {Mart{\'{\i}}nez-Pais} I.~G.,  {Casares} J.,  {Dhillon}
  V.~S.,  {Marsh} T.~R.,  {Cornelisse} R.,  {Steeghs} D.,    {Charles} P.~A.,
  2007, MNRAS, 379, 1637

\bibitem[\protect\citeauthoryear{{Pedersen}, {Lub}, {Inoue}, {Koyama},
  {Makishima}, {Matsuoka}, {Mitsuda}, {Murakami}, {Oda}, {Ogawara}, {Ohashi},
  {Shibazaki}, {Tanaka}, {Hayakawa}, {Kunieda}, {Makino}, {Masai}, {Nagase} \&
  {Tawara}}{{Pedersen} et~al.}{1982}]{1982ApJ...263..325P}
{Pedersen} H.,  {Lub} J.,  {Inoue} H.,  {Koyama} K.,  {Makishima} K.,
  {Matsuoka} M.,  {Mitsuda} K.,  {Murakami} T.,  {Oda} M.,  {Ogawara} Y.,
  {Ohashi} T.,  {Shibazaki} N.,  {Tanaka} Y.,  {Hayakawa} S.,  {Kunieda} H.,
  {Makino} F.,  {Masai} K.,  {Nagase} F.,    {Tawara} Y.,  1982, ApJ, 263, 325

\bibitem[\protect\citeauthoryear{{Raiteri}, {Villata}, {Larionov}, {Aller},
  {Bach}, {Gurwell}, {Kurtanidze}, {L{\"a}hteenm{\"a}ki}, {Nilsson}, {Volvach},
  {Aller}, {Arkharov}, {Bachev}, {Berdyugin}, {B{\"o}ttcher} \&
  {Buemi}}{{Raiteri} et~al.}{2008}]{2008A&A...480..339R}
{Raiteri} C.~M.,  {Villata} M.,  {Larionov} V.~M.,  {Aller} M.~F.,  {Bach} U.,
  {Gurwell} M.,  {Kurtanidze} O.~M.,  {L{\"a}hteenm{\"a}ki} A.,  {Nilsson} K.,
  {Volvach} A.,  {Aller} H.~D.,  {Arkharov} A.~A.,  {Bachev} R.,  {Berdyugin}
  A.,  {B{\"o}ttcher} M.,    {Buemi} C.~S.,  2008, A\&A, 480, 339

\bibitem[\protect\citeauthoryear{{Smith} \& {Vaughan}}{{Smith} \&
  {Vaughan}}{2007}]{2007MNRAS.375.1479S}
{Smith} R.,  {Vaughan} S.,  2007, MNRAS, 375, 1479

\bibitem[\protect\citeauthoryear{{Spruit} \& {Kanbach}}{{Spruit} \&
  {Kanbach}}{2002}]{2002A&A...391..225S}
{Spruit} H.~C.,  {Kanbach} G.,  2002, A\&A, 391, 225

\bibitem[\protect\citeauthoryear{{Uttley}, {McHardy} \& {Vaughan}}{{Uttley}
  et~al.}{2005}]{2005MNRAS.359..345U}
{Uttley} P.,  {McHardy} I.~M.,    {Vaughan} S.,  2005, MNRAS, 359, 345

\bibitem[\protect\citeauthoryear{{van der Klis}}{{van der
  Klis}}{2004}]{2004astro.ph.10551V}
{van der Klis} M.,  2004, ArXiv:astro-ph/0410551

\bibitem[\protect\citeauthoryear{{van Paradijs} \& {McClintock}}{{van Paradijs}
  \& {McClintock}}{1994}]{1994A&A...290..133V}
{van Paradijs} J.,  {McClintock} J.~E.,  1994, A\&A, 290, 133

\bibitem[\protect\citeauthoryear{{Wu}, {Soria}, {Hunstead} \& {Johnston}}{{Wu}
  et~al.}{2001}]{2001MNRAS.320..177W}
{Wu} K.,  {Soria} R.,  {Hunstead} R.~W.,    {Johnston} H.~M.,  2001, MNRAS,
  320, 177

\bibitem[\protect\citeauthoryear{{Yuan}, {Cui} \& {Narayan}}{{Yuan}
  et~al.}{2005}]{2005ApJ...620..905Y}
{Yuan} F.,  {Cui} W.,    {Narayan} R.,  2005, ApJ, 620, 905

\end{thebibliography}
\bsp

\appendix
\section{Cross correlations}\label{data}
\begin{figure*}
\includegraphics[width=0.9\hsize]{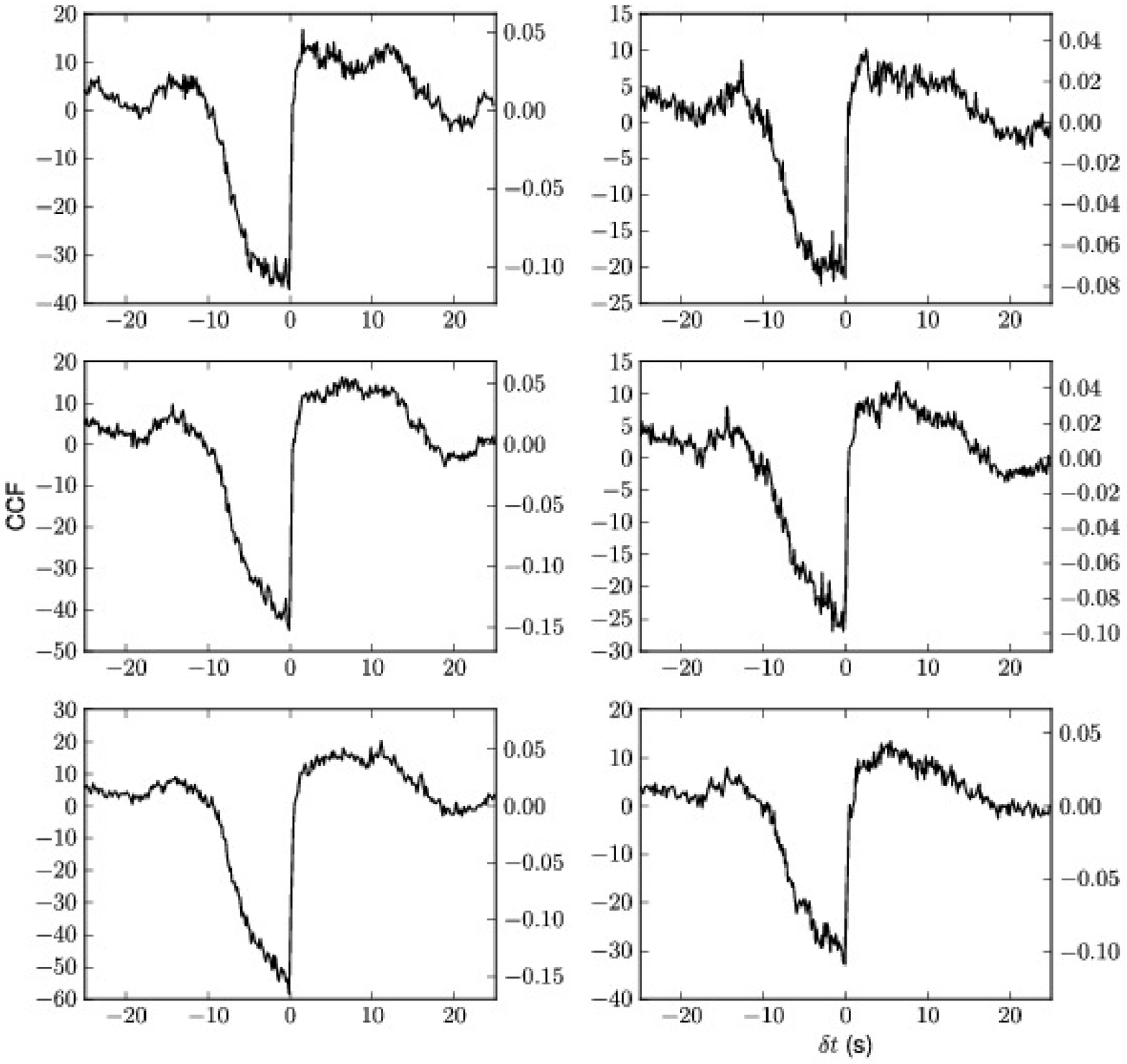}
\caption{\swift, 2007-06-12, 93119-02-02-00}\label{first_ccf}
\end{figure*}

\begin{figure*}
\includegraphics[width=0.9\hsize]{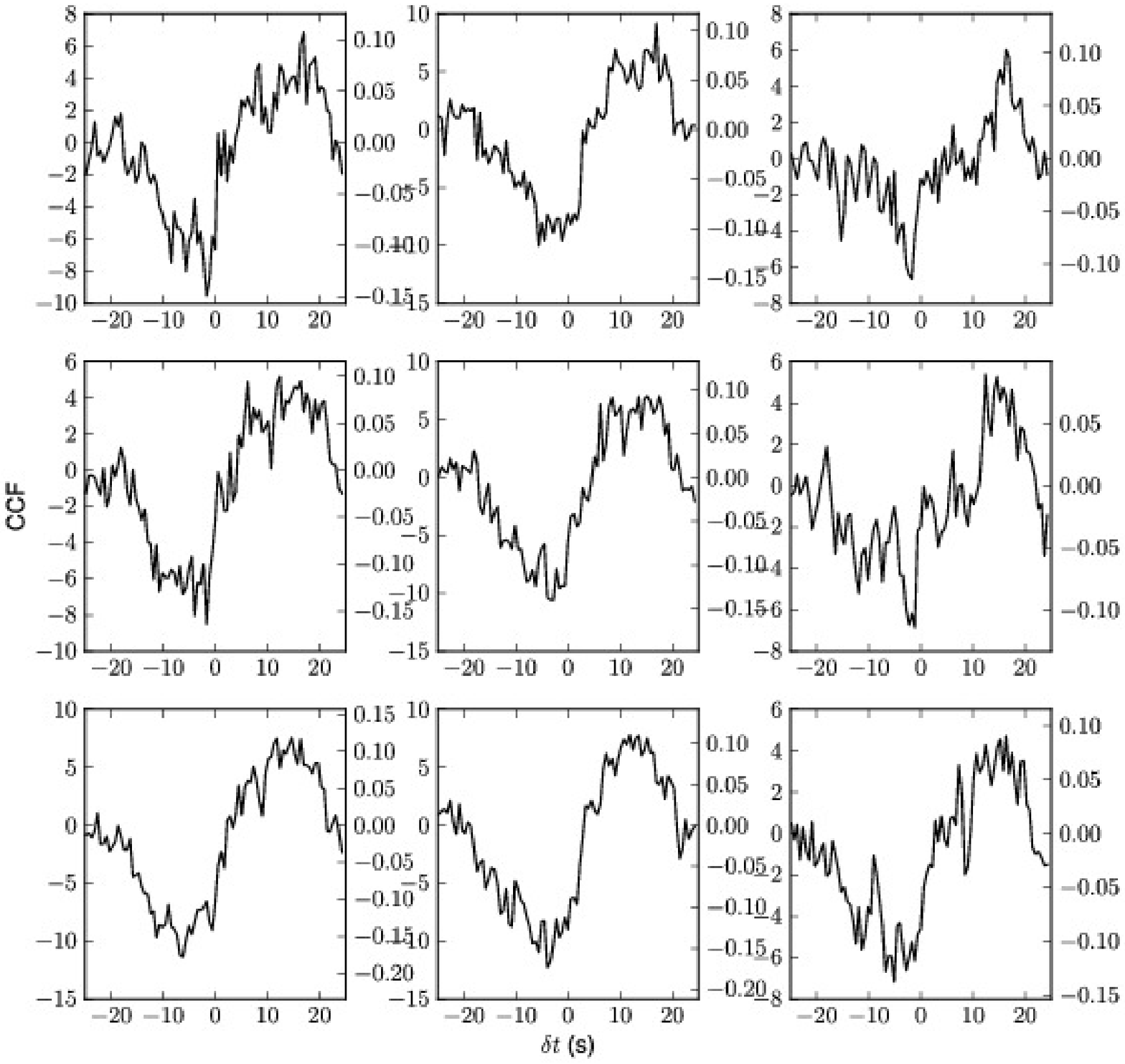}
\caption{\sco, 2004-05-18, 90020-01-02-01}\label{duplicate_ave}
\end{figure*}

\begin{figure*}
\includegraphics[width=0.9\hsize]{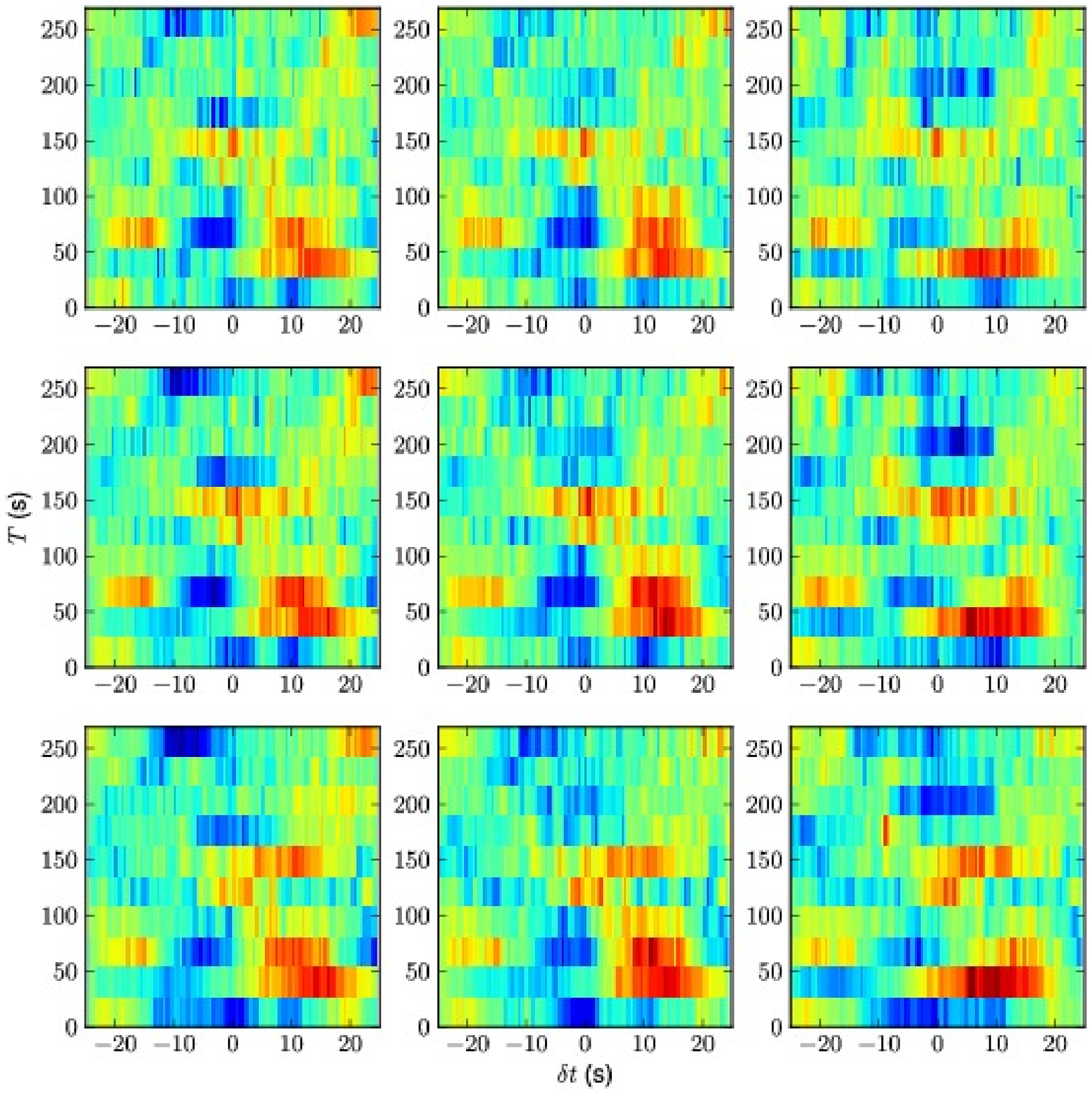}
\caption{\sco, 2004-05-18, 90020-01-02-01}\label{duplicate_dyn}
\end{figure*}

\begin{figure*}
\includegraphics[width=0.9\hsize]{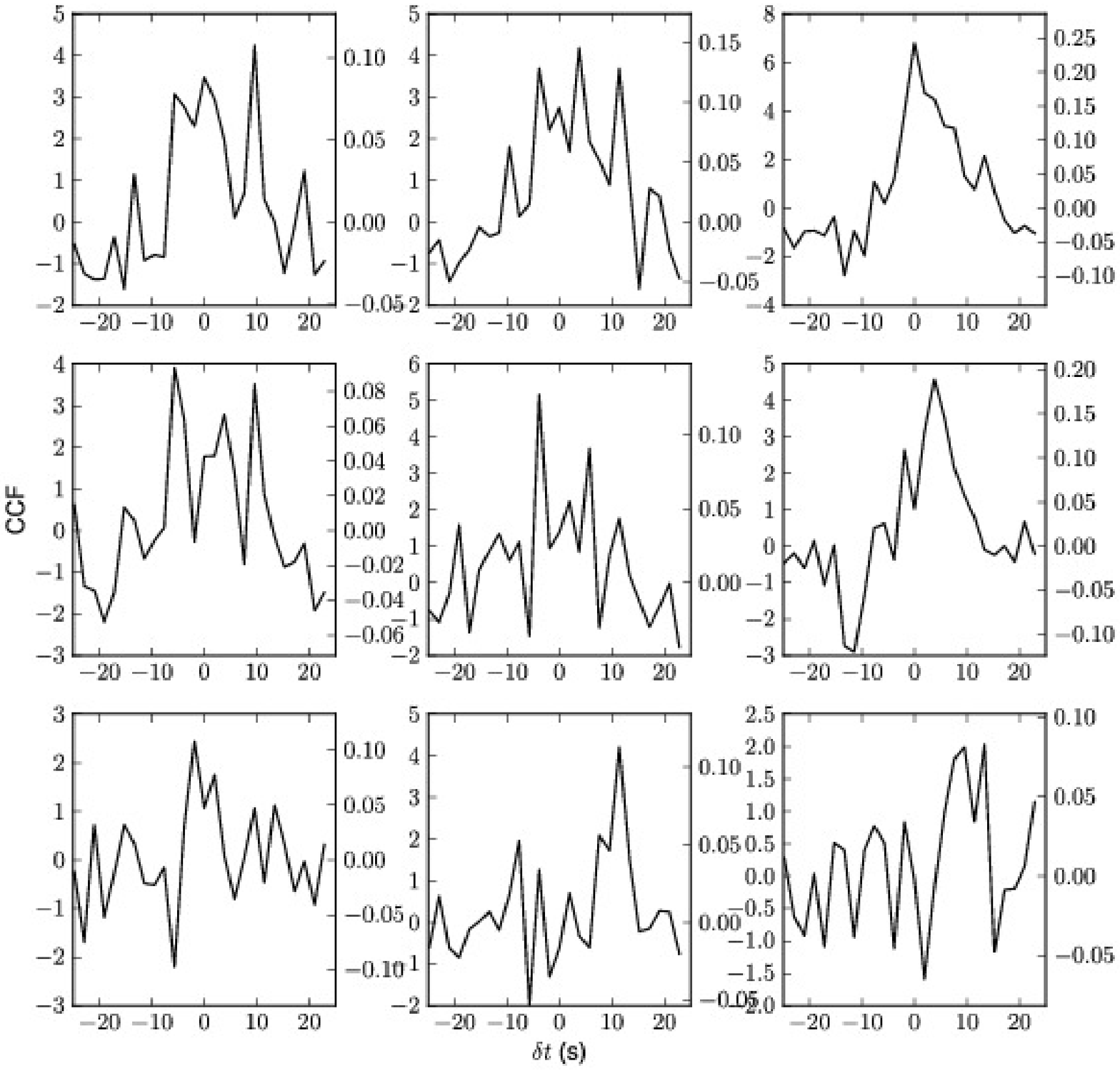}
\caption{\cyg, 2007-10-20, 92038-01-06-00}\label{dup_2_ave}
\end{figure*}

\begin{figure*}
\includegraphics[width=0.9\hsize]{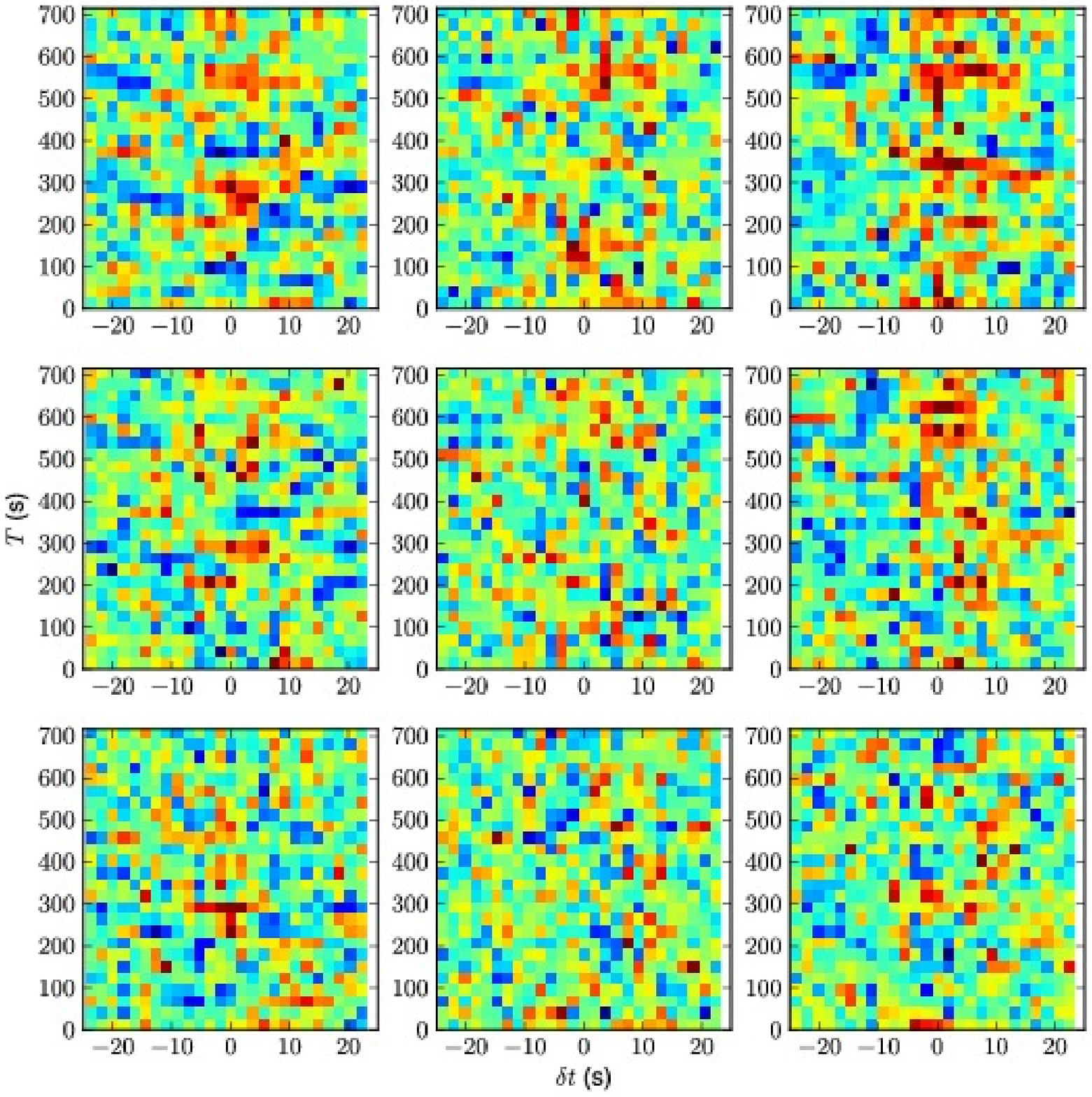}
\caption{\cyg, 2007-10-20, 92038-01-06-00}\label{dup_2_dyn}
\end{figure*}

\begin{figure*}
\includegraphics[width=0.9\hsize]{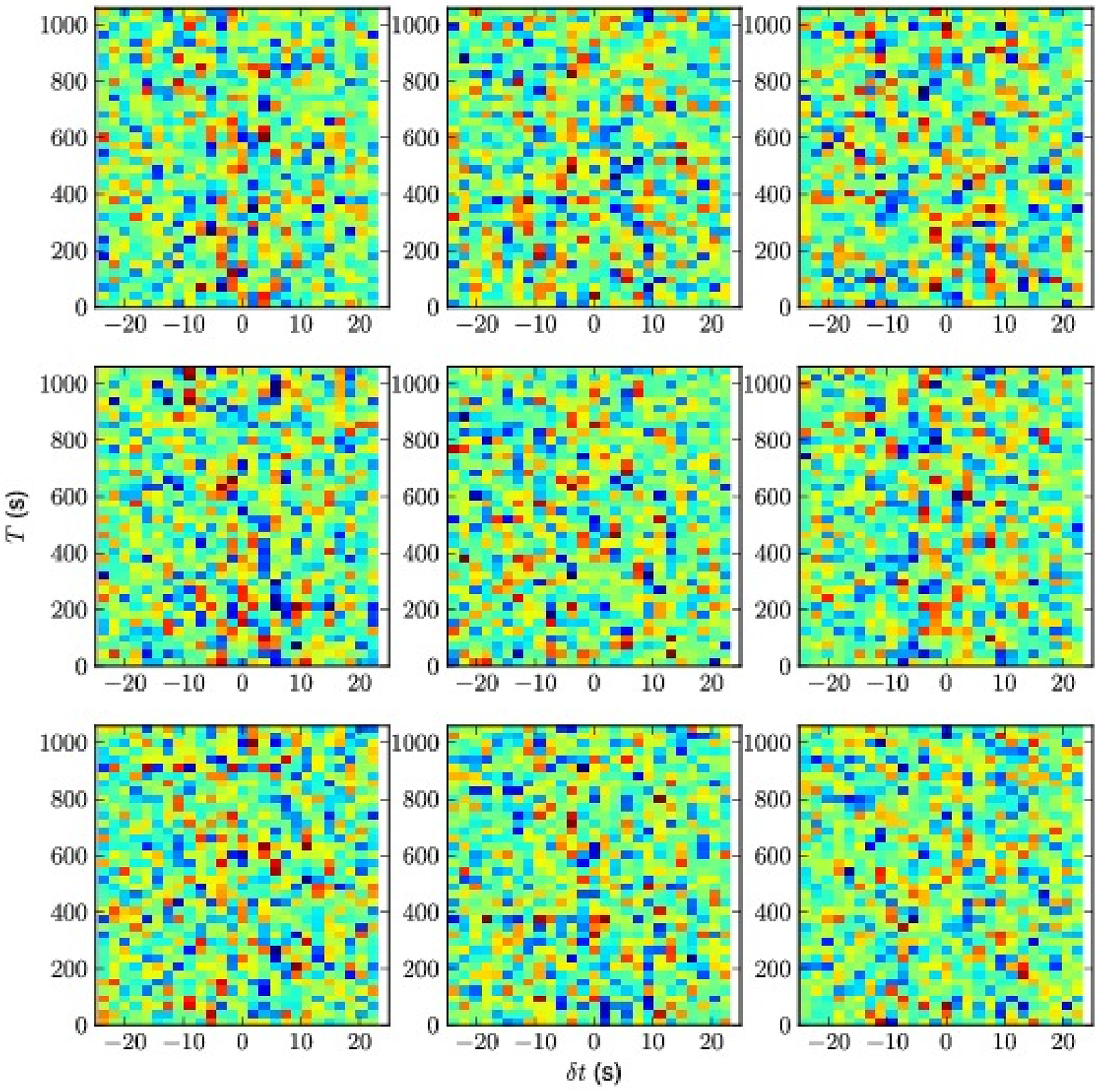}
\caption{\cyg, 2007-10-21, 92038-01-07-01}\label{last_ccf}
\end{figure*}

In Figures \ref{first_ccf} to \ref{last_ccf} we show all the CCFs generated in our work, in both averaged and dynamic form. There are nine panels per figure, with the optical bands red, green, UV in columns left-to-right and X-ray bands 2--4\,keV, 4--8\,keV, 8--20\,keV in rows top to bottom. The exception is \swift, which was too faint in the UV band for us to attempt accurate photometry. Each figure is captioned by the Object name, the date, and the RXTE dataset analysed. 

For the average CCF panels, the left-hand y-axis scales are in terms of the uncertainty estimate $\sigma$ (see Section 3.2), and can be thought of as a rough measure of significance. We have kept the y-axis scale the same for all the CCFs of a particular object, to enable a fair comparison of changes in the CCFs. In each case, the absolute CCF value is also given on the right-hand y-axis scale (where the range now depends on the noise in each data-set). 

For the dynamic CCF plots, we have kept the same colour scaling throughout, since the noise is not necessarily a constant throughout the observation, and the concept of significance becomes somewhat muddied. Note, however, that the y-axis scale for the dynamic CCF plots are {\em not} the same, as this reflects the length in time of each simultaneous observation.

\newpage

\begin{figure*}
\includegraphics[width=0.9\hsize]{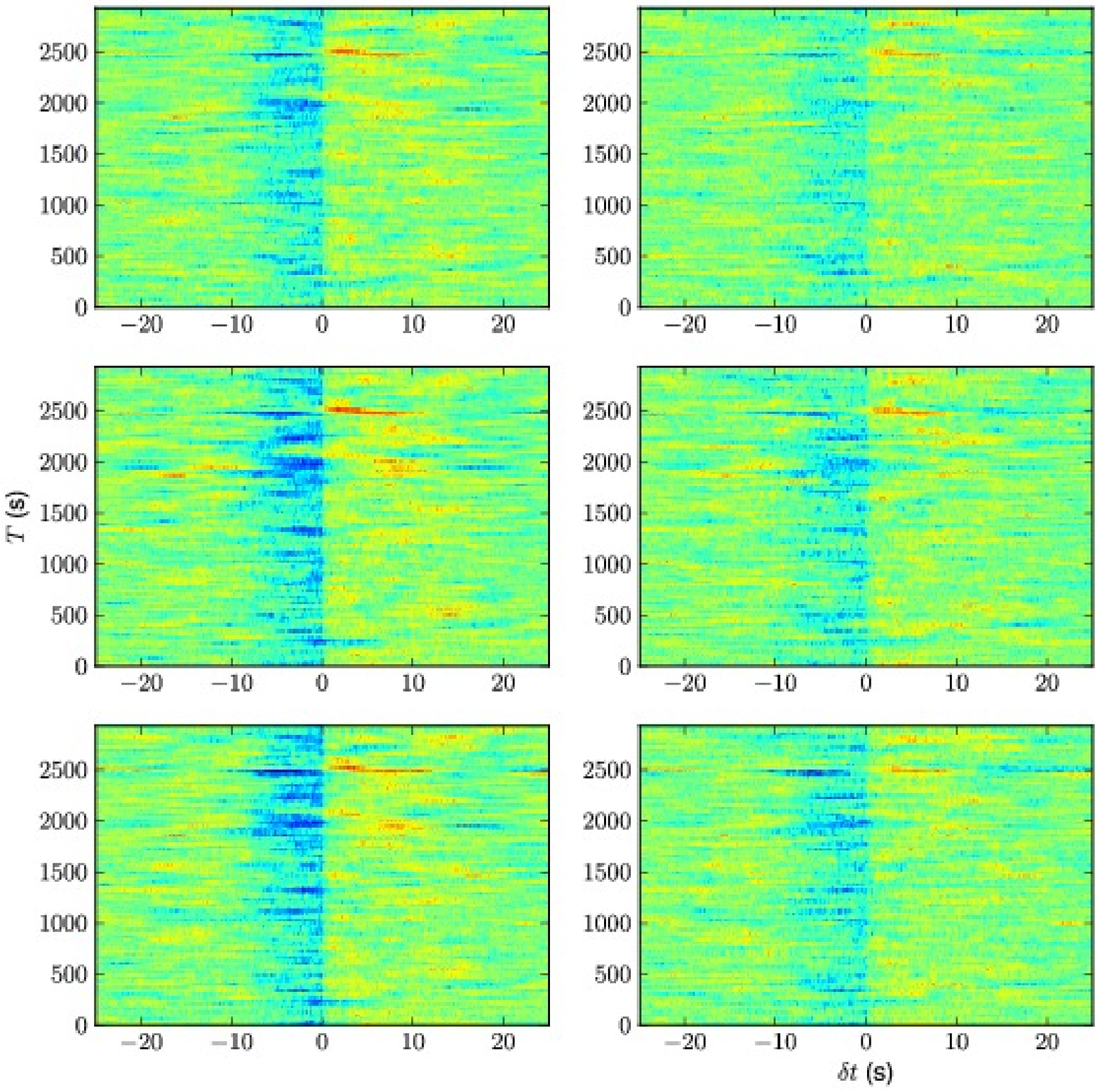}
\caption{\swift, 2007-06-12, 93119-02-02-00}
\end{figure*}

\begin{figure*}
\includegraphics[width=0.9\hsize]{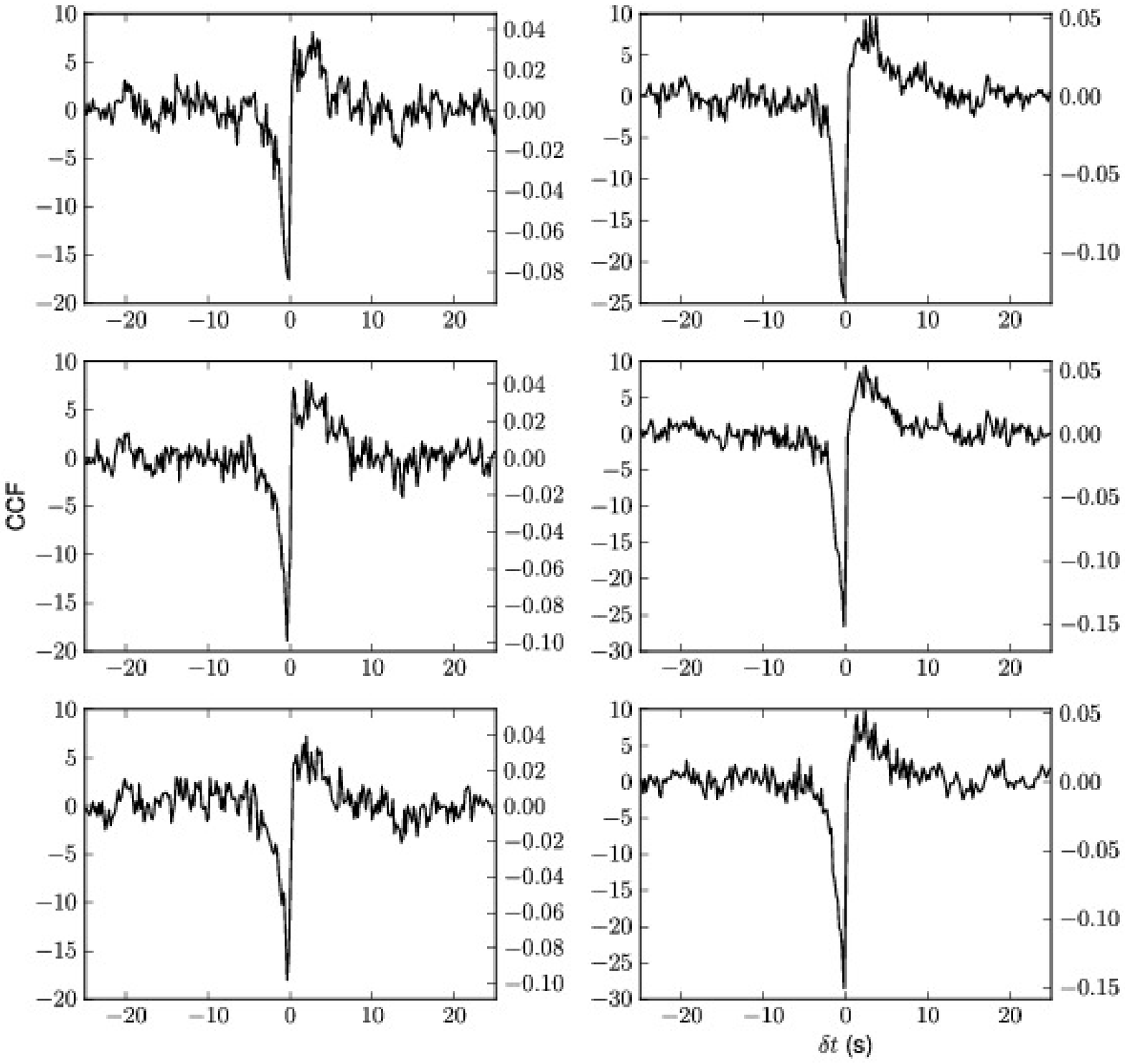}
\caption{\swift, 2008-08-20, 93105-01-56-00}
\end{figure*}

\begin{figure*}
\includegraphics[width=0.9\hsize]{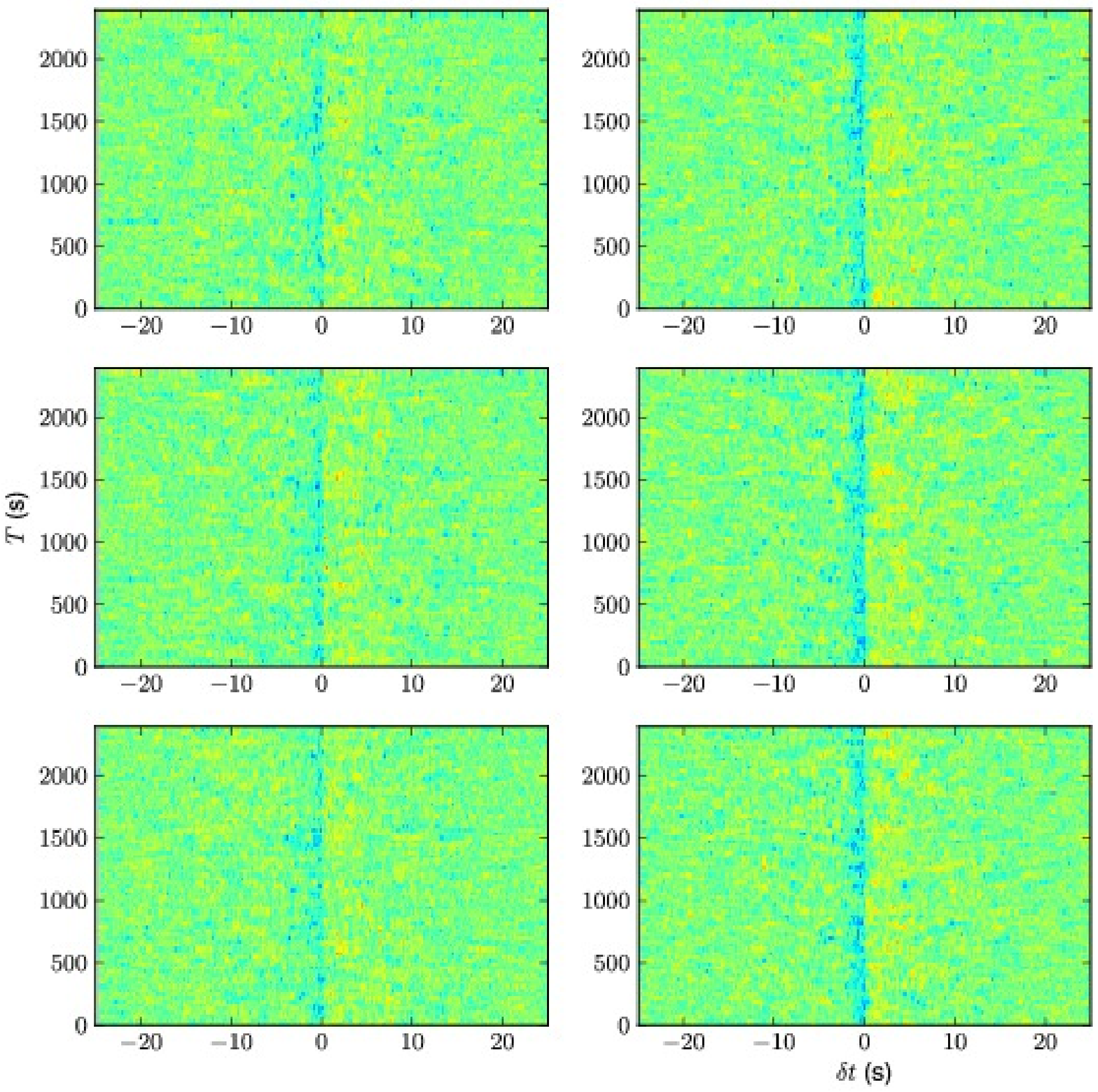}
\caption{\swift, 2008-08-20, 93105-01-56-00}
\end{figure*}

\begin{figure*}
\includegraphics[width=0.9\hsize]{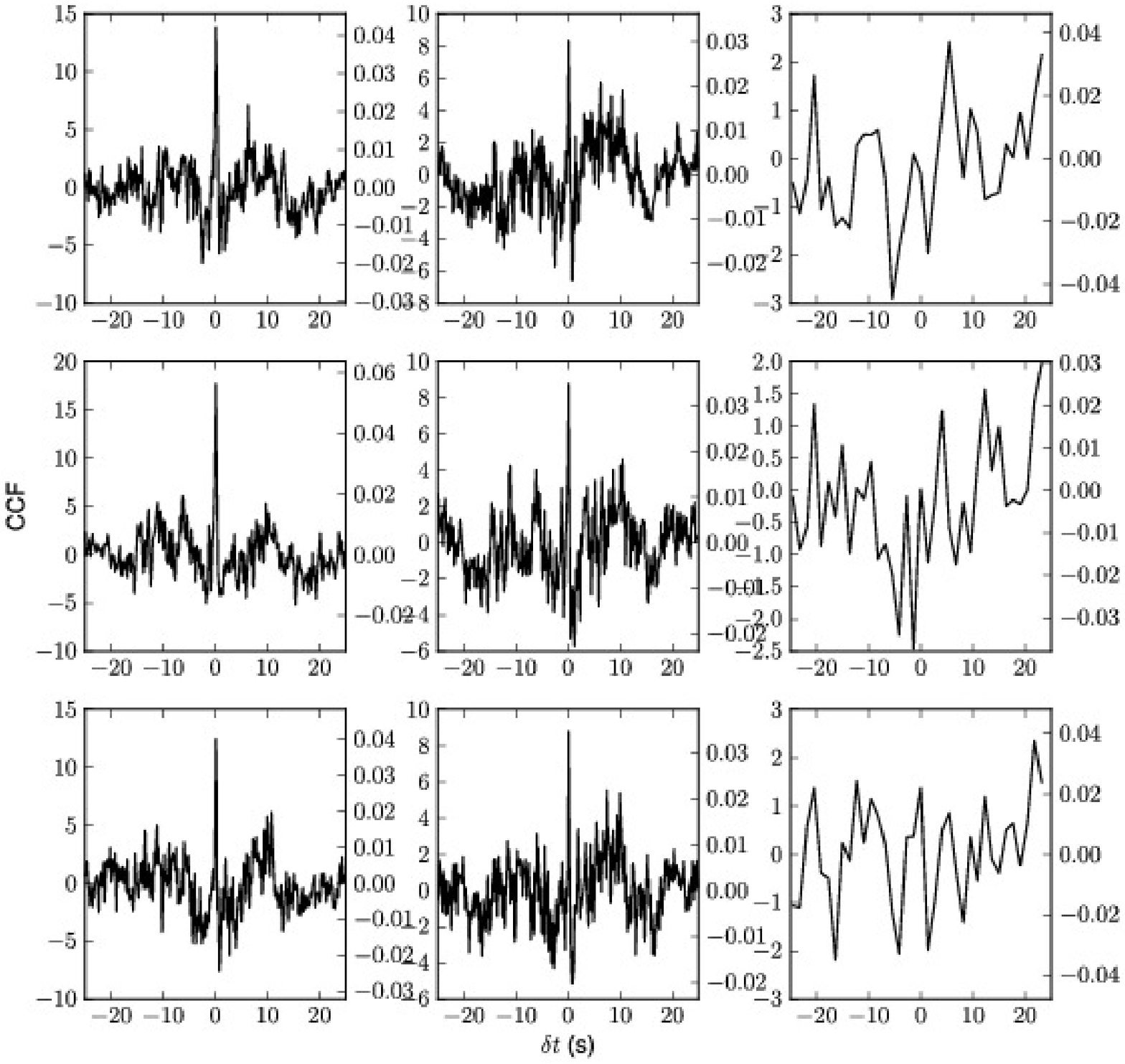}
\caption{\gx, 2007-06-13, 93119-01-02}
\end{figure*}

\begin{figure*}
\includegraphics[width=0.9\hsize]{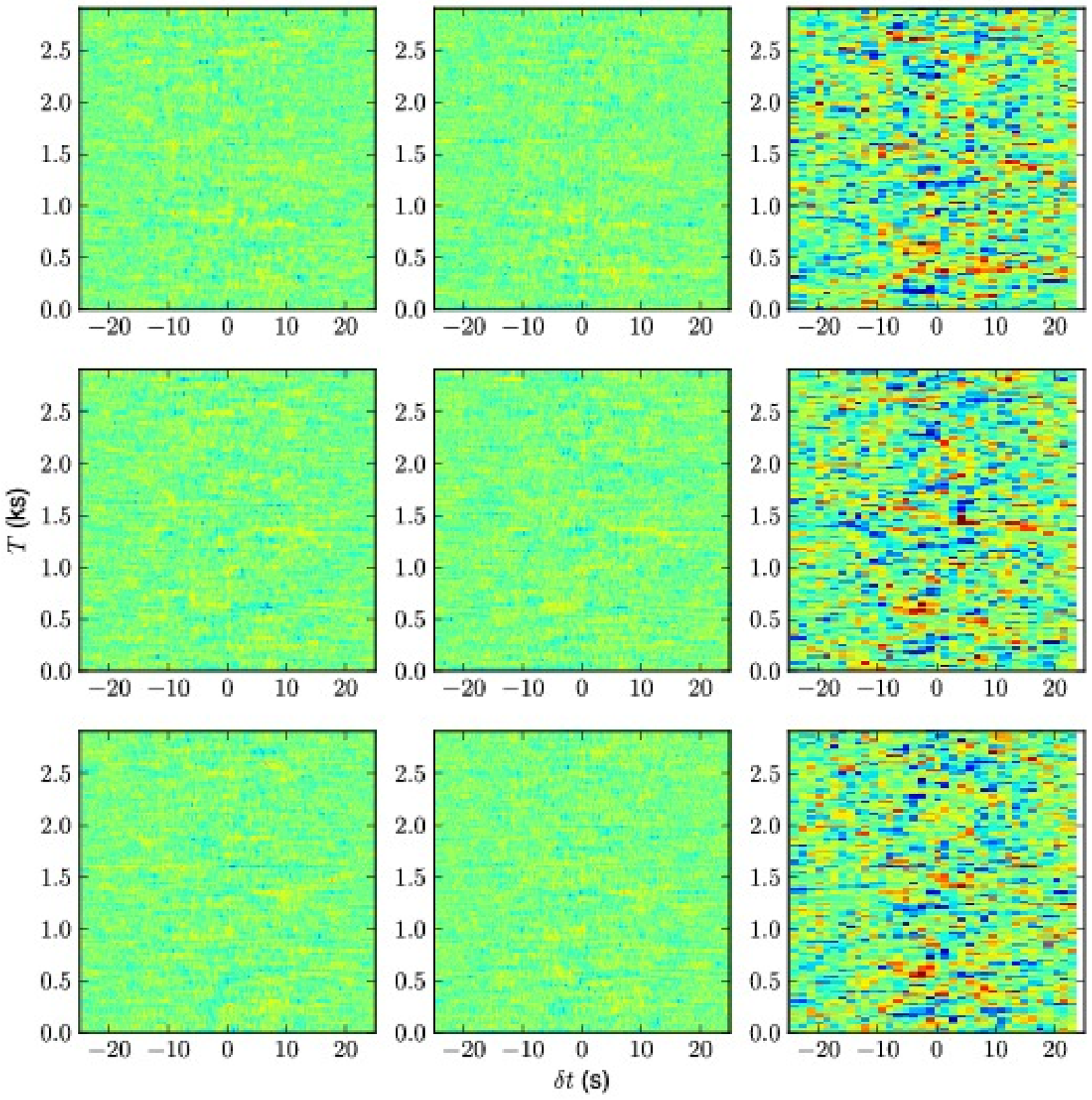}
\caption{\gx, 2007-06-13, 93119-01-02}
\end{figure*}

\begin{figure*}
\includegraphics[width=0.9\hsize]{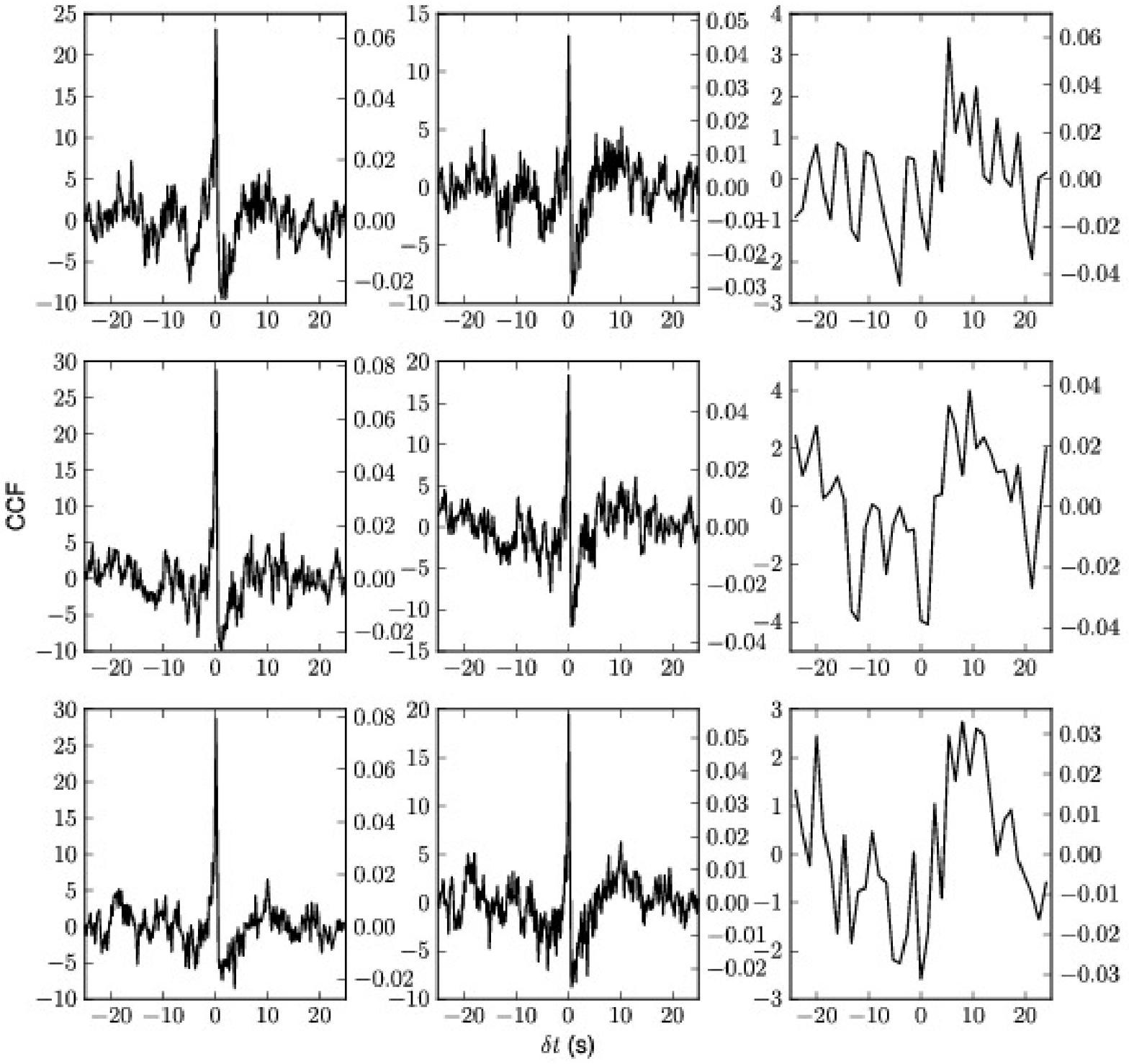}
\caption{\gx, 2007-06-16, 93119-01-03}
\end{figure*}

\begin{figure*}
\includegraphics[width=0.9\hsize]{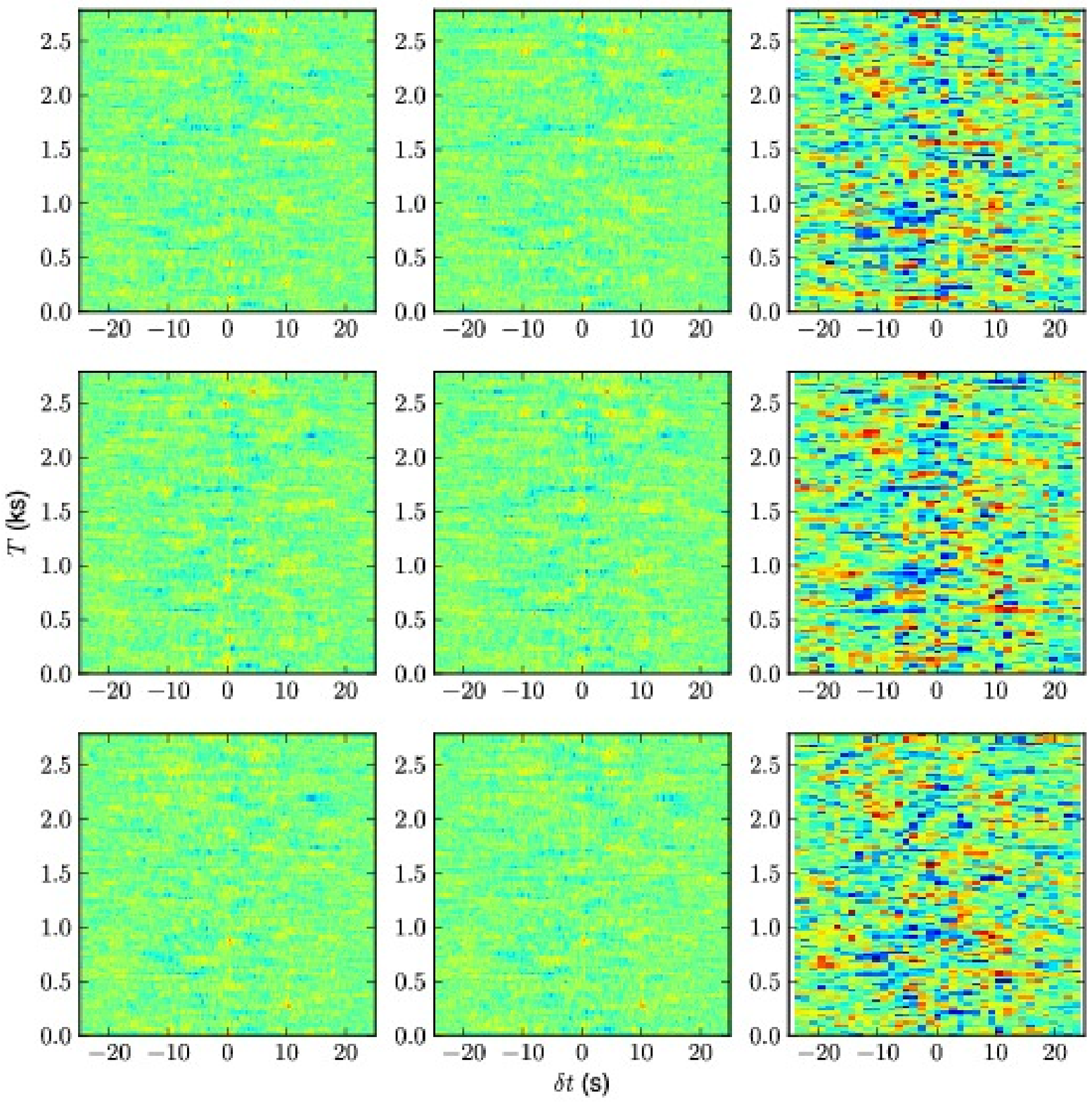}
\caption{\gx, 2007-06-16, 93119-01-03}
\end{figure*}

\begin{figure*}
\includegraphics[width=0.9\hsize]{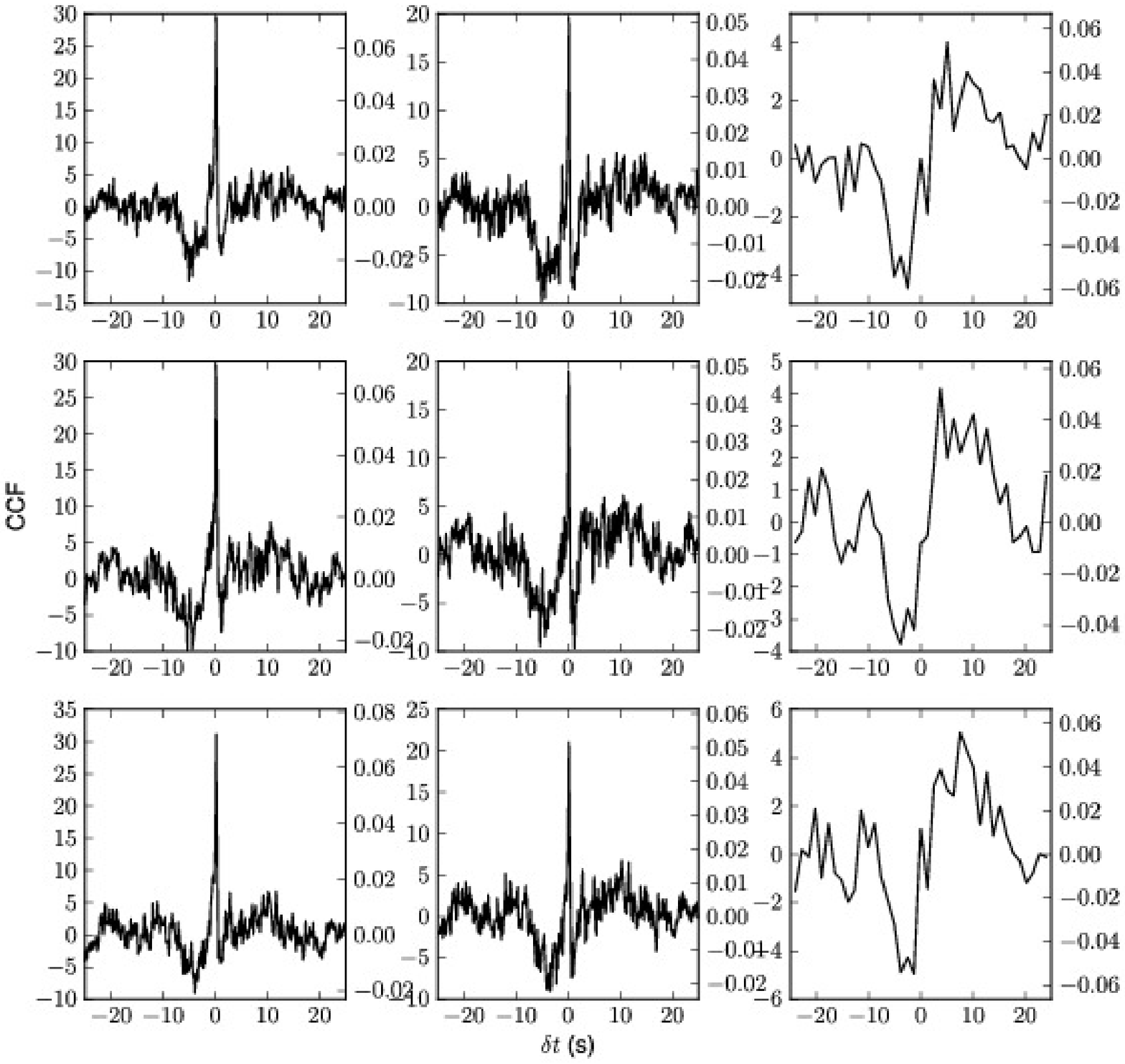}
\caption{\gx, 2007-06-18, 93119-01-04}
\end{figure*}

\begin{figure*}
\includegraphics[width=0.9\hsize]{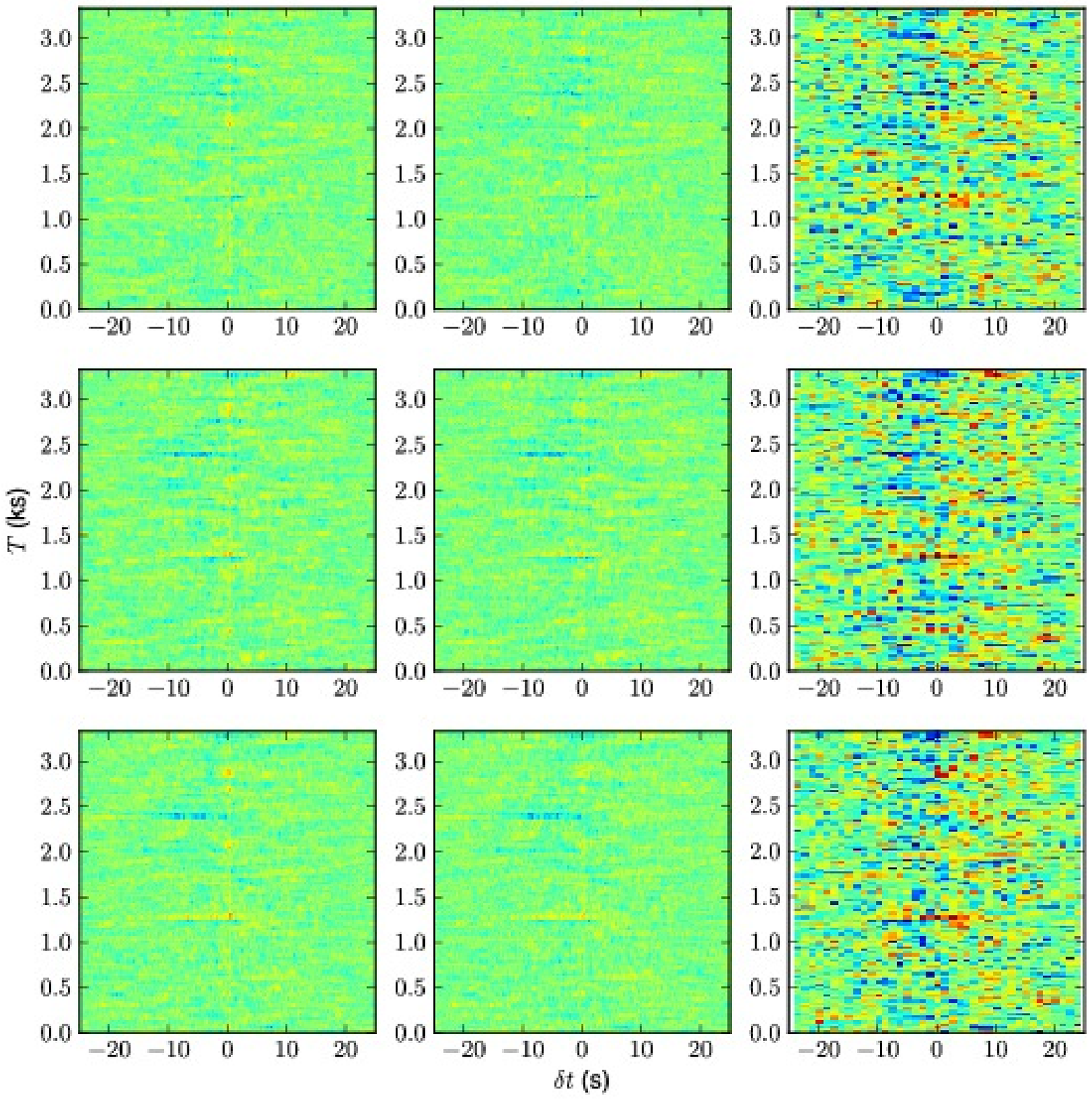}
\caption{\gx, 2007-06-18, 93119-01-04}
\end{figure*}

\begin{figure*}
\includegraphics[width=0.9\hsize]{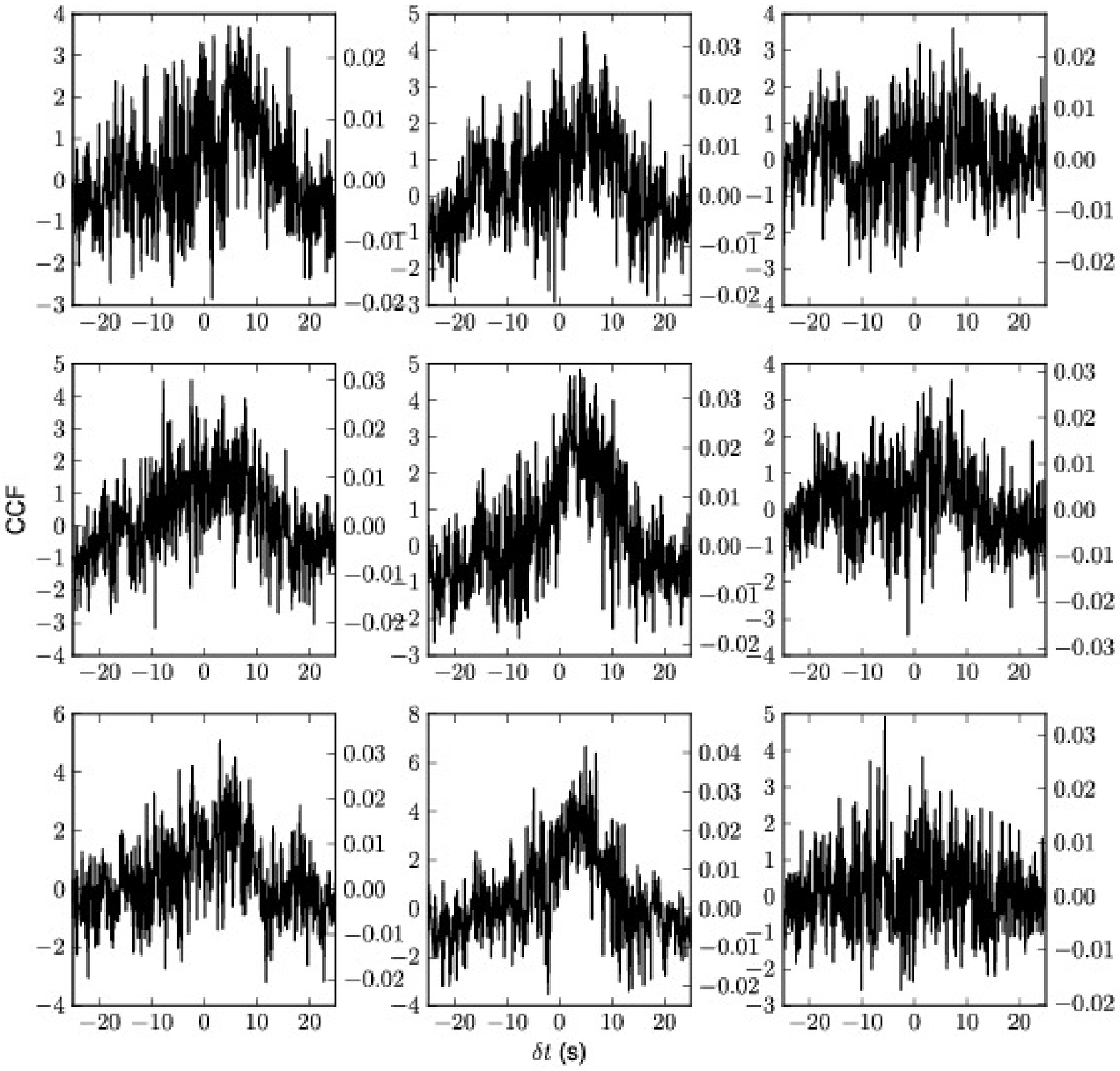}
\caption{\sco, 2004-05-17, 90020-01-01-01}
\end{figure*}

\begin{figure*}
\includegraphics[width=0.9\hsize]{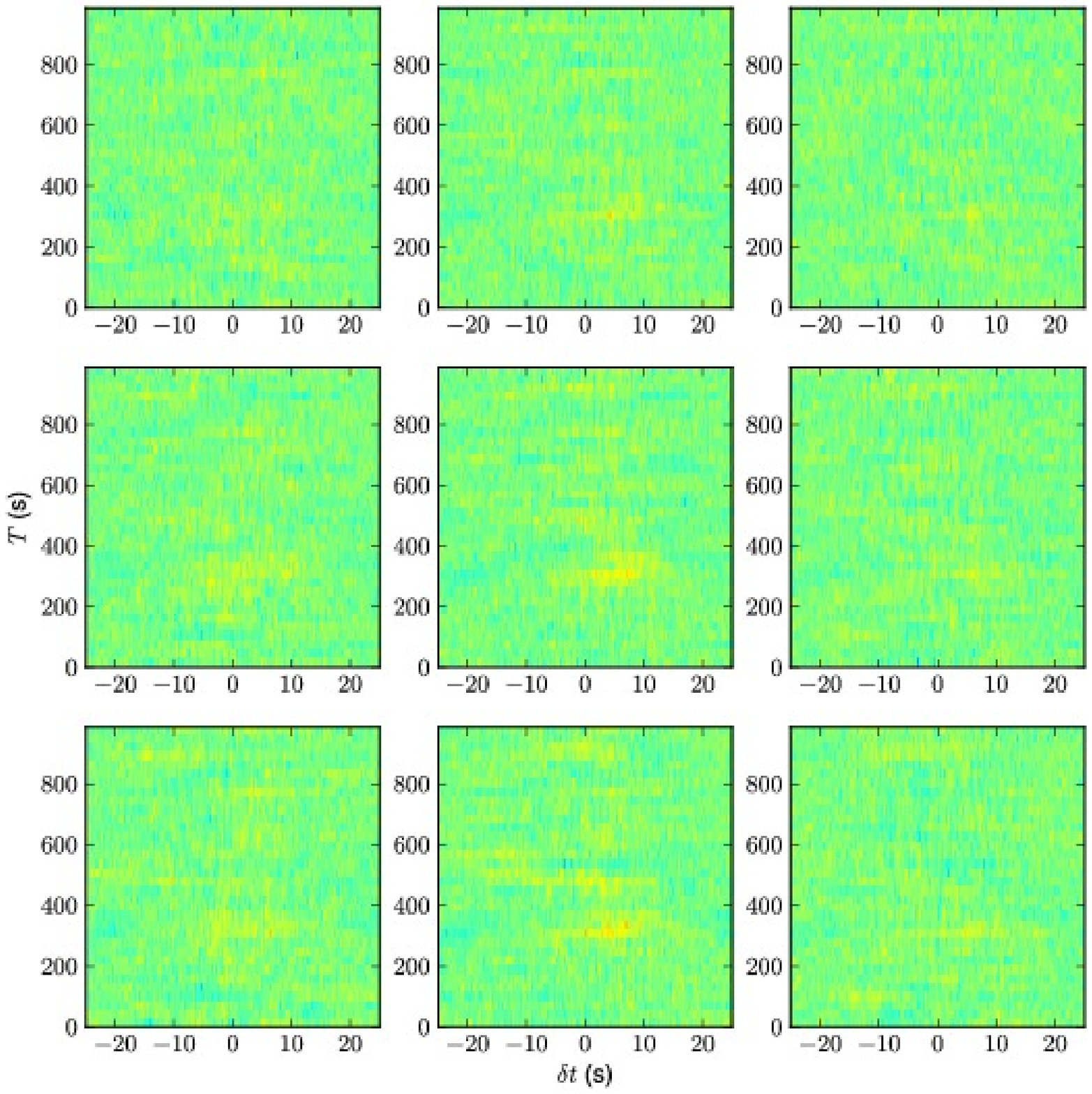}
\caption{\sco, 2004-05-17, 90020-01-01-01}
\end{figure*}

\begin{figure*}
\includegraphics[width=0.9\hsize]{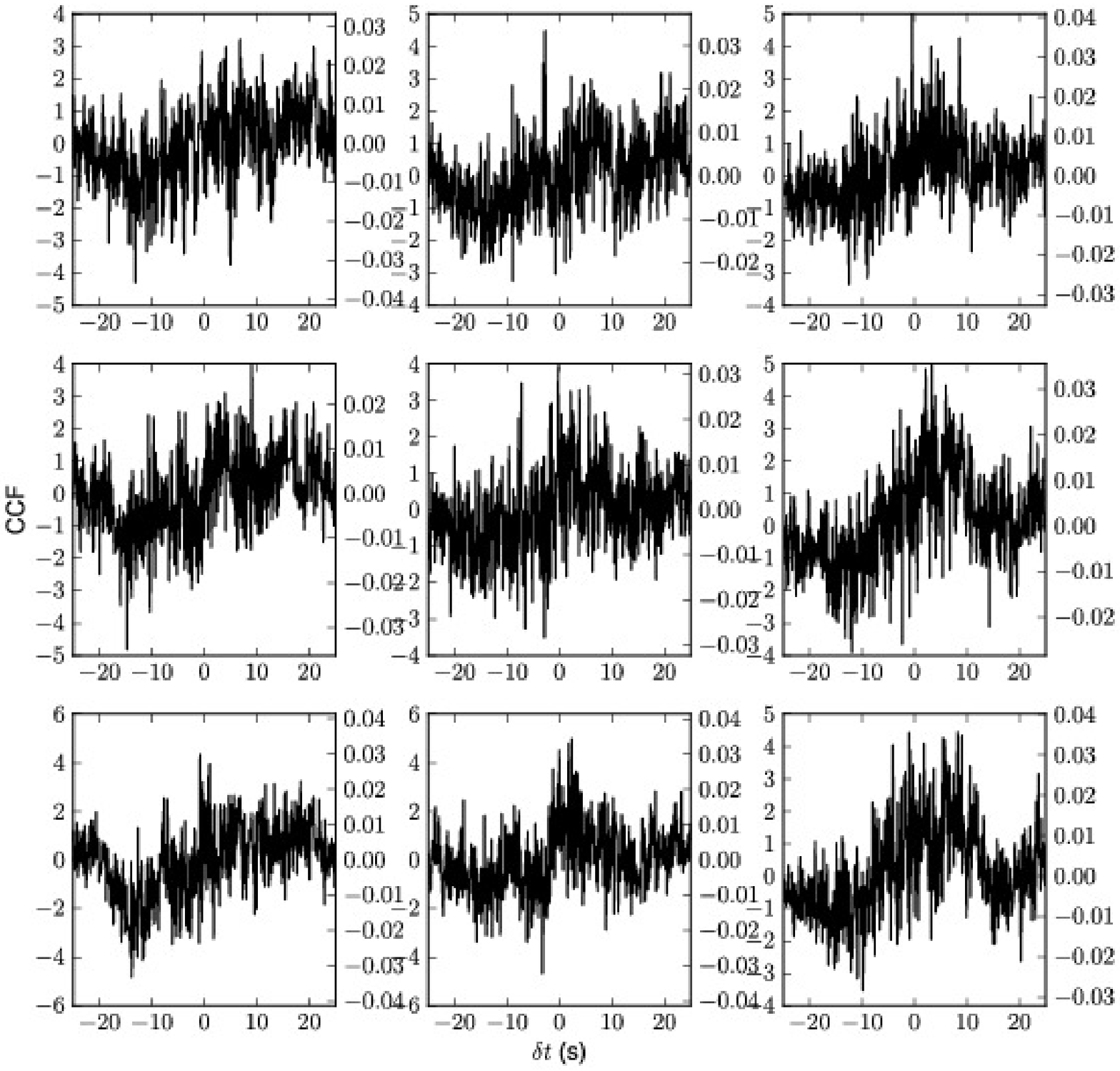}
\caption{\sco, 2004-05-17, 90020-01-01-02}
\end{figure*}

\begin{figure*}
\includegraphics[width=0.9\hsize]{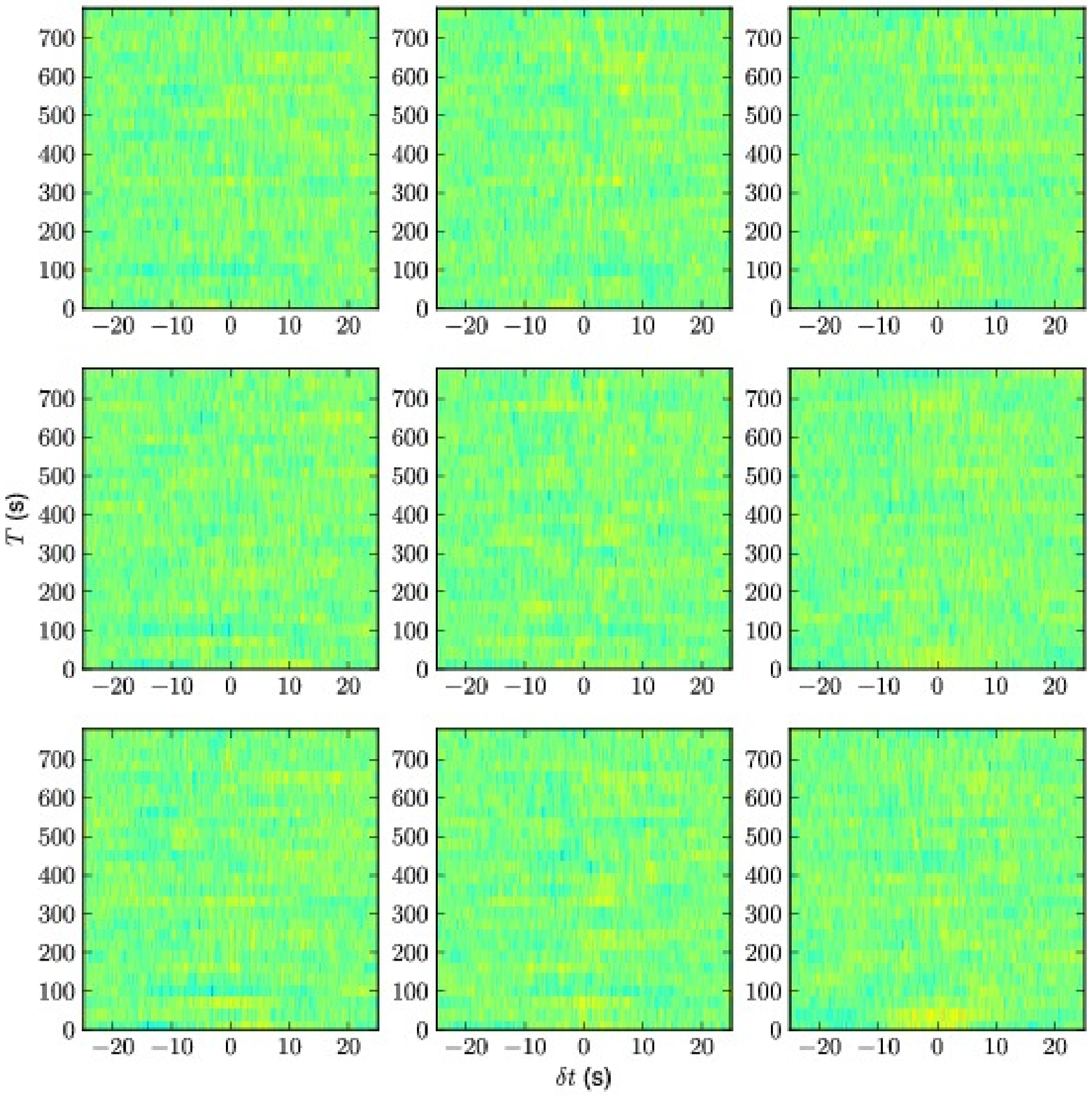}
\caption{\sco, 2004-05-17, 90020-01-01-02}
\end{figure*}

\begin{figure*}
\includegraphics[width=0.9\hsize]{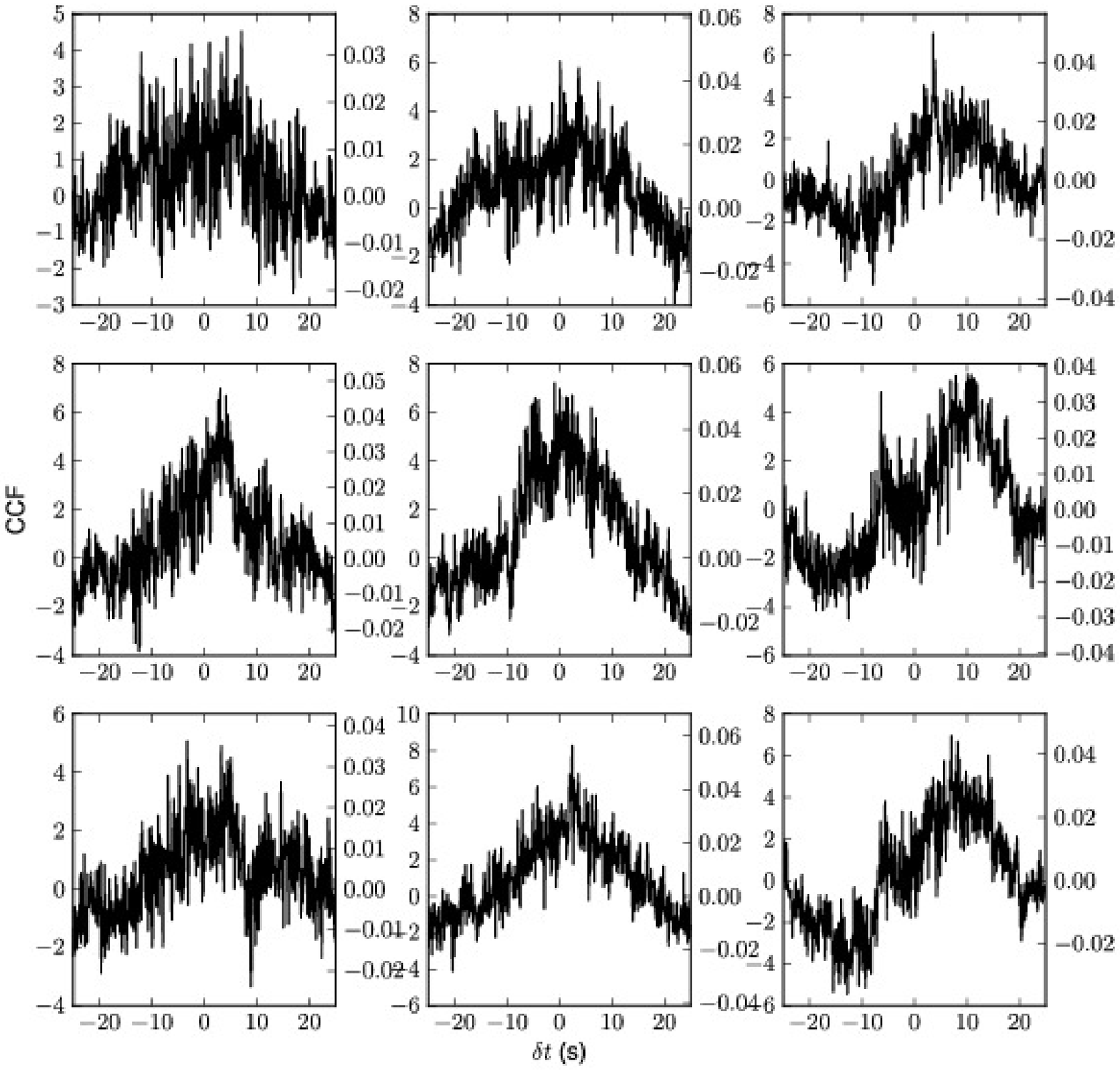}
\caption{\sco, 2004-05-17, 90020-01-01-03}
\end{figure*}

\begin{figure*}
\includegraphics[width=0.9\hsize]{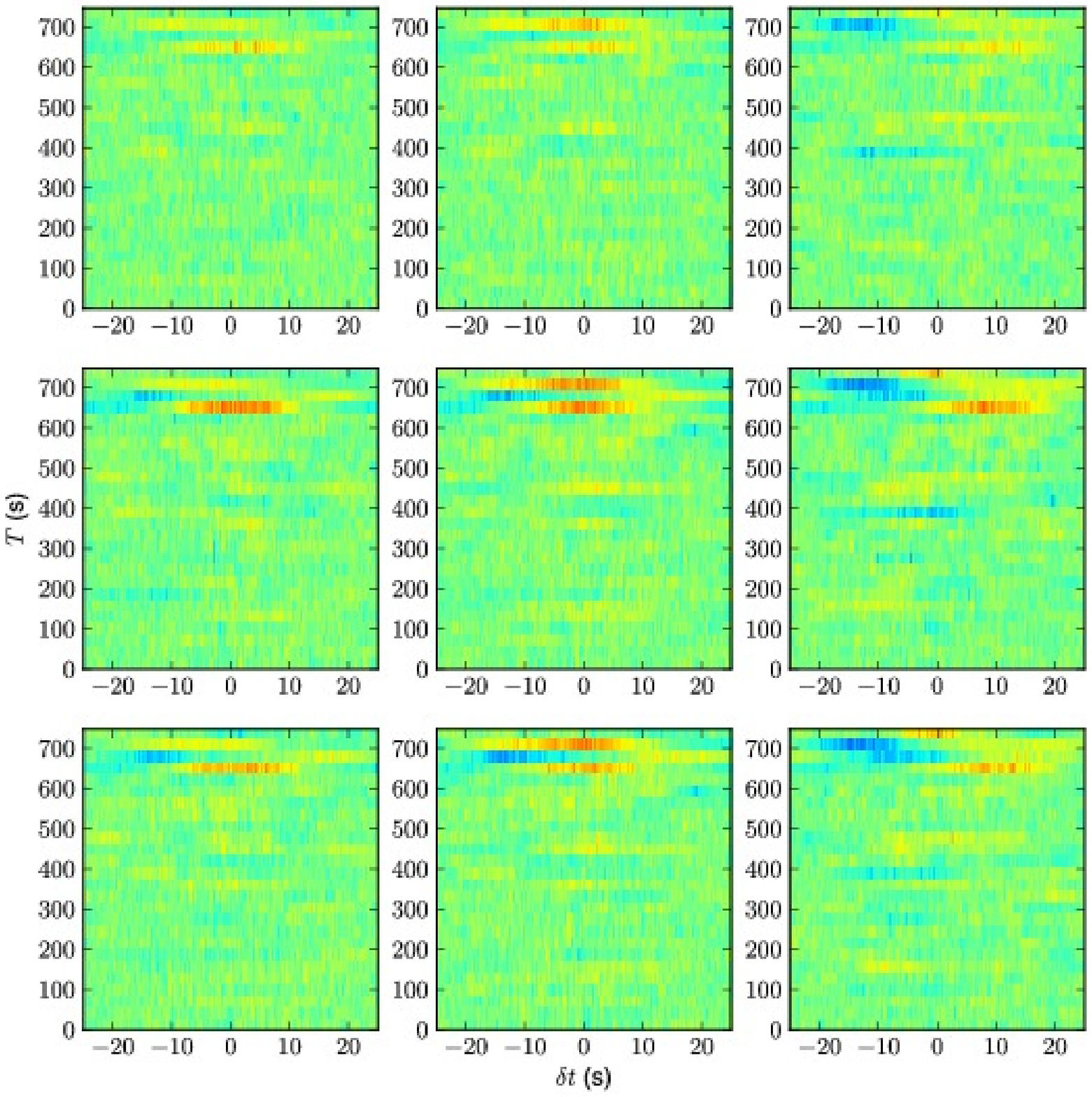}
\caption{\sco, 2004-05-17, 90020-01-01-03}
\end{figure*}

\clearpage
\begin{figure*}
\includegraphics[width=0.9\hsize]{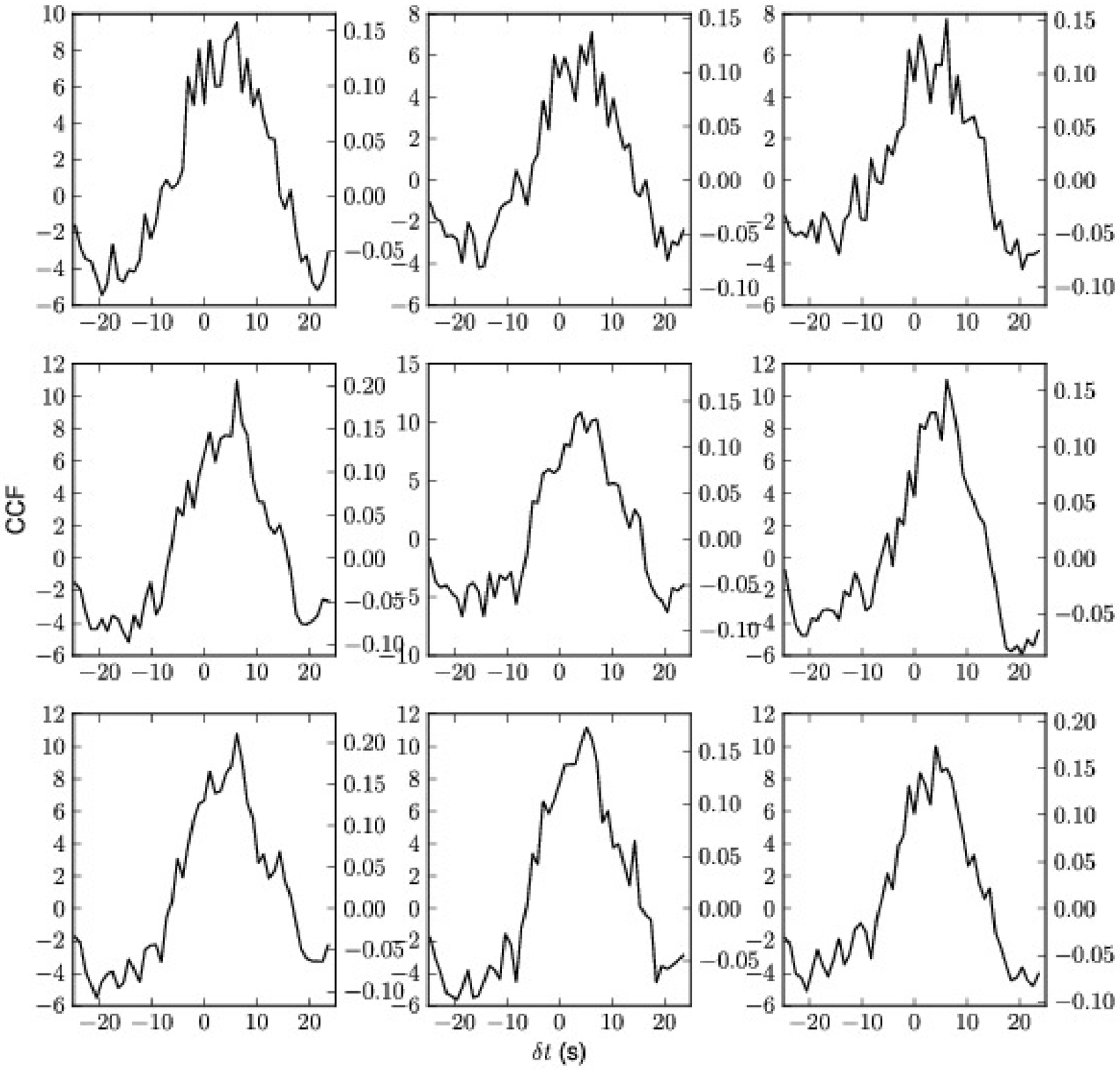}
\caption{\sco, 2004-05-18, 90020-01-02-01; immediately succeeds data in Figure A17.}
\end{figure*}

\begin{figure*}
\includegraphics[width=0.9\hsize]{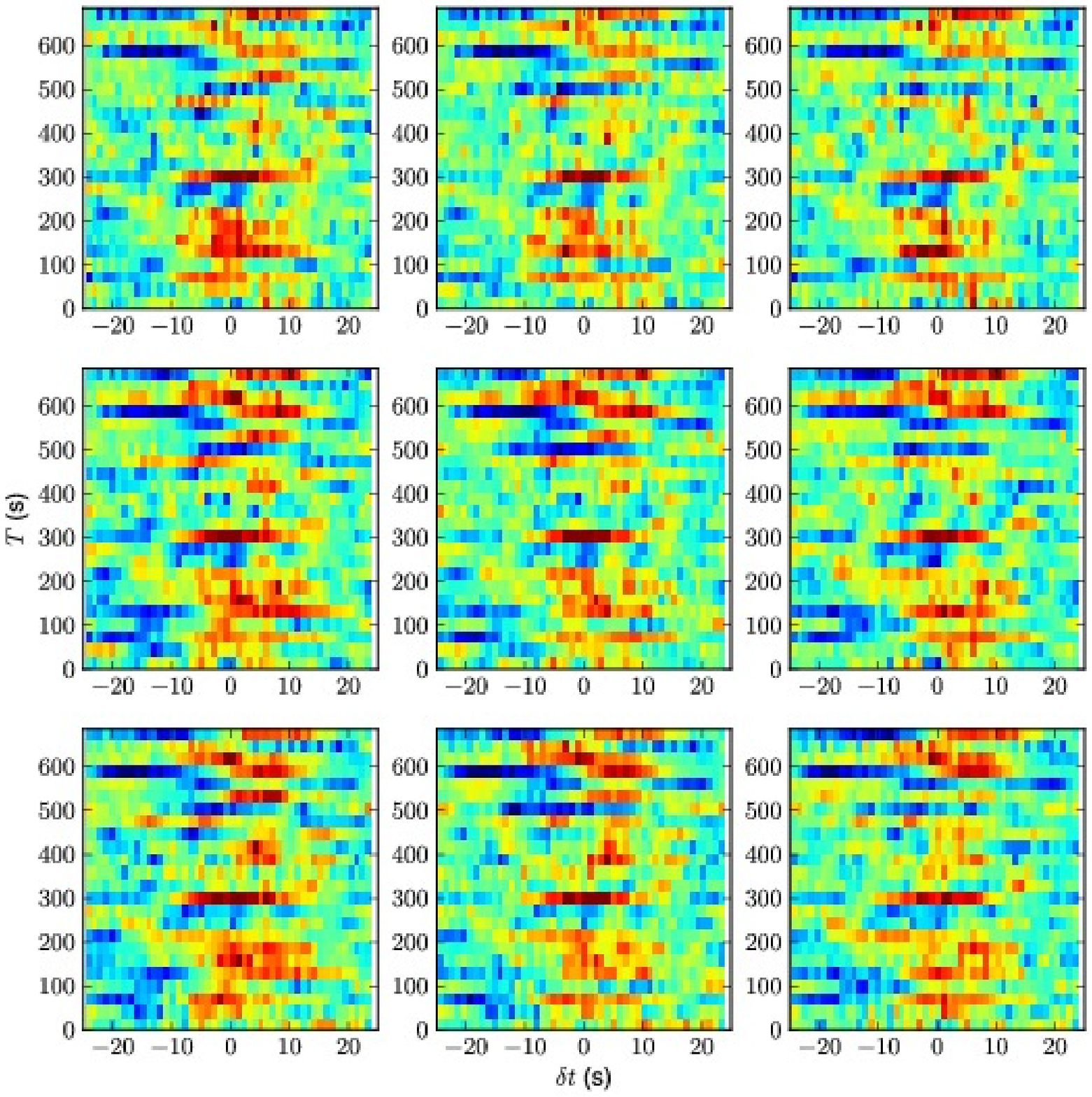}
\caption{\sco, 2004-05-18, 90020-01-02-01; immediately succeeds data in Figure A18.}
\end{figure*}

\begin{figure*}
\includegraphics[width=0.9\hsize]{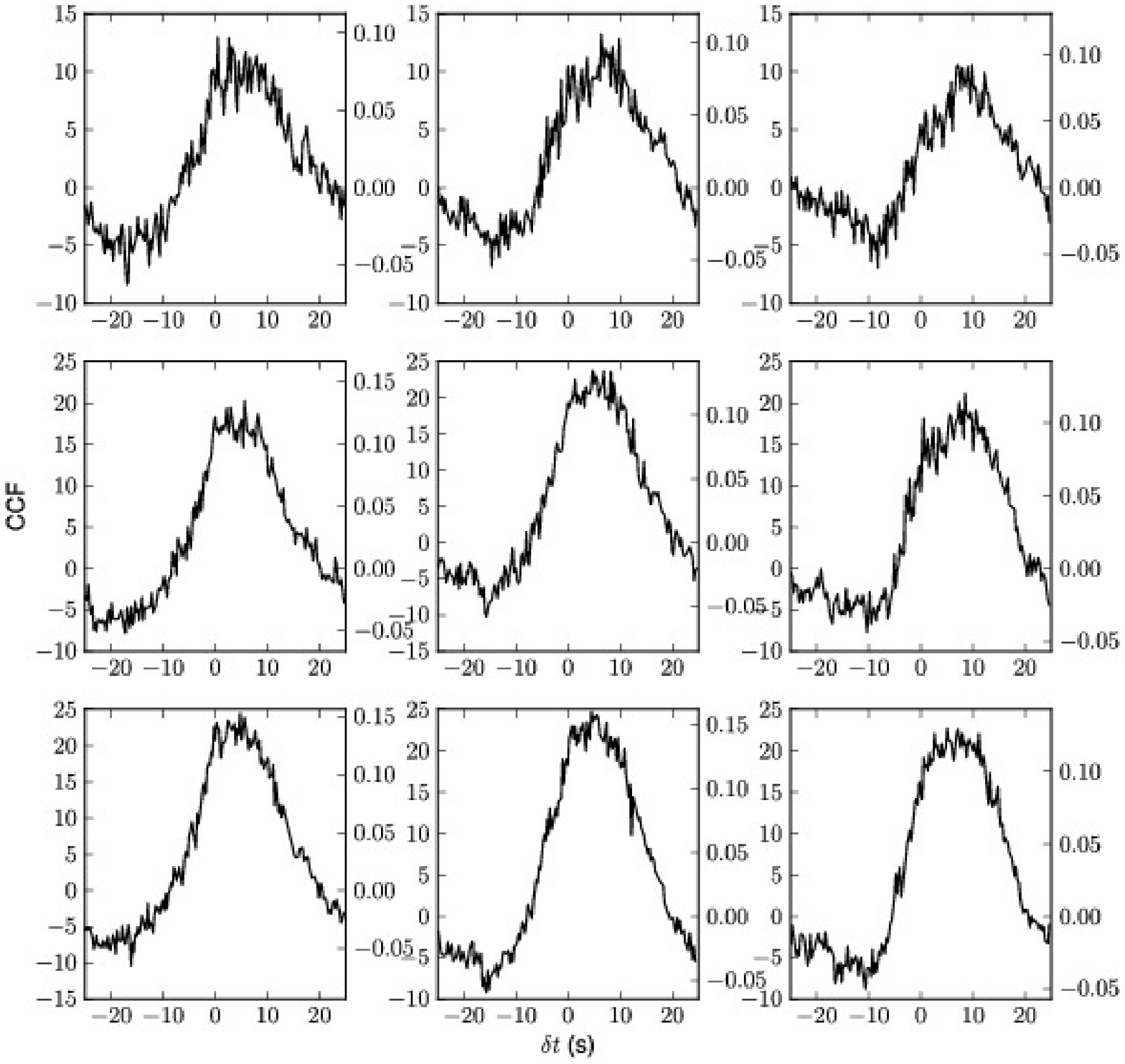}
\caption{\sco, 2004-05-18, 90020-01-02-02}
\end{figure*}

\begin{figure*}
\includegraphics[width=0.9\hsize]{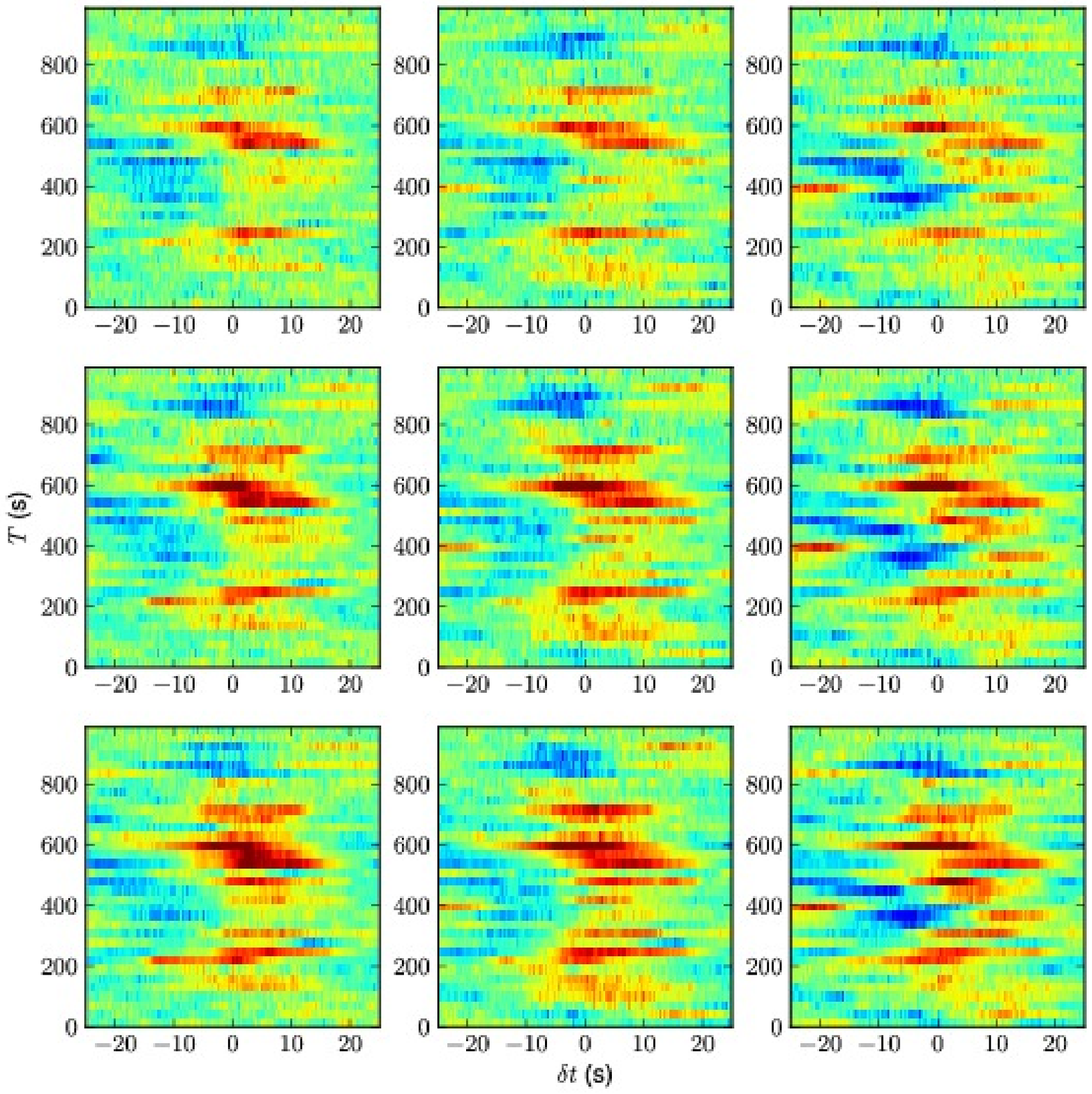}
\caption{\sco, 2004-05-18, 90020-01-02-02}
\end{figure*}

\begin{figure*}
\includegraphics[width=0.9\hsize]{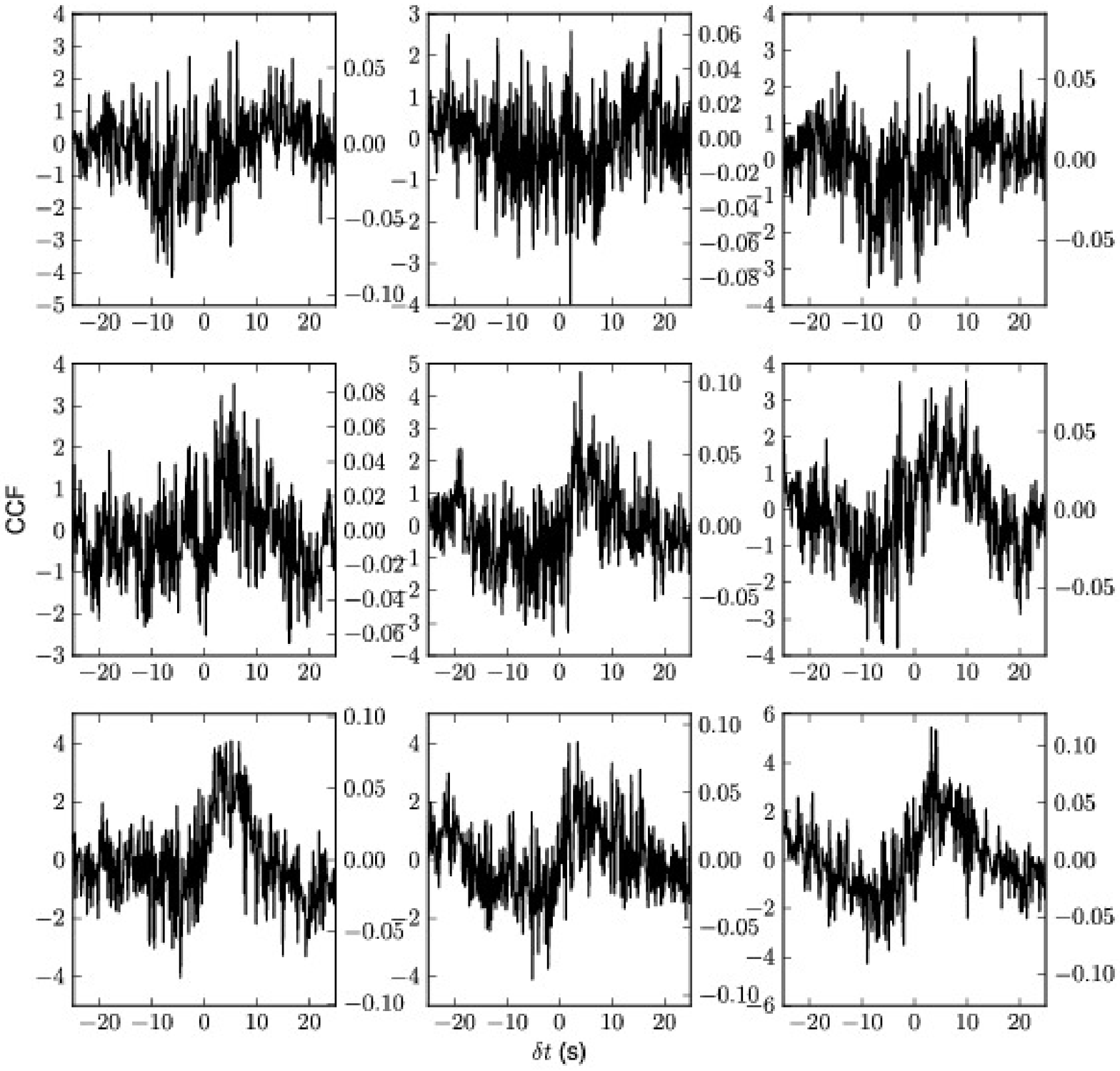}
\caption{\sco, 2004-05-19, 90020-01-03-02}
\end{figure*}

\begin{figure*}
\includegraphics[width=0.9\hsize]{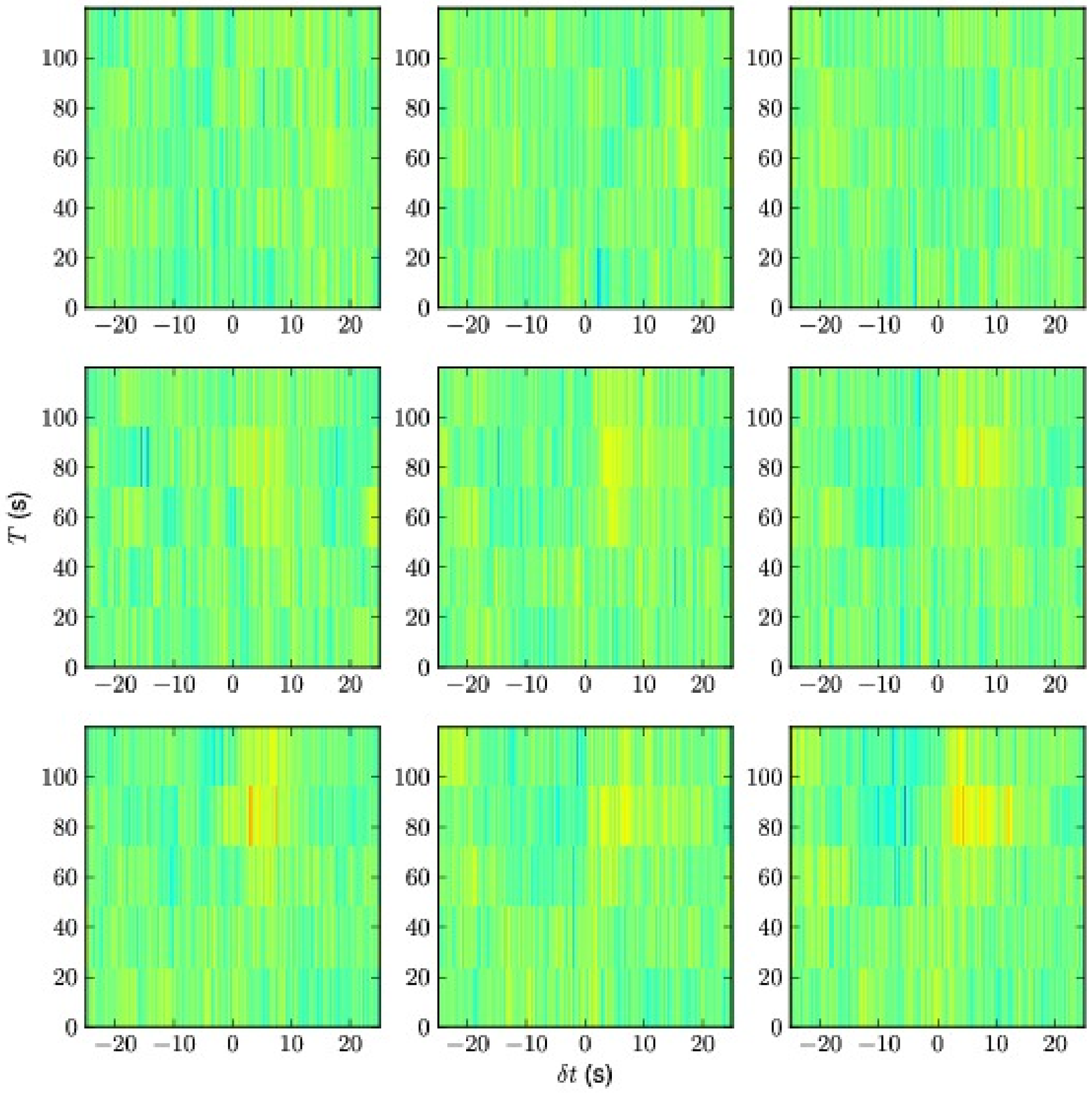}
\caption{\sco, 2004-05-19, 90020-01-03-02}
\end{figure*}

\clearpage
\begin{figure*}
\includegraphics[width=0.9\hsize]{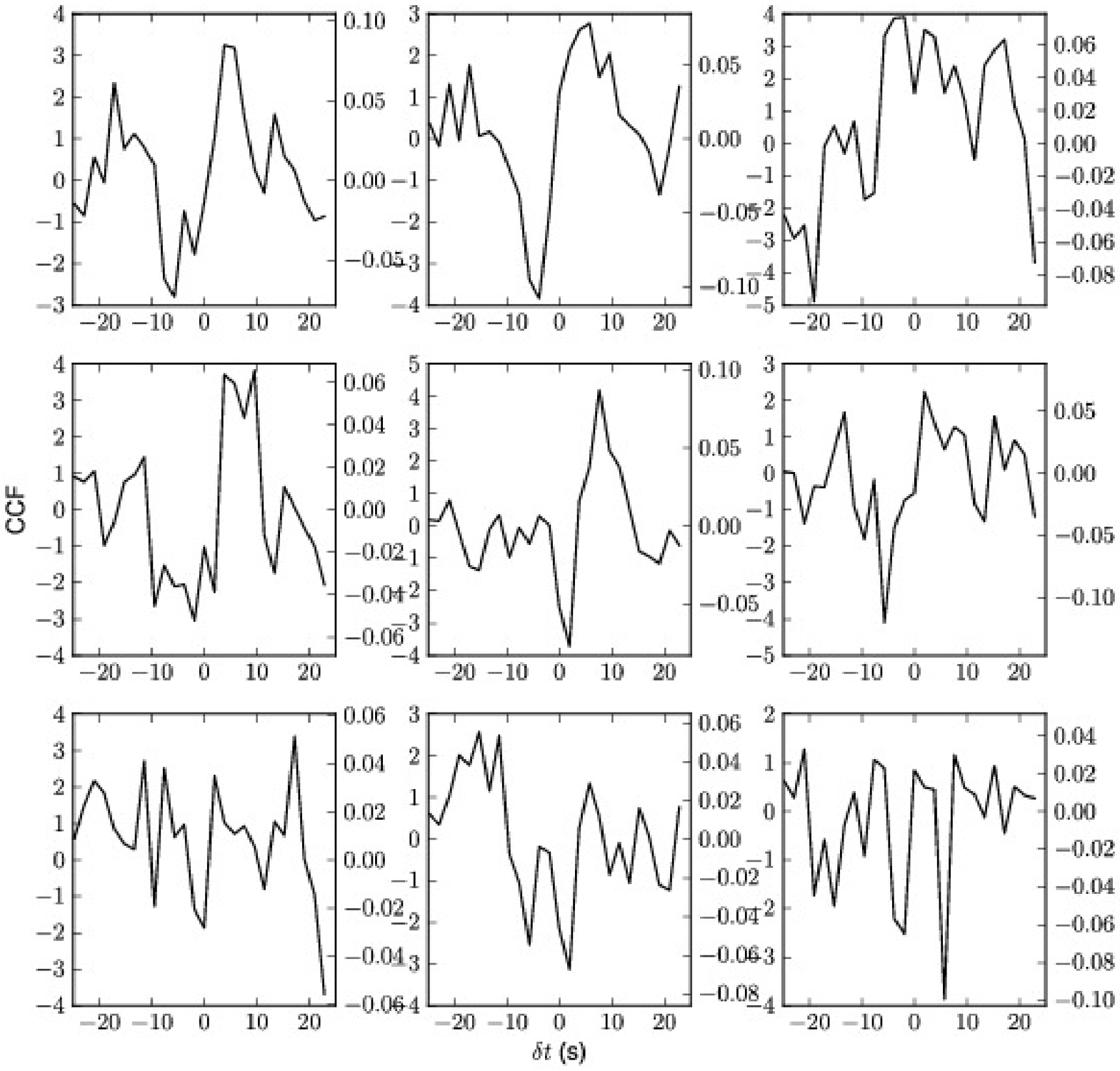}
\caption{\cyg, 2007-10-16, 92038-01-02-00}
\end{figure*}

\begin{figure*}
\includegraphics[width=0.9\hsize]{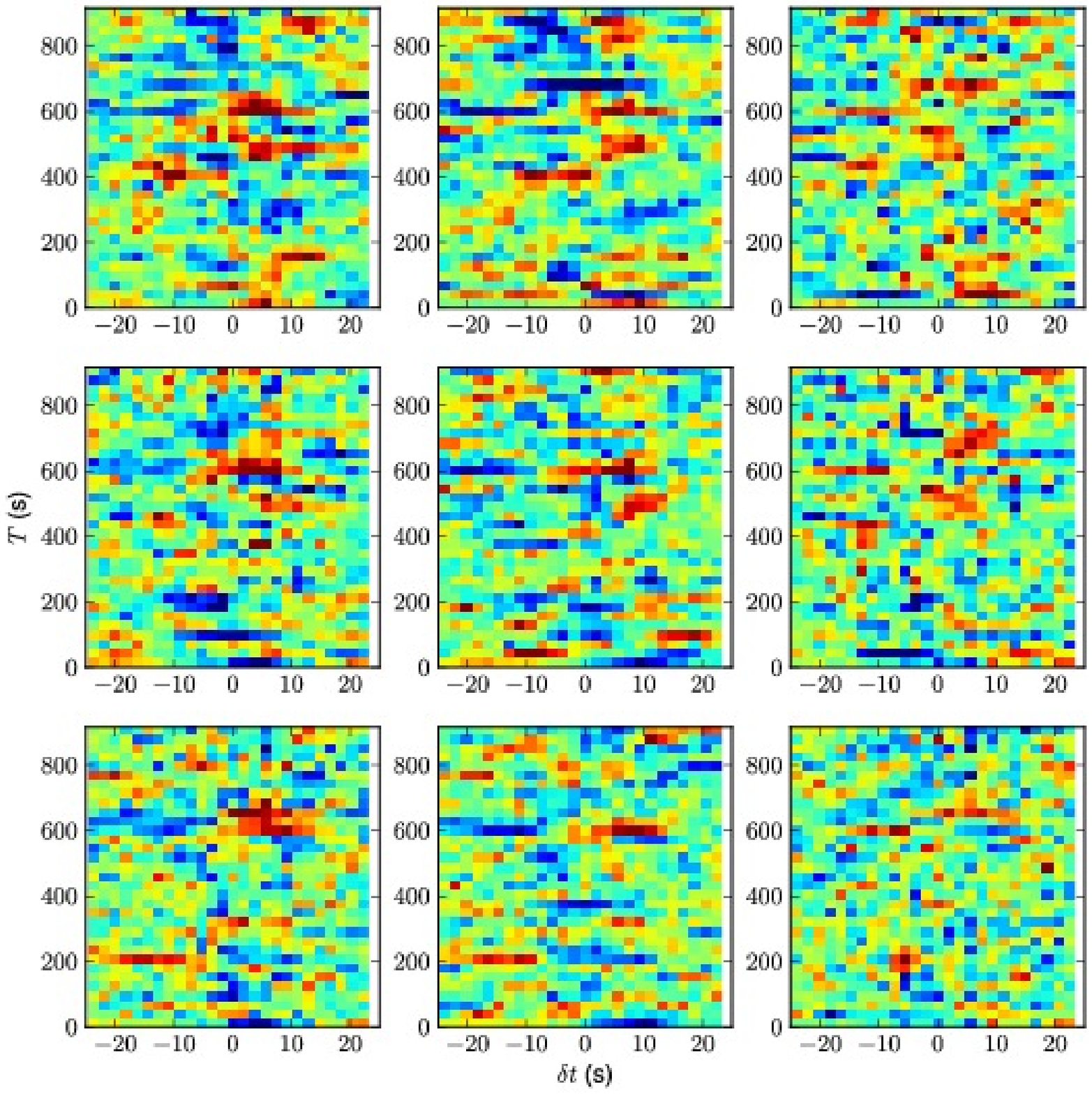}
\caption{\cyg, 2007-10-16, 92038-01-02-00}
\end{figure*}

\begin{figure*}
\includegraphics[width=0.9\hsize]{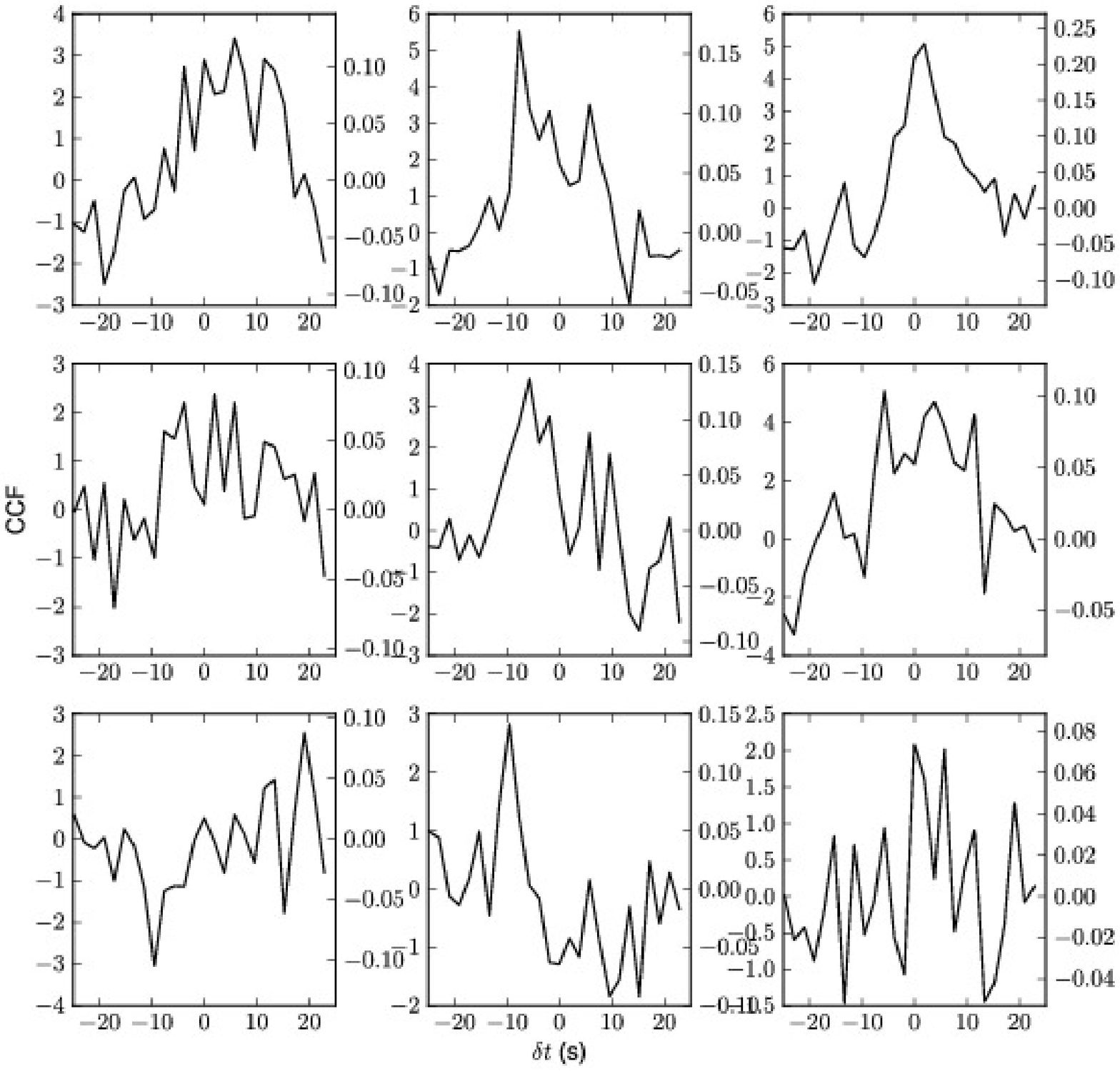}
\caption{\cyg, 2007-10-19, 92038-01-05-03}
\end{figure*}

\begin{figure*}
\includegraphics[width=0.9\hsize]{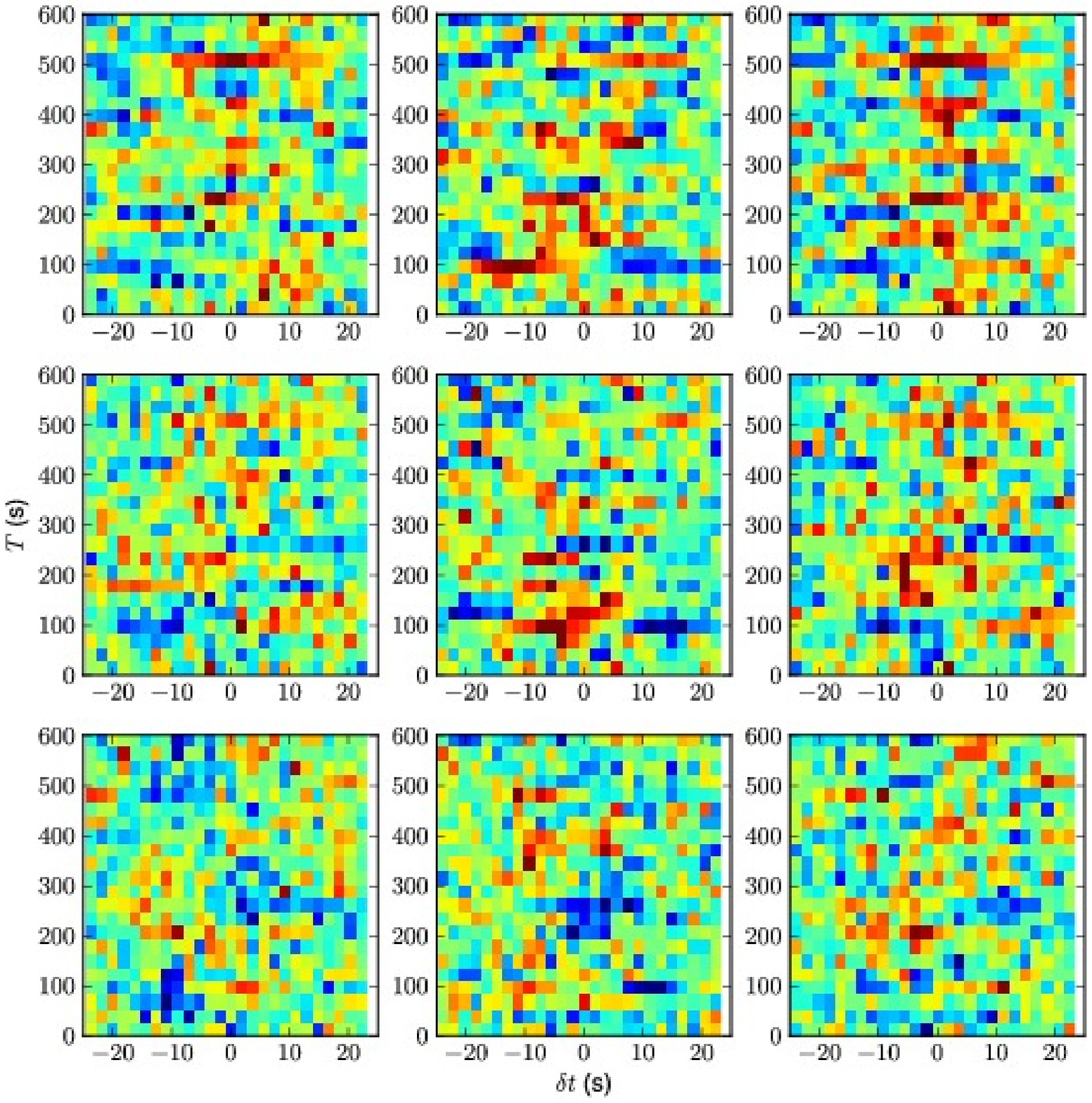}
\caption{\cyg, 2007-10-19, 92038-01-05-03}
\end{figure*}

\begin{figure*}
\includegraphics[width=0.9\hsize]{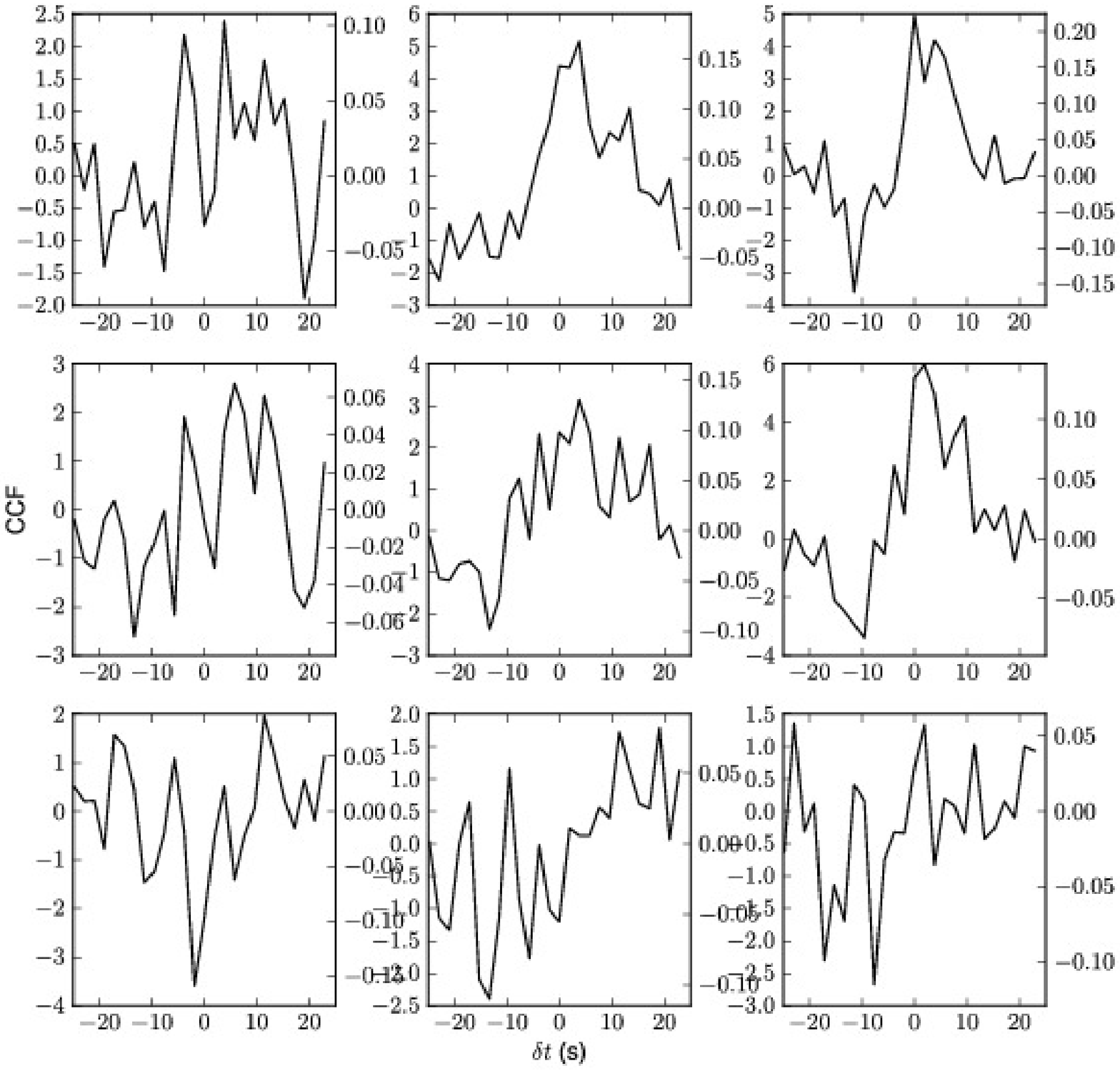}
\caption{\cyg, 2007-10-20, 92038-01-06-00; follows data in Figure A29.}
\end{figure*}

\begin{figure*}
\includegraphics[width=0.9\hsize]{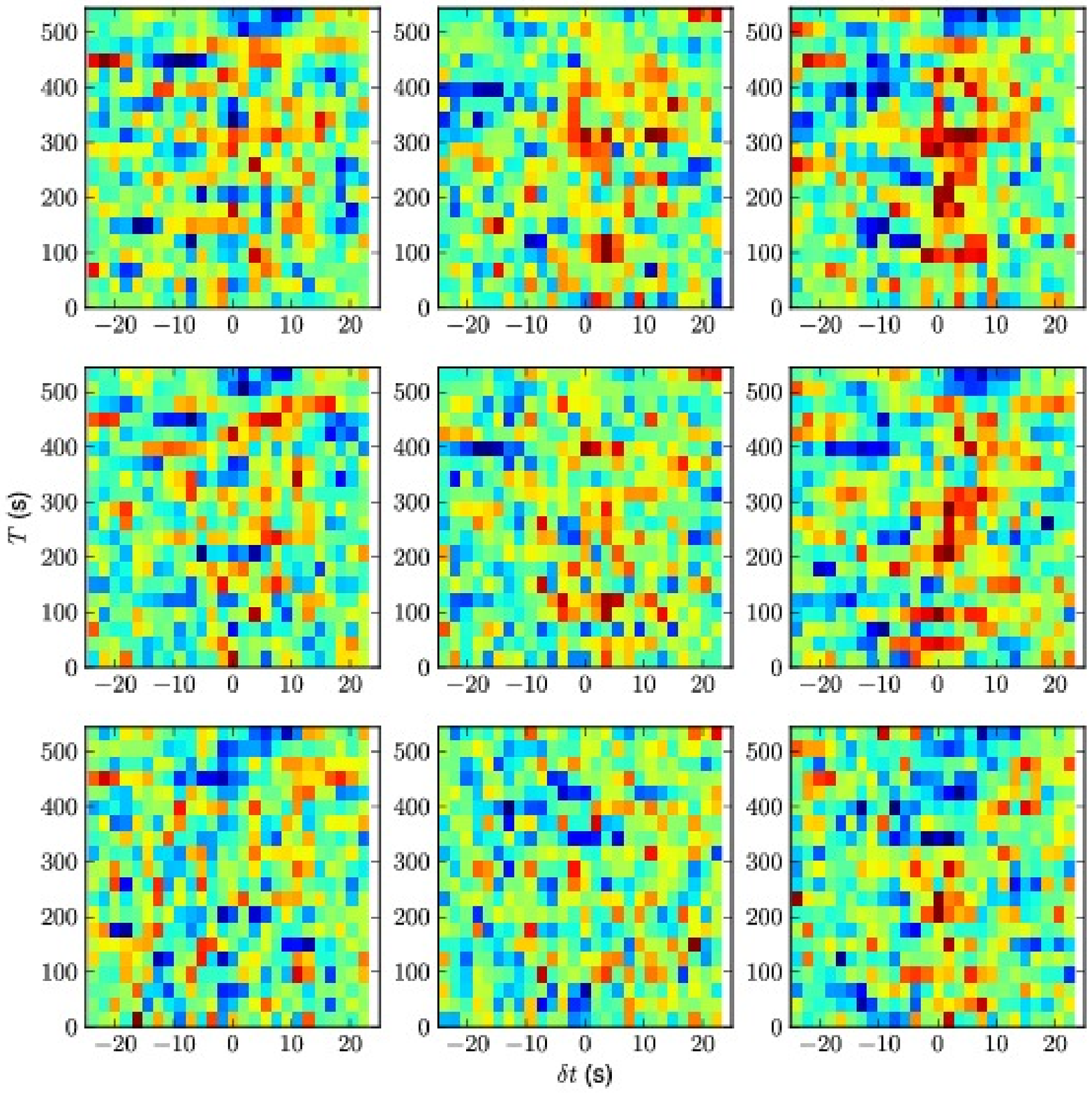}
\caption{\cyg, 2007-10-20, 92038-01-06-00; follows data in Figure A30.}
\end{figure*}

\begin{figure*}
\includegraphics[width=0.9\hsize]{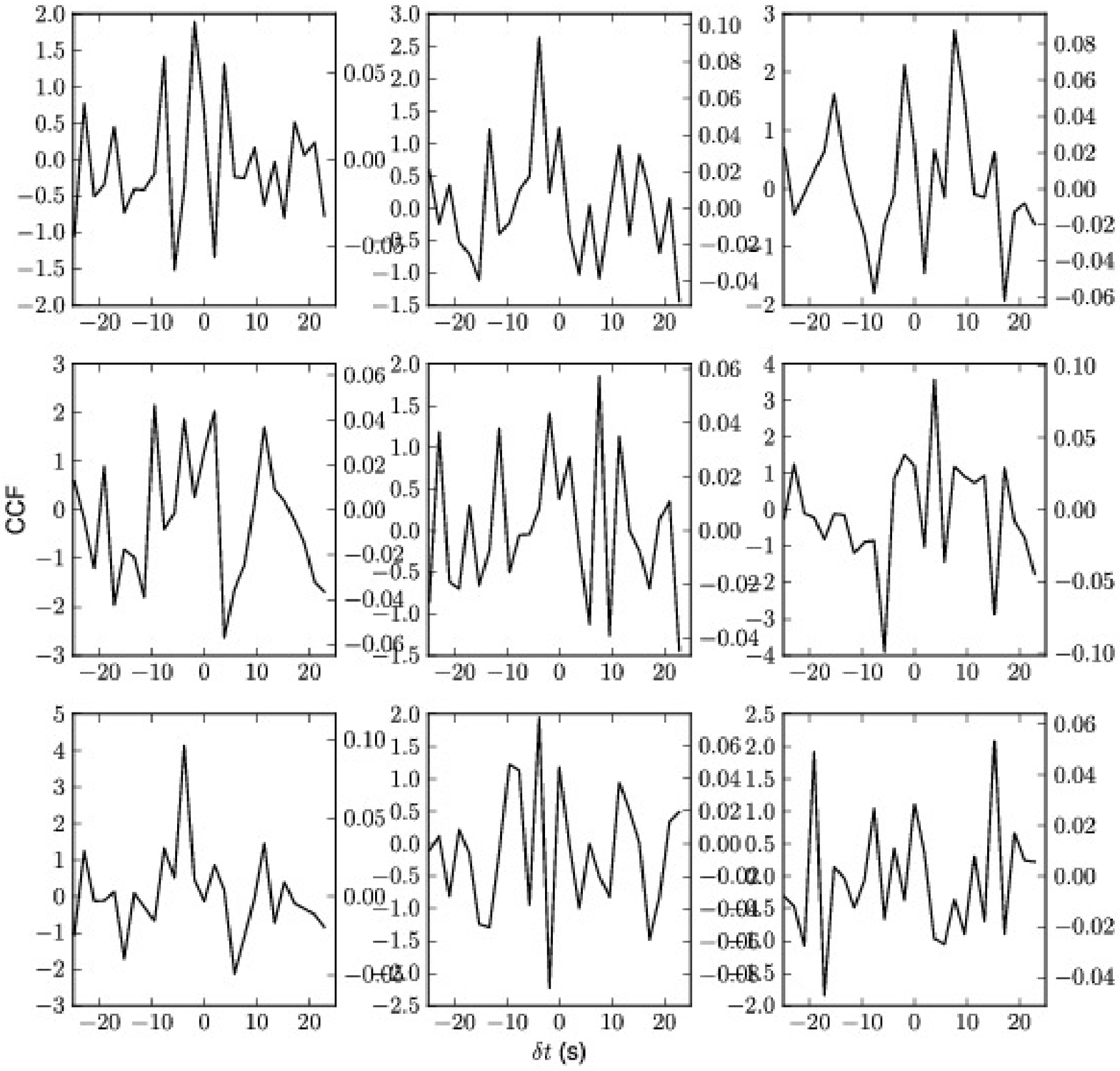}
\caption{\cyg, 2007-10-21, 92038-01-07-01}
\end{figure*}

\clearpage

\clearpage
~
\label{lastpage}

\end{document}